\documentclass[11pt,onecolumn,nofootinbib,aps,superscriptaddress,prd]{revtex4}
\usepackage{graphicx,dcolumn,bm,amsfonts,amsmath,color,xcolor,epsfig}
\usepackage{slashed}
\usepackage{graphics}
\usepackage{tikz}
\usepackage{multirow}
\usepackage{amsmath}
\usepackage{gensymb}
\usepackage{epstopdf}
\usepackage{hyperref}
\usepackage{amssymb}
\allowdisplaybreaks

\begin{document}

\title{Investigation of the mass spectra of singly heavy baryons $\Sigma_{Q}$, $\Xi^{\prime}_{Q}$ and $\Omega_{Q}$ $(Q=c, b)$ in the Regge trajectory model}

\author{Ji-Hai Pan}
\email{Tunmnwnu@outlook.com}
\address{College of Mathematics and Physics, Liuzhou Institute of Technology, Liuzhou 545000, China}
\author{Jisi Pan\footnote{Corresponding author}}
\email{panjisi@gxstnu.edu.cn}
\address{School of Mechanical and Electrical Engineering, Guangxi Science $\&$ Technology Normal University,
Laibin 546199, China}

\begin{abstract}
Very recently, LHCb Collaboration observed that two new $\Omega_{c}^{0}$ states decay into $\Xi^{+}_{c}K^{-}$ with masses of about $3185$ MeV and $3327$ MeV. However, their spin parity quantum numbers $J^{P}$ have not been determined. In this paper, we exploit the quark-diquark model, the linear Regge trajectory and the perturbation treatment method to analyze the mass spectra of the discovered experimental data for the singly heavy baryons $\Sigma_{c}/\Sigma_{b}$, $\Xi^{\prime}_{c}/\Xi^{\prime}_{b}$ and $\Omega_{c}/\Omega_{b}$. In addition, we further predict the mass spectra of several unobserved $\Sigma_{c}/\Sigma_{b}$, $\Xi^{\prime}_{c}/\Xi^{\prime}_{b}$ and $\Omega_{c}/\Omega_{b}$ baryons. In the
case of the $\Omega_c(3185)^{0}$ and $\Omega_c(3327)^{0}$ states, we determine $\Omega_{c}(3185)^{0}$ as $2S$ state and $\Omega_{c}(3327)^{0}$ as $1D$ state with $J^{P}=1/2^{+}$ and $J^{P}=3/2^{+}$, respectively. An overall good agreement of the obtained predictions with available experimental data are found.
\end{abstract}

\maketitle

\section{introduction}

With the discovery of more and more highly excited strongly interacting particles in experiments, such as LHCb, Belle, BaBar, and CLEO, a deeper understanding of the singly heavy baryons has been gained. In the quark-diquark picture, singly heavy baryons are composed of an anti-color triplet ($\overline{3}_c$) diquark with spin one ($S_d=1$), formed by two light quarks, and a heavy quark $(S_{Q}=1/2)$. The latest review of particle physics by PDG can shed new light on the singly heavy baryons $\Sigma_{c}/\Sigma_{b}$, $\Xi^{\prime}_{c}/\Xi^{\prime}_{b}$ and $\Omega_{c}/\Omega_{b}$.

From PDG \cite{Workman:A11} in 2022, the establishment of $S$, $P$ and $D$-wave excited states are gradually improved providing valuable insights into the fundamental structure and behavior for the $\Sigma_{c}/\Sigma_{b}$, $\Xi^{\prime}_{c}/\Xi^{\prime}_{b}$ and $\Omega_{c}/\Omega_{b}$ baryons. In the $\Sigma_{c}/\Sigma_{b}$ baryons, the $\Sigma_{c}(2455)^{0,+,++}$ and $\Sigma_{c}(2520)^{0,+,++}$ states can be well interpreted as $S$-wave charmed baryons with $J^{P}=1/2^{+}$, $J^{P}=3/2^{+}$, respectively. The triplet of the excited $\Sigma_{c}(2800)^{0,+,++}$ states decaying to $\Lambda_{c}^{+} \pi$ were observed by Belle Collaboration in 2005 \cite{Mizuke:PPP888}. The four ground states $\Sigma_b(5815)^{-+}$ and $\Sigma^*_b(5835)^{-+}$ of $\Sigma_b$ have been observed by the CDF Collaboration in Ref. \cite{collaborationT:PPP888} with $J^{P}=1/2^{+}$, $J^{P}=3/2^{+}$, respectively. In the $\Xi^{\prime}_{c}/\Xi^{\prime}_{b}$ baryons, the neutral state $\Xi^{0}_{c}$ and its charged partner $\Xi_{c}(2645)^{+}$ were reported by CLEO in the decay channels $\Xi^{+}_{c}\pi^{-}$ \cite{Averyet:PPP888} and $\Xi^{0}_{c}\pi^{+}$ \cite{Gibbonset:PPP888} as $S$-wave states with $J^{P}=1/2^{+}$, $J^{P}=3/2^{+}$, respectively. However, the $J^{P}$ of $\Xi_{c}(2923)$ \cite{Chistov:PPP888} and $\Xi_{c}(2930)^{+}$ \cite{Aubert:PPP888}, which are good candidates for $P$-wave states, are yet to be determined. Similarly, LHCb observed two new charged  $\Xi^{\prime}_{b}(5935)^{-}$ and $\Xi^{\ast}_{b}(5955)^{-}$ states of the $\Xi^{\prime}_{b}$ baryons in Ref. \cite{Aaijolla:PPP888}. They were proposed to be the $J^{P}=1/2^{+}$, $J^{P}=3/2^{+}$ ground states. Note that the $\Xi^{\prime}_{b}$ baryon has only one neutral state $\Xi_{b}(5945)^{0}$ with $J^{P}=3/2^{+}$ based on quark model expectations. Therefore, the discovery of these singly heavy baryons have great significance for research.

In the study of $\Omega_{c}$, only two ground states $\Omega_{c}^{0}$ and $\Omega_{c}(2770)^{0}$ have been discovered experimentally with $J^{P}=1/2^{+}$, $J^{P}=3/2^{+}$, respectively. In 2017, the LHCb Collaboration reported five new narrow excited states of $\Omega_{c}$ in the decay channel $\Xi^{+}_{c}K^{-}$ \cite{Aaij:A11} which later confirmed by Belle \cite{Yelton:A11} with interesting spin-parity properties and inner structures. For a discussion of the excited
$\Omega_c$ states we refer to Refs. \cite{EFG:C10, EFG:A12, MI:A11, RP:A11, GV:A11, MMP:A11, VGVT:A11, YOHH:A11, PCB:A11, ShahK:PP888, YHHHS:A11, CCCHLZ:A11} or recent explorations given in Refs. \cite{PM:A11, CL:A11, WZ:A11, ZGW:A11, KarlinerR:11}. In 2020, the LHCb Collaboration reported the discovery of four narrow excited $\Omega_{b}$ states in the decay channel $\Xi^{0}_{b}K^{-}$ \cite{Aaij:A12}. In Refs. \cite{KarlinerR:12, JiaPan:PP888}, the authors used the constituent quark model to obtain masses compatible with the experiment. Very recently, LHCb observed two new narrow $\Omega_{c}$ states decaying into $\Xi^{+}_{c}K^{-}$ \cite{Aaij:A13} with masses of $\Omega_{c}(3185)^{0}$ and $\Omega_{c}(3327)^{0}$ about $3185$ MeV and $3327$ MeV. The value of $J^P$ for the newly discovered states remains unclear.

In this paper, we study the mass spectra of the singly heavy baryons $\Sigma_{c}/\Sigma_{b}$, $\Xi^{\prime}_{c}/\Xi^{\prime}_{b}$ and $\Omega_{c}/\Omega_{b}$ from the Regge trajectory and the spin-dependent potential. By analyzing the Regge trajectory formula, we get the spin-average masses of the baryons. In addition, to obtain the mass shifts, we exploit new scaling relations to calculate the spin coupling parameters. In the end, the properties of the charmed baryons and the bottom baryons will be discussed.

This paper is organized as follows. We analyze the Regge trajectory formula to give the spin-average mass $\bar M$ of excited states of the $\Sigma_{c}/\Sigma_{b}$, $\Xi^{\prime}_{c}/\Xi^{\prime}_{b}$ and $\Omega_{c}/\Omega_{b}$ baryons in Sec. II. In Sec. III, we review about the spin-dependent Hamiltonian and the scaling relations. We calculate the mass spectra of the $\Omega_{c}/\Omega_{b}$ baryons in Sec. IV. In Sec. V, we discuss the mass spectra of the $\Sigma_{c}/\Sigma_{b}$ baryons. In Sec. VI, a similar mass analysis is given for the $\Xi^{\prime}_{c}/\Xi^{\prime}_{b}$ baryons. Finally, we outline our conclusion in Section VII.

\section{The Regge trajectory and the spin-average masses}\label{Sec.II}

In the QCD rotating string model \cite{Susskind:S11, Nambu:N11}, the strong interaction binds the heavy and light quark inside the hadron, where one end of the string is a heavy quark and the other is a light antiquark or light diquark moving around the heavy quark. Based on this model, it is interesting to investigate the Regge trajectory behavior of the
hadronic system.

For the orbital excitations of the baryons, we obtain the spin-average mass $\bar M$ and angular momentum $L$ following the equations given by Refs. \cite{LaCourse:A13, ChenWei:A13}
\begin{equation}
\bar M=\frac{m_{\text{cur}Q}}{\sqrt{1-{v_{Q}^{2}}}}+\frac{\alpha}{\omega}\int_{0}^{v_{Q}}\frac{d u}{\sqrt{1-{u}^{2}}}+\frac{m_{\text{cur} d}}{\sqrt{1-{v_{d}^{2}}}}+\frac{\alpha}{\omega}\int_{0}^{v_{d}}\frac{d u}{\sqrt{1-{u}^{2}}} , \label{ppu1}
\end{equation}
\begin{equation}
L=\frac{m_{\text{cur}Q}v_{Q}^{2}}{\sqrt{1-{{v_{Q}^{2}}}}}+\frac{\alpha}{\omega^{2}}\int_{0}^{v_{Q}}\frac{u^{2}d u}{\sqrt{1-{u}^{2}}}+\frac{m_{\text{cur}d}v_{d}^{2}}{\sqrt{1-{{v_{d}^{2}}}}}+\frac{\alpha}{\omega^{2}}\int_{0}^{v_{d}}\frac{u^{2}d u}{\sqrt{1-{u}^{2}}} , \label{ppu2}
\end{equation}
where $\alpha$ is the QCD string tension coefficient, and $v_{Q}$, $v_{d}$ the velocity of the string end tied to between the heavy quark $Q$ and light diquark $d$. We define the velocity $v_{i}=\omega r_{i}$ $(i = Q, d)$, where $\omega$ and $r_{i}$ are the angular velocity and the position from the centre of mass, respectively. For simplicity, we have chosen the velocity of light c = 1. The light diquark is ultrarelativistic, we take the velocity of light diquark $v_{d}\approx 1$ for approximation. Then $m_{\text{cur}Q}$ and $m_{\text{cur}d}$ can be regarded as current mass of the heavy quark and light diquark, respectively. Including relativistic effects, one can obtain the constituent quark masses
\begin{equation}
M_{Q}=\frac{m_{\text{cur}Q}}{\sqrt{1-v_{Q}^{2}}} , m_{d}=\frac{m_{\text{cur}d}}{\sqrt{1-v_{d}^{2}}}. \label{ppuu1}
\end{equation}
Eqs. (\ref{ppu1}) and (\ref{ppu2}) can be integrated to give
\begin{equation}
\bar M=M_{Q}+m_{d}+M_{Q}v_{Q}^{2}+\frac{\pi \alpha}{2\omega} , \label{pppuuu1}
\end{equation}
\begin{equation}
L=\frac{1}{\omega}(m_{d}+M_{Q}v_{Q}^{2}+\frac{\pi \alpha}{4\omega}), \label{pppuuu2}
\end{equation}
where for the string ending at the heavy quark we use the boundary condition
\begin{equation}
\frac{\alpha}{\omega}=\frac{m_{\text{cur}Q}v_{Q}}{1-v^{2}_{Q}}\approx M_{Q}v_{Q}. \label{pppu1}
\end{equation}
Substituting Eq. (\ref{pppu1}) into Eqs. (\ref{pppuuu1}) and (\ref{pppuuu2}) eliminating the angular velocity $\omega$ gives the spin-averaged mass formula \cite{JiaD:PP888, JiaDH:PP888, ChenDong:PP888, Jia:PP88} for the orbital excited states,
\begin{equation}
(\bar M-M_{Q})^{2}=\pi \alpha L+a_{0}, \label{ppp7uu1}
\end{equation}
here, the intercept factor $a_{0}=(m_{d}+M_{Q}v_{Q}^{2})^{2}$ depends on the diquark mass $m_{d}$ and the non-relativistic 3-kinematic energy $M_{Q}v_{Q}^{2}=P^{2}_{Q}/M_{Q}$ for the heavy quark. Note that the non-relativistic kinematic 3-momentum $P_{Q}$ is conserved in the heavy quark limit, which has been associated with both $M_{Q}$ and $v_{Q}$. Using a variant of Eq. (\ref{ppuu1}), the velocity $v_{Q}$ is
\begin{eqnarray}
 v_{Q}=\left(1-\frac{m_{\text{cur}Q}^{2}}{M_{Q}^{2}}\right)^{\frac{1}{2}}, \label{pp3}
\end{eqnarray}
and the spin-averaged mass formula (\ref{ppp7uu1}) becomes
\begin{equation}
\bar M=M_{Q}+\sqrt{\alpha\pi L+\left( m_{d}+M_{Q}\left( 1-\frac{m_{\text{cur}Q}^{2}}{M_{Q}^{2}}\right) \right) ^{2}}\text{.} \label{pp421}
\end{equation}
Here, $M_{Q}$ and $m_{d}$ are the constituent masses of the heavy quark and the diquark, respectively. $L$ is the orbital angular momentum of the baryon systems ($L=0, 1, 2, \cdot\cdot\cdot$). Accordingly, the current masses, the constituent quark masses and the string tension are applied in Eq. (\ref{pp421}) as listed in Table \ref{tab:Eff-mass}, which were previously determined in Refs. \cite{Jia:PP88, JiaPan:PP888} via matching the measured mass spectra of the singly heavy baryons.

\renewcommand{\tabcolsep}{0.08cm}
\renewcommand{\arraystretch}{1.0}
\begin{table}[tbh]
\caption{ The current masses and the constituent quark masses (in GeV) of the quark and the string
tensions $\alpha$ (in GeV$^{2}$) of the  singly heavy baryons. }%
\label{tab:Eff-mass}
\begin{tabular}
[c]{ccccccccccccccc}\hline\hline
\text{Parameters} & \text{$M_{c}$} & \text{$M_{b}$} & \text{$m_{\text{cur} c}$} & \text{$m_{\text{cur} b}$} & \text{$m_{nn}$} &\text{$m_{ns}$}&\text{$m_{ss}$} & \text{$\alpha(cnn)$} & \text{$\alpha({cns})$} & \text{$\alpha({css})$} & \text{$\alpha({bnn})$} &\text{$\alpha({bns)}$} & \text{$\alpha({bss})$}\\\hline
Input & $1.44$ & $4.48$ & $1.275$ & $4.18$ & $0.745$ & $0.872$ & $0.991$ & $0.212$ & $0.255$ & $0.316$ & $0.246$ & $0.307$  & $0.318$    \\\hline\hline
\end{tabular}
\end{table}

To obtain the spin-average masses of the orbital and radial excited states $\Sigma_{c}/\Sigma_{b}$, $\Xi^{\prime}_{c}/\Xi^{\prime}_{b}$ and $\Omega_{c}/\Omega_{b}$, we re-examine the Regge-like mass relation Eq. (\ref{pp421}). By an analysis of the experimental data given by PDG \cite{Workman:A11} we suggest that the slope ratio of the Regge trajectory between the radial and angular momentum is 1.37 : 1. Accordingly, $\pi\alpha L$ in Eq. (\ref{pp421}) is replaced by $\pi\alpha(L+1.37n)$,
\begin{equation}
\bar M=M_{Q}+\sqrt{\alpha\pi(L+1.37 n)+\left( m_{d}+M_{Q}\left( 1-\frac{m_{\text{cur}Q}^{2}}{M_{Q}^{2}}\right) \right) ^{2}}\text{,}  \label{pp4}
\end{equation}
where $n$ is a radial quantum number ($n=0, 1, 2, \cdot\cdot\cdot$). We use Eq. (\ref{pp4}) to calculate the spin-average masses of the $\Sigma_{c}/\Sigma_{b}$, $\Xi^{\prime}_{c}/\Xi^{\prime}_{b}$ and $\Omega_{c}/\Omega_{b}$ baryons. The results are listed in Table \ref{tab:Eff-mass11}.

Accordingly, the squared mass difference $(M-\bar M)^{2}$ of the heavy-light hadronic system is related to $L$ and $n$ by
\begin{equation}
(M-\bar M)^{2}=\alpha\pi(L+1.37 n)+\left( m_{d}+M_{Q}\left( 1-\frac{m_{\text{cur}Q}^{2}}{M^{2}_{Q}}\right)\right)^{2}.
\end{equation}
The squared mass difference $(M-\bar M)^{2}$ for the charm baryons is calculated and plotted against $L$ in Fig. 1, 3, 5 with $n = 0$, $1$, $2$, $3$ and $4$. Similarly, the results of the bottom baryons are shown in Fig. 2, 4, 6. The (red) solid circles correspond to the observed (mean) masses and the empty circles indicate the predicted value in Fig. 1-6. It can be seen that $(M-\bar M)^{2}$ increases with both $L$ and $n$.
\renewcommand\tabcolsep{0.6cm}
\renewcommand{\arraystretch}{1.2}
\begin{table*}[!htbp]
\caption{Spin-average masses (MeV) of the $\Sigma_{Q}$, $\Xi^{\prime}_{Q}$ and $\Omega_{Q}$ $(Q=c, b)$ baryons predicted by Eq. (\ref{pp4}).   \label{Table:4}} \label{tab:Eff-mass11}
\resizebox{\textwidth}{12cm}{\begin{tabular}{cccccc}
\hline\hline
State(\text{MeV}) & $\bar{M}(L=0)$ &$ \bar{M}(L=1)$ &$ \bar{M}(L=2) $& $\bar{M}(L=3)$ & $\bar{M}(L=4)$ \\ \hline
$\Sigma _{c}$($n=0$) & 2496.09 & 2774.67 & 3004.41 & 3204.48& 3384.07 \\
$\Sigma _{c}$($n=1$) & 2864.00& 3081.28& 3272.98& 3446.45& 3606.07 \\
$\Sigma _{c}$($n=2$) &3154.71& 3339.01& 3506.94& 3662.22& 3807.34 \\
$\Sigma _{c}$($n=3$) &3402.82& 3565.72& 3716.99& 3858.83& 3992.79 \\
$\Sigma _{c}$($n=4$) &3622.91& 3770.48& 3909.24& 4040.61& 4165.65 \\
\hline
$\Sigma _{b}$($n=0$) &5819.26& 6082.84& 6308.71& 6509.53& 6692.16 \\
$\Sigma _{b}$($n=1$) &6169.94& 6385.49& 6578.96& 6756.02& 6920.23 \\
$\Sigma _{b}$($n=2$) &6459.29& 6646.17& 6818.12& 6978.25& 7128.70 \\
$\Sigma _{b}$($n=3$) &6711.34& 6878.62& 7034.95& 7182.24& 7321.88\\
$\Sigma _{b}$($n=4$) &6937.62& 7090.42& 7234.73& 7371.84& 7502.73 \\
\hline
$\Xi^{\prime} _{c}$($n=0$) &2623.09& 2923.52& 3172.61& 3390.14& 3585.72 \\
$\Xi^{\prime} _{c}$($n=1$) &3020.26& 3256.13& 3464.71& 3653.72& 3827.81\\
$\Xi^{\prime} _{c}$($n=2$) &3335.98& 3536.63& 3719.68& 3889.09& 4047.52\\
$\Xi^{\prime} _{c}$($n=3$) &3606.16& 3783.79& 3948.88& 4103.75& 4250.10 \\
$\Xi^{\prime} _{c}$($n=4$) &3846.19& 4007.27& 4158.82& 4302.36& 4439.03 \\
\hline
$\Xi^{\prime} _{b}$($n=0$) &5945.40& 6244.84& 6500.28& 6726.81& 6932.46\\
$\Xi^{\prime} _{b}$($n=1$) &6343.45& 6586.94& 6805.02& 7004.30& 7188.93 \\
$\Xi^{\prime} _{b}$($n=2$) &6670.17& 6880.69& 7074.14& 7254.12& 7423.09 \\
$\Xi^{\prime} _{b}$($n=3$) &6954.03& 7142.16& 7317.82& 7483.20& 7639.93 \\
$\Xi^{\prime} _{b}$($n=4$) &7208.47& 7380.11& 7542.13& 7695.98& 7842.80 \\
\hline
$\Omega _{c}$($n=0$) &2742.09& 3081.49& 3361.85& 3606.22& 3825.70 \\
$\Omega _{c}$($n=1$) &3190.46& 3455.72& 3689.93& 3901.95& 4097.11 \\
$\Omega _{c}$($n=2$) &3545.42& 3770.62& 3975.91& 4165.78& 4343.25 \\
$\Omega _{c}$($n=3$) &3848.62& 4047.77& 4232.76& 4406.23& 4570.10 \\
$\Omega _{c}$($n=4$) &4117.71& 4298.17& 4467.90& 4628.60& 4781.59\\
\hline
$\Omega _{b}$($n=0$) &6053.78& 6344.51& 6595.64& 6819.96& 7024.57\\
$\Omega _{b}$($n=1$) &6441.18& 6681.30& 6897.68& 7096.22& 7280.72\\
$\Omega _{b}$($n=2$) &6763.76& 6972.99& 7165.96& 7345.97& 7515.32 \\
$\Omega _{b}$($n=3$) &7046.08& 7233.94& 7409.77& 7575.63& 7733.04 \\
$\Omega _{b}$($n=4$) &7300.27& 7472.21& 7634.78& 7789.38& 7937.07 \\
\hline\hline
\end{tabular}}
\end{table*}
\begin{figure*}[htbp]
\centering
\begin{minipage}[ht]{0.49\textwidth}
\includegraphics[width=6.5cm]{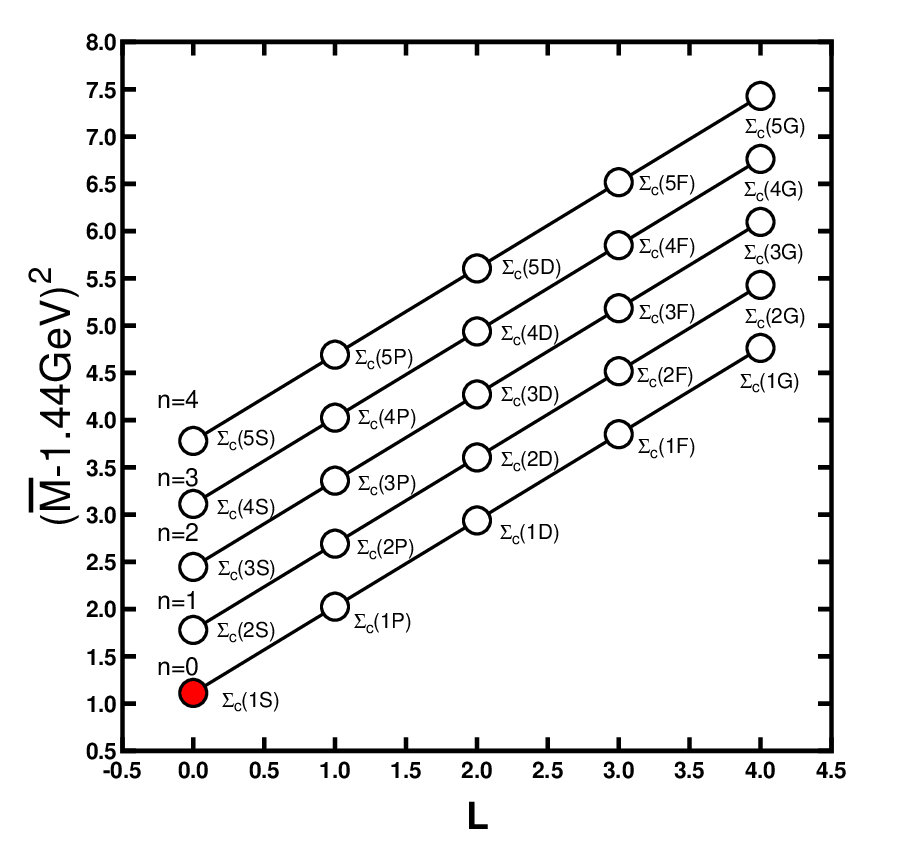}
\caption{Spin-average mass of $\Sigma_{c}$ baryons}
\end{minipage}
\begin{minipage}[ht]{0.49\textwidth}
\includegraphics[width=6.5cm]{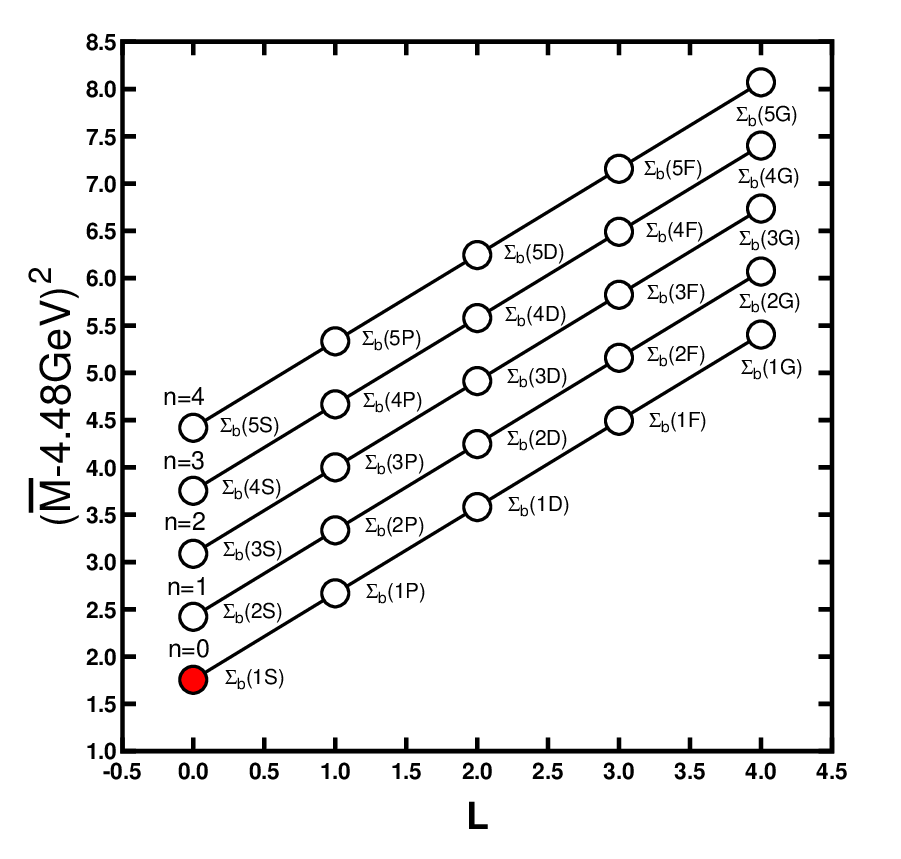}
\caption{Spin-average mass of $\Sigma_{b}$ baryons}
\end{minipage}
\begin{minipage}[ht]{0.49\textwidth}
\includegraphics[width=6.5cm]{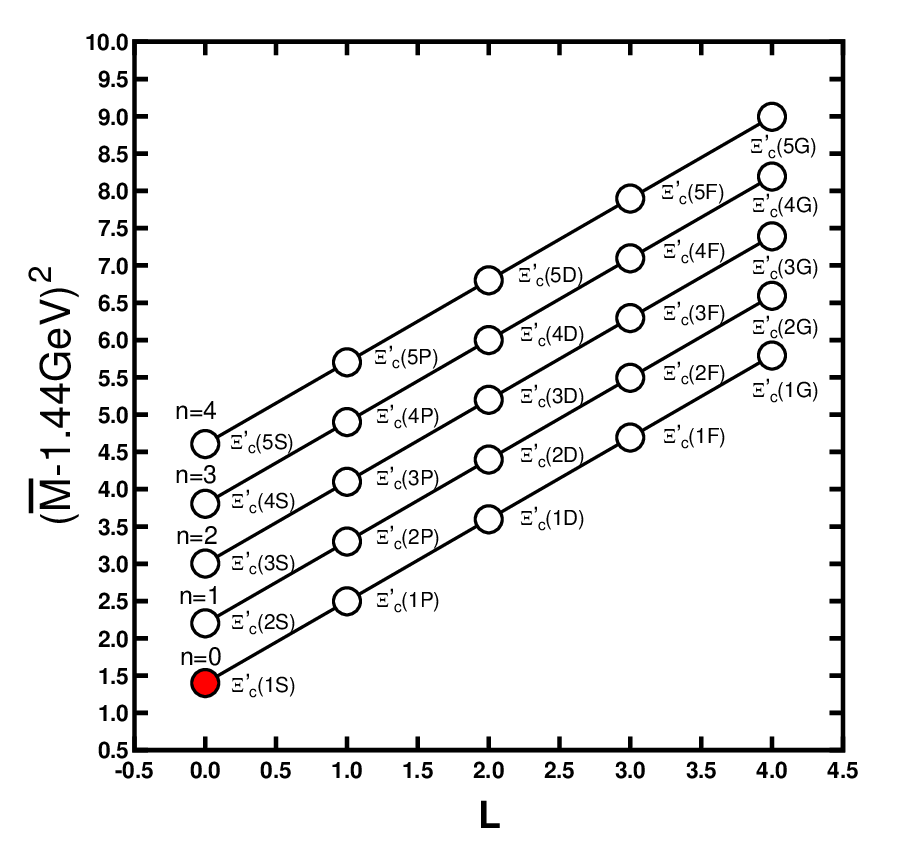}
\caption{Spin-average mass of $\Xi^{\prime}_{c}$ baryons}
\end{minipage}
\begin{minipage}[ht]{0.49\textwidth}
\includegraphics[width=6.5cm]{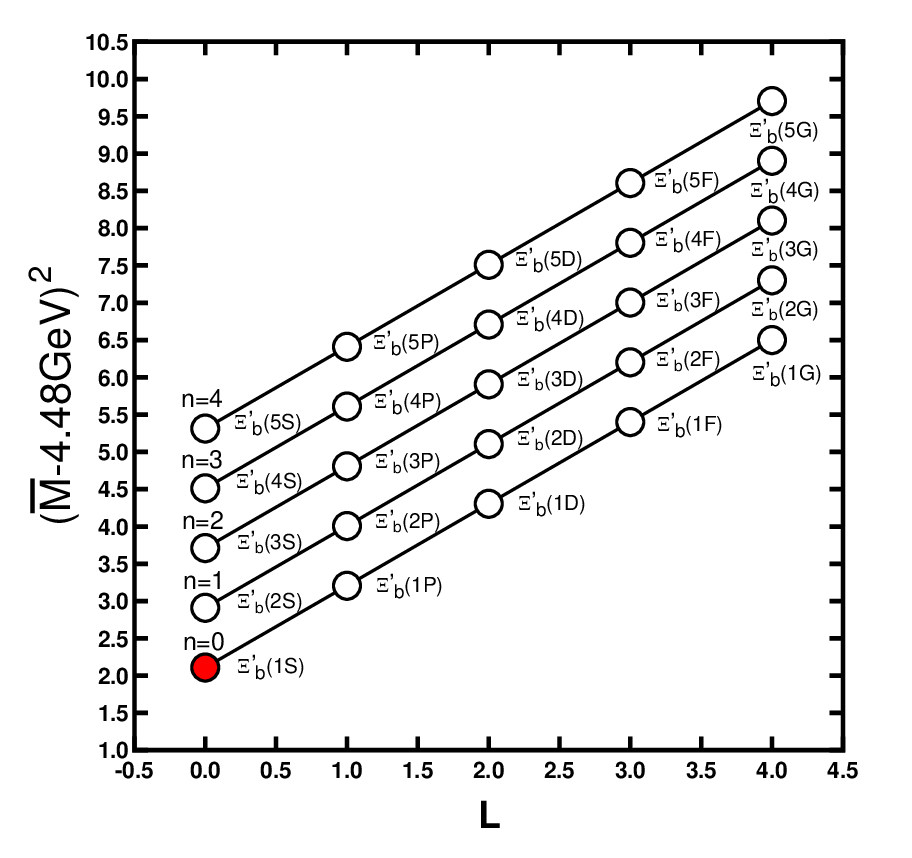}
\caption{Spin-average mass of $\Xi^{\prime}_{b}$ baryons}
\end{minipage}
\begin{minipage}[ht]{0.49\textwidth}
\includegraphics[width=6.5cm]{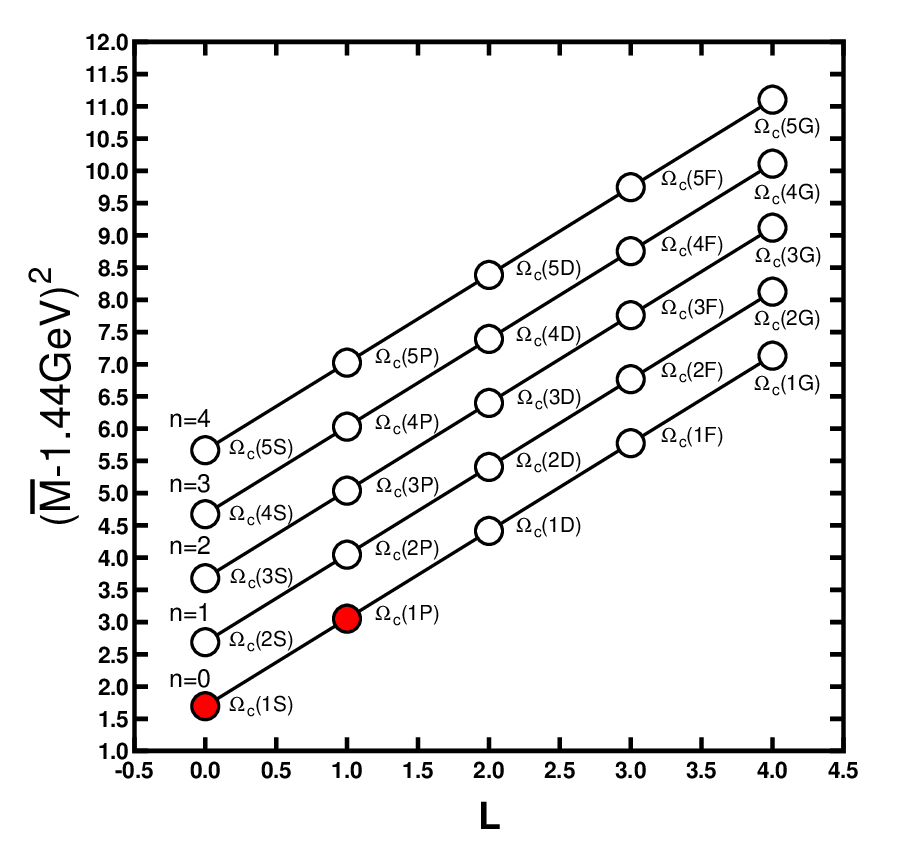}
\caption{Spin-average mass of $\Omega_{c}$ baryons}
\end{minipage}
\begin{minipage}[ht]{0.49\textwidth}
\includegraphics[width=6.5cm]{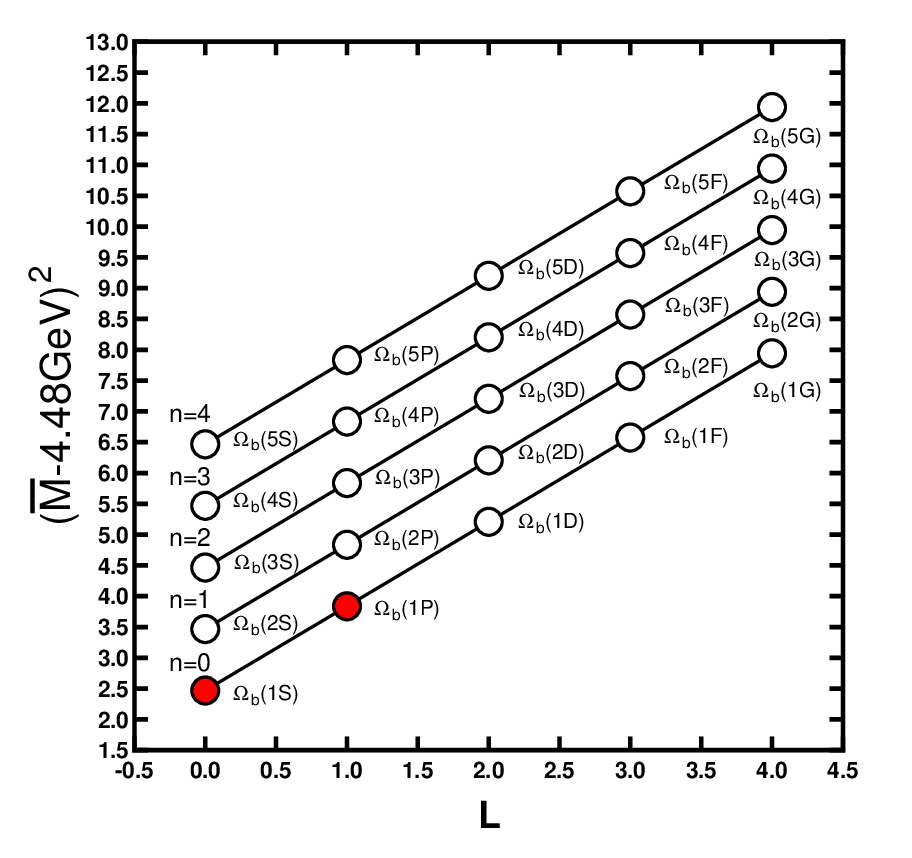}
\caption{Spin-average mass of $\Omega_{b}$ baryons}
\end{minipage}
\end{figure*}

\section{The spin-dependent potential and the scaling relations}\label{Sec.III}

Even though the baryon is a three-body system under the strong interaction, it is helpful to understand the measured mass data of the excited baryons using a simple heavy quark-diquark picture. To estimate the mass splitting for the singly heavy baryons, we consider the spin-dependent Hamiltonian $H^{SD}$ \cite{EFG:C10, KarlinerRP:PP888} between the heavy quark $(Q)$ and the spin-1 diquark $(d)$ as
\begin{equation}
H^{SD}=a_{1}\mathbf{L}\cdot \mathbf{S}_{d}+a_{2}\mathbf{L}\cdot \mathbf{S}%
_{Q}+b_{1}S_{12}+c_{1}\mathbf{S}_{d}\cdot \mathbf{S}_{Q},  \label{PP5}
\end{equation}%
where $a_{1}$, $a_{2}$, $b_{1}$, $c_{1}$ are the spin coupling parameters. The first two terms are spin-orbit interactions, the third is the tensor energy, and the last is the contact interaction between the heavy quark spin $\mathbf{S}_{Q}$ and the diquark spin $\mathbf{S}_{d}$. For the particular choice $L=0$ for the $S$-wave baryons in appendix A, the first three terms of Eq. (\ref{PP5}) can be eliminated and only the last term survives, see Eq. (\ref{PP6}). Here, $S_{12}=3(\mathbf{S}_{d}\cdot \mathbf{\hat{r}})(\mathbf{S}_{Q}\cdot \mathbf{\hat{r}})/r^{2}-%
\mathbf{S}_{d}\cdot \mathbf{S}_{Q}$ in Ref. \cite{KarlinerR:11} with $L=1$ and $L=2$ can be given by
\begin{equation}
L=1:  S_{12}=-\frac{3}{5}[(\mathbf{L}\cdot \mathbf{S}_{d})(\mathbf{L}\cdot \mathbf{S}_{Q})+(\mathbf{L}\cdot \mathbf{S}_{Q})(\mathbf{L}\cdot \mathbf{S}_{d})-\frac{4}{3}(\mathbf{S}_{d}\cdot \mathbf{S}_{Q})],
\end{equation}%
\begin{equation}
L=2: S_{12}=-\frac{1}{7}[(\mathbf{L}\cdot \mathbf{S}_{d})(\mathbf{L}\cdot \mathbf{S}_{Q})+(\mathbf{L}\cdot \mathbf{S}_{Q})(\mathbf{L}\cdot \mathbf{S}_{d})-4(\mathbf{S}_{d}\cdot \mathbf{S}_{Q})].
\end{equation}%

Combined with the experimental data \cite{Aaij:A11} of the $\Omega_{c}$(css), we used the Regge trajectory Eq. (\ref{pp421}) to fit the constituent quark masses of the charm quark $(c)$ and two strange quarks $(ss)$ in Ref. \cite{JiaPan:PP888}, the results are $M_{c}=1.44$ GeV and $m_{ss}=0.991$ GeV. In the case of doubly strange $Qss$ baryons with the mass of the diquark $ss$ comparable with the mass of the heavy quark $Q$ ($m_{ss}\approx M_c$), the finite mass effect of the heavy quark may become important and makes it appropriate to go beyond the $jj$ coupling. Therefore, in contrast to the scheme used in Ref. \cite{KarlinerRP:PP888}, we proposed a new scheme of state classification named the $JLS$ coupling \cite{JiaPan:PP888}. The first three terms are treated as operators $H^{SD}_{1}$ defining representations and the last term $H^{SD}_{2} = c_{1}S_{d}\cdot S_{Q}$ in Eq. (\ref{PP5}) as a perturbation.
The operator $H^{SD}_{1}$ is given by
\begin{equation}
H^{SD}_{1}=a_{1}(\mathbf{L}\cdot \mathbf{S}_{d})+a_{2}(\mathbf{L}\cdot \mathbf{S}_{Q})+b_{1}S_{12}.
\end{equation}
Using the bases $|J,j\rangle$ in terms of eigenvalues $J,j$ of the total angular momentum $\mathbf{J}$ and total light-quark angular momentum $\mathbf{j}$, respectively, in order to diagonalize the mass operators $H^{SD}_{1}$ and $H^{SD}_{2}$, we can obtain the mass shifts $\Delta M$ of $P$-wave in Eq. (\ref{MM111}) and $D$-wave in Eq. (\ref{MM222}) for the singly heavy baryons, see appendix B and C. In this scheme, the $P$-wave states of the baryons may be classified as $^{2S+1}P_{J}$ = $^{2}P_{1/2}$, $^{4}P_{1/2}$, $^{2}P_{3/2}$, $^{4}P_{3/2}$, $^{4}P_{5/2}$ and the $D$-wave states as $^{2S+1}D_{J}$ = $^{4}D_{1/2}$, $^{2}D_{3/2}$, $^{4}D_{3/2}$, $^{2}D_{5/2}$, $^{4}D_{5/2}$, $^{4}D_{7/2}$.

Next, it is necessary to estimate the four spin coupling parameters $a_{1}$, $a_{2}$, $b_{1}$, $c_{1}$ in the heavy-light quark system. If Eq. (\ref{PP5}) is taken as a spin-relevant relativistic correction, the parameters $a_{1}$, $a_{2}$, $b_{1}$, $c_{1}$ are related to the magnetic moment $\mathbf{S}_{Q}/M_{Q}$ of the heavy quark. Therefore, these parameters can be considered roughly inversely proportional to the heavy quark mass $(M_{Q})$. In Ref. \cite{KarlinerRP:PP888}, the authors calculated the parameters of the partner in baryons using the scaling relations
\begin{eqnarray}
a_{1}(b)&=&a_{1}(c), \notag\\
a_{2}(b)&=&\frac{M_{c}}{M_{b}}a_{2}(c), \label{ppbb111} \\
b_{1}(b)&=&\frac{M_{c}}{M_{b}}b_{1}(c), \notag
\end{eqnarray}
with the constituent quark masses ($M_{c}, M_{b}$) of the heavy quark in baryons. The parameter $c_{1}$ is expected to be negligible, because it should be very small in the $P$-wave states of the baryons.

In order to calculate the mass splitting of all excited states, we utilize the scaling relations based on the similarity between a baryon and its the partner baryons in the color configurations to study the spin coupling parameters. In this subsection, we need to generalize Eq. (\ref{ppbb111}) and consider the parameter $c_1$ which should include the effect of the principal quantum number $N$ together with the radial quantum number $n$ and orbital quantum number $L$ \cite{EFG:C10, EbertFGM:PP888, EichtenFei:PP888, GodfreyIsg:PP888, ChenLiuen:PP888}. The parameters $a_{1}$, $a_{2}$, $b_{1}$ are obtained by following the scaling rules:

(i) The parameter $a_{1}$ is proportional to $\frac{1}{M_{Q}m_{d}}\langle\frac{1}{r}\rangle$.

(ii) The parameter $a_{2}$ is proportional to $\frac{1}{M_{Q}m_{d}}\langle\frac{1}{r}\rangle$.

(iii) The tensor parameter $b_{1}$ is proportional to $\frac{1}{M_{Q}m_{d}}\langle\frac{1}{r^{3}}\rangle$. \\
Here, $\langle 1/r\rangle=1/((n+L+1)^{2}a_{B})$, $\langle 1/r^{3}\rangle=1/(L(L+1/2)(L+1)(n+L+1)^{3}a^{3}_{B})$ and $a_{B}$ is the Bohr radius. According to the scaling rules, $a_{2}$ can be of the same order as $a_{1}$ with the same $n$, $L$ in the excited states, while the parameter $b_{1}$ should be smaller than the $a_{1}$, $a_{2}$, as $b_{1}$ scales with $\langle\frac{1}{r^{3}}\rangle$.

In order to obtain the parameter $c_{1}$ in Eq. (\ref{PP5}), we need a scaling rule similar to (i)-(iii). Considering that $c_1$ becomes dominant in determining the mass splitting Eq. (\ref{pp99}), we can estimate $c_1$ based on the hyperfine structure term given by \cite{Eberale:PP888, PVB:PPP888}
\begin{equation}
H^{hp}=\frac{8}{9M_{Q}m_{d}} \bigtriangledown^{2}V \mathbf{S}_{d}\cdot\mathbf{S}_{Q}= \frac{32\pi \alpha_{s}}{9M_{Q}m_{d}} \mathbf{S}_{d}\cdot\mathbf{S}_{Q} \delta^{3}(\mathbf{r}),   \label{ppdd312}
\end{equation}
where $\nabla^{2}$ is the Laplace operator and $\delta^3(\mathbf{r})$ is the three-dimensional delta distribution. The derivative of the Coulomb potential $V$ gives $\bigtriangledown^{2}V=4\pi\alpha_{s}\delta^{3}(\mathbf{r})$ with the strong coupling $\alpha_{s}$. By taking the average $\langle \delta^{3}(\mathbf{r})\rangle=\mathbf{|}\psi(0)\mathbf{|}^{2}$ established for the hydrogen-like atoms wave function $\psi(\mathbf{r})$ of $S$-wave ($L=0$) \cite{Griffiths:PP888}, Eq. (\ref{ppdd312}) becomes
\begin{eqnarray}
\langle H^{hp}\rangle =\frac{32\pi \alpha_{s}}{9M_{Q}m_{d}} \frac{1}{N^{3}a_{B}^{3}} \langle\mathbf{S}_{d}\cdot\mathbf{S}_{Q}\rangle, \label{ppdd3312}
\end{eqnarray}
with $N=n+L+1$. To extend Eq. (\ref{ppdd3312}) further to the excited states of the baryons, we introduce a parameter $\lambda$ as follows,
\begin{eqnarray}
\langle H^{hp}\rangle =\frac{32\pi \alpha_{s}}{9M_{Q}m_{d}} \frac{1}{(L+\lambda)N^{3}a_{B}^{3}} \langle\mathbf{S}_{d}\cdot\mathbf{S}_{Q}\rangle. \label{ppdd11}
\end{eqnarray}
Based on the systematic analysis of experimental values, we find the parameter $\lambda=3.3$. Analyzing the coefficient in Eq. (\ref{ppdd11}), the parameter $c_1$ is inversely proportional to $M_{Q}$, $m_{d}$ and $(L+\lambda)N^{3}$. Thus, the scaling rule of $c_{1}$ can be determined as follows:

(iv) The parameter $c_{1}$ is proportional to $\frac{1}{M_{Q}m_{d}} \frac{1}{(L+\lambda)N^{3}}$.

Eventually, the scaling relations of the spin coupling parameters in Eq. (\ref{PP5}) for the baryon system are
\begin{equation}
\left\{
\begin{array}{rrrr} \vspace{1ex}
a_{1}(B_{a}, (n+1)L)&=&\frac{M^{\prime}_{Q}m^{\prime}_{d}}{M_{Q}m_{d}}\frac{{N^{\prime}_{a_{1}}}}{{N_{a_{1}}}}a_{1}(B_{a}^{\prime}, (n^{\prime}+1)L^{\prime}), \\ \vspace{1ex}
a_{2}(B_{a}, (n+1)L)&=&\frac{M^{\prime}_{Q}m^{\prime}_{d}}{M_{Q}m_{d}}\frac{{N^{\prime}_{a_{2}}}}{{N_{a_{2}}}}a_{2}(B_{a}^{\prime}, (n^{\prime}+1)L^{\prime}), \\  \vspace{1ex}
b_{1}(B_{a}, (n+1)L)&=&\frac{M^{\prime}_{Q}m^{\prime}_{d}}{M_{Q}m_{d}}\frac{{N^{\prime}_{b_{1}}}}{{N_{b_{1}}}}b_{1}(B_{a}^{\prime}, (n^{\prime}+1)L^{\prime}), \\
c_{1}(B_{a}, (n+1)L)&=&\frac{M^{\prime}_{Q}m^{\prime}_{d}}{M_{Q}m_{d}}\frac{{N^{\prime}_{c_{1}}}}{{N_{c_{1}}}}c_{1}(B_{a}^{\prime}, (n^{\prime}+1)L^{\prime}),\\
\end{array}%
\right.   \label{scr:pp888}
\end{equation}%
where $n, n^{\prime}=0, 1 , 2, \cdots$; $L, L^{\prime}= S, P, D, F, \cdots$; and $B_{a}, B_{a}^{\prime}$ are baryons with $N_{a_{1}}=(n+L+1)^{2}=N_{a_{2}}$, $N_{b_{1}}=L(L+1/2)(L+1)(n+L+1)^{3}$, $N_{c_{1}}=(L+\lambda)(n+L+1)^{3}$ corresponding to the similar form of $N^{\prime}_{a_{1}}$, $N^{\prime}_{a_{2}}$, $N^{\prime}_{b_{1}}$, $N^{\prime}_{c_{1}}$ with $L^{\prime}$ and $n^{\prime}$, respectively. The prime denotes the quantities of the baryon $B_{a}^{\prime}$ obtained from experiments, distinguishing them from that of an unobserved baryon $B_{a}$.

\section{The baryons $\Omega _{c}$ and $\Omega _{b}$}\label{Sec.IV}

For $\Omega _{c}$ baryon family, it was a pleasant surprise that the LHCb Collaboration recently discovered five new narrow $\Omega_{c}$ states observed in decay channel $\Xi^{+}_{c}K$ \cite{Aaij:A11}: $\Omega_{c}(3000)^{0}$, $\Omega_{c}(3050)^{0}$, $\Omega_{c}(3065)^{0}$, $\Omega_{c}(3090)^{0}$, $\Omega_{c}(3120)^{0}$, the measured masses are
\begin{eqnarray}\notag
\Omega_{c}(3000)^{0}:M&=&3000.4\pm0.2\pm0.1\ \text{MeV}, \\ \notag
\Omega_{c}(3050)^{0}:M&=&3050.2\pm0.1\pm0.1\ \text{MeV}, \\ \notag
\Omega_{c}(3065)^{0}:M&=&3065.6\pm0.1\pm0.3\ \text{MeV}, \\ \notag
\Omega_{c}(3090)^{0}:M&=&3090.2\pm0.3\pm0.5\ \text{MeV}, \\ \notag
\Omega_{c}(3120)^{0}:M&=&3119.1\pm0.3\pm0.9\ \text{MeV}.  \notag
\end{eqnarray}
Later, the Belle Collaboration confirmed the existence of these states \cite{YeltonBel:PP888}. In Ref. \cite{JiaPan:PP888}, the authors employ the quark model to analyze the narrow $\Omega_{c}$ states, and suggested that the parity was negative for all of five states. These can be interpreted as $1P$-wave charmed baryons candidates. Correspondingly, the masses $M(1/2, 0)$, $M(1/2, 1)$, $M(3/2, 1)$, $M(3/2, 2)$, $M(5/2, 2)$ are
\begin{eqnarray}
M(\Omega_{c},1P): 3000.4\  \text{MeV}, 3050.2\  \text{MeV}, 3065.6\  \text{MeV}, 3090.2\  \text{MeV}, 3119.1\  \text{MeV},  \label{Vq11}
\end{eqnarray}
which can give good results for the $\Omega_{c}$ states, and are consistent with the experimental data of the LHCb Collaboration. At the same time, by fitting, the spin coupling parameters $a_1$,
$a_2$, $b_1$, $c_1$ are also obtained in Ref. \cite{JiaPan:PP888},
\begin{eqnarray}
&a_{1}(\Omega_{c},1P)=26.96\ \text{MeV},\quad a_{2}(\Omega_{c},1P)=25.76\ \text{MeV}, \notag \\
&b_{1}(\Omega_{c},1P)=13.51\ \text{MeV},\quad c_{1}(\Omega_{c},1P)=4.04\ \text{MeV}.      \label{VVq11}
\end{eqnarray}
These results are the same as those in both of Refs. \cite{KarlinerR:11, Ali:PP888}. For more information of the $\Omega_{c}$ baryons, we recommend interested readers to see Refs. \cite{EFG:C10, EFG:A12, MI:A11, RP:A11, GV:A11, MMP:A11, VGVT:A11, YOHH:A11, PCB:A11, ShahK:PP888, YHHHS:A11, CCCHLZ:A11, CL:A11, Katoijima:PP888}.

To elaborate on the mass shifts $\Delta M(J, j)$ for the entire baryon systems, we utilize the parameters (\ref{VVq11}) of the $1P$-wave $\Omega_{c}$ states as the object of the scaling relations in Eq. (\ref{scr:pp888}) to calculate the parameters of the other states. Adding the spin-average mass $\bar M$, the baryon mass becomes $M(J, j)= \bar M + \Delta M(J, j)$, where details of calculating $\Delta M(J, j)$ and $M(J, j)$ are presented in the Appendix. Therefore, the mass spectra of the singly heavy baryons can be predicted.

In earlier times, the observed $1S$-wave states $\Omega_{c}^{0}$ and $\Omega_{c}(2770)^{0}$ with $J^{P}=1/2^{+}$ and $J^{P}=3/2^{+}$, corresponding to the masses $M(\Omega_{c},1/2^{+})=2695.2$ MeV and $M(\Omega_{c},3/2^{+})=2765.9$ MeV, respectively, had already been established. As seen in our model calculations, by using Eqs. (\ref{pp4}),  (\ref{scr:pp888}) and the parameters (\ref{VVq11}) with $L^{\prime}=1, n^{\prime}=0$ for the $1P$-wave $\Omega_{c}$ states, the calculation of the spin-averaged mass for the $1S$-wave $\Omega_{c}$ states with $L=0,n=0$ gives
\begin{eqnarray}
\bar M(\Omega_{c}, 1S)=M_{c}+\left( m_{ss}+M_{c}\left( 1-\frac{m_{\text{cur}c}^{2}}{M_{c}^{2}}\right) \right)=2742.09\  \text{MeV} , \label{wep123}
\end{eqnarray}
and the parameter is
\begin{eqnarray}
c_{1}(\Omega_{c}, 1S)=\frac{{N^{\prime}_{c}}}{{N_{c}}}c_{1}(\Omega_{c}, 1P)&=&\frac{(L^{\prime}+3.3)(n^{\prime}+L^{\prime}+1)^{3}}{(L+3.3)(n+L+1)^{3}}c_{1}(\Omega_{c}, 1P) \notag\\
&=&\frac{(1+3.3)(0+1+1)^{3}}{(0+3.3)(0+0+1)^{3}}4.04\ \text{MeV} \notag\\
&=&42.11\  \text{MeV}. \label{wepp123}
\end{eqnarray}
Substituting Eq. (\ref{wep123}) and Eq. (\ref{wepp123}) into Eq. (\ref{pp99}), we obtain the masses $M(\Omega_{c},1/2^{+})=2699.98$\ MeV and $M(\Omega_{c},3/2^{+})=2763.15$\ MeV as shown in Table \ref{ppdm5} of two ground states for $\Omega_{c}^{0}$, which are in good agreement with the experimental values.

The mixed state $\Omega_c(3327)^0$ has been speculated as a $2S$ state in Ref. \cite{KaerRoter:PP888} and as a $1D$ state in Refs. \cite{LuoLiu:PP888, FengYang:PP888, JakhadGK:PP888}. However, we still need more observable objects to get clarity about the internal structure. In addition, in Ref. \cite{CL:A11} the authors suggested that $\Omega_{c}(3185)^{0}$ may be regarded as a $2S$ state with $J^{P}=1/2^{+}$ or $J^{P}=3/2^{+}$, or their overlapping structure, and $\Omega_{c}(3185)^{0}$ is interpreted as a $P$-wave state in Ref. \cite{PachecoBijker:PP888}. Very recently, the $\Omega_{c}(3185)^{0}$ and $\Omega_{c}(3327)^{0}$ states of $\Omega _{c}$ baryons were observed by LHCb Collaboration \cite{Aaij:A13} with masses 3185.1 MeV and 3327.1 MeV, respectively. The quantum numbers of these states remain to be determined. According to our model, the calculation of the spin-averaged mass and the parameter for the $\Omega _{c}$ states in $2S$-wave ($L=0, n=1$) are obtained by using Eqs. (\ref{pp4}), (\ref{scr:pp888}) and (\ref{VVq11}),
\begin{eqnarray}
\bar M(\Omega_{c},2S)&=&M_{c}+\sqrt{\pi\alpha(\Omega _{c})\times1.37+\left( m_{ss}+M_{c}\left( 1-\frac{m_{\text{cur}c}^{2}}{M_{c}^{2}}\right) \right)^{2}}=3190.46\  \text{MeV} , \\ \label{we123}
c_{1}(\Omega_{c},2S)&=&\frac{(L^{\prime}+3.3)(n^{\prime}+L^{\prime}+1)^{3}}{(L+3.3)(n+L+1)^{3}}c_{1}(\Omega_{c},1P)=\frac{(1+3.3)(0+1+1)^{3}}{(0+3.3)(1+0+1)^{3}}4.04\ \text{MeV} \notag \\
&=&5.26\ \text{MeV}.
\end{eqnarray}
Hence, the masses of the $2S$-wave $\Omega_c$ states are 3185.20 MeV and 3193.09 MeV as listed in Table \ref{ppdm5} with $J^{P}=1/2^{+}$ and $J^{P}=3/2^{+}$, respectively. The $\Omega_{c}(3185)^{0}$ can be grouped into the $2S$ state. We assign $J^{P}=1/2^{+}$ for $\Omega_{c}(3185)^{0}$. On the other hand, we also have calculated the spin-average of the $1D$-wave ($L=2, n=0$) for $\Omega_{c}$ states,
\begin{eqnarray}
\bar M(\Omega_{c}, 1D)&=&M_{c}+\sqrt{2\pi\alpha(\Omega _{c})+\left( m_{ss}+M_{c}\left( 1-\frac{m_{\text{cur}c}^{2}}{M_{c}^{2}}\right) \right)^{2}}=3361.85\ \text{MeV}, \label{weee123}
\end{eqnarray}
the parameters are
\begin{eqnarray}
a_{1}(\Omega_{c},1D)&=&\frac{(n^{\prime}+L^{\prime}+1)^{2}}{(n+L+1)^{2}}a_{1}(\Omega_{c},1P)=\frac{(0+1+1)^{2}}{(0+2+1)^{2}}26.96\ \text{MeV}=11.98\ \text{MeV},\\
a_{2}(\Omega_{c},1D)&=&\frac{(n^{\prime}+L^{\prime}+1)^{2}}{(n+L+1)^{2}}a_{2}(\Omega_{c},1P)=\frac{(0+1+1)^{2}}{(0+2+1)^{2}}25.76\ \text{MeV}=11.45\ \text{MeV},\\
b_{1}(\Omega_{c},1D)&=&\frac{L^{\prime}(L^{\prime}+\frac{1}{2})(L^{\prime}+1)(n^{\prime}+L^{\prime}+1)^{3}}{L(L+\frac{1}{2})(L+1)(n+L+1)^{3}}b_{1}(\Omega_{c},1P)=\frac{(1+\frac{1}{2})(1+1)(0+1+1)^{3}}{2(2+\frac{1}{2})(2+1)(0+2+1)^{3}}13.51\text{MeV} \notag \\
&=&0.80\ \text{MeV},\\
c_{1}(\Omega_{c},1D)&=&\frac{(L^{\prime}+3.3)(n^{\prime}+L^{\prime}+1)^{3}}{(L+3.3)(n+L+1)^{3}}c_{1}(\Omega_{c},1P)=\frac{(1+3.3)(0+1+1)^{3}}{(2+3.3)(0+2+1)^{3}}4.04\ \text{MeV} \notag \\
&=&0.97\ \text{MeV}, \label{we1132}
\end{eqnarray}
and the masses are
\begin{equation}
M(\Omega_{c},1D):3308.41\ \text{MeV}, 3326.92\ \text{MeV}, 3342.64\ \text{MeV}, 3356.96\ \text{MeV}, 3373.08\ \text{MeV}, 3397.52\ \text{MeV}. \label{we1133}
\end{equation}
Using experimental values in Ref. \cite{Aaij:A13}, $\Omega_c(3327)^0$ is assigned by us as a $1D$ state with $J^P=3/2^+$ rather than a $2S$ state, as the hyperfine splitting $3327.1\mbox{\,MeV}-3185.1\mbox{\,MeV}=142.0$ $\mbox{\,MeV}$ between $\Omega_c(3185)^0$ and $\Omega_c(3327)^0$ is much larger than the result $5.26\mbox{\,MeV}$ of our model calculation. In this work, by using the scaling relations Eq. (\ref{scr:pp888}) we get the spin coupling parameters $a_{1}$, $a_{2}$, $b_{1}$, $c_{1}$ as shown in Table \ref{Table:PP88}. We use the mass splitting Eqs. (\ref{pp99}), (\ref{MMM111}) and (\ref{MMM222}) to calculate the mass spectra of the $\Omega_{c}$ states. The results are given in Table \ref{ppdm5} for the $\Omega_{c}$ baryons and compared with other models.

For $\Omega _{b}$ baryon family, in the quark model $\Omega^{-}_{b}$ is the ground state of $\Omega_{b}$, where the system $(bss)$ consists of a bottom quark ($b$) and a spin-1 diquark ($ss$). The mass of the $\Omega^{-}_{b}$ state is $M({\Omega^{-}_{b}})=6045.2$ MeV with $J^{P}=1/2^{+}$ identified in Ref. \cite{Workman:A11}. As a result, our calculation agrees very well with PDG for the $\Omega^{-}_{b}$ state. Recently, the LHCb experiment reported four extremely narrow $\Omega_{b}$ states in the decay channel $\Xi^{0}_{b}K$ \cite{Aaij:A12}. According to our discussion with the observations, the four states $\Omega_{b}(6316)^{-}$, $\Omega_{b}(6330)^{-}$, $\Omega_{b}(6340)^{-}$, $\Omega_{b}(6350)^{-}$ may be assigned as $1P$-wave excitations around the spin-average mass $\bar M=6344.51$\ MeV with $J^{P}=1/2^{-}$, $1/2^{-}$, $3/2^{-}$ and $3/2^{-}$, respectively. Based on the masses of the $1P$-wave $\Omega_{b}$ states, we predict that there exists another excited $\Omega_{b}$ state with $J^{P} = 5/2^{?}$ in addition to the four $\Omega_{b}$ states observed by the LHCb Collaboration. The spin coupling parameters $a_{1}$, $a_{2}$, $b_{1}$, $c_{1}$ and the masses $M(1/2, 0)$, $M(1/2, 1)$, $M(3/2, 1)$, $M(3/2, 2)$, $M(5/2, 2)$ are given by
\begin{eqnarray}
&&\bar M=6344.51\ \text{MeV}, a_{1}=8.67\ \text{MeV}, a_{2}=8.28\ \text{MeV}, b_{1}=4.34\ \text{MeV}, c_{1}=1.30\ \text{MeV},\qquad  \label{Vx2q11}\\
&&M(\Omega_{b},1P): 6318.95\ \text{MeV}, 6334.95\ \text{MeV}, 6339.90\ \text{MeV}, 6347.80\ \text{MeV}, 6357.09\ \text{MeV}. \label{Vxx21q11}
\end{eqnarray}
Therefore, the mass of the excited $\Omega_{b}$ state with $J^P=5/2^{?}$ is about 6357\ MeV. This can be compared with values from Ref. \cite{KarlinerR:12}. For the $\Omega_{b}$ baryons, we calculate the parameters $a_{1}$, $a_{2}$, $b_{1}$, $c_{1}$ as shown in Table \ref{pdm5}, while our mass results are compared to results of other models in Table \ref{ppdm6}. These mass predictions presented in Table \ref{ppdm5} and Table \ref{ppdm6} for the $\Omega_{Q}$ $(Q=c, b)$ baryons will be helpful for future experimental searches.
\renewcommand\tabcolsep{1.0cm}
\renewcommand{\arraystretch}{0.8}
\begin{table*}[!htbp]
\caption{The spin coupling parameters (MeV) of the $\Omega_{c}$ baryons.   \label{Table:PP88}}
\begin{tabular}{ccccc}
\hline\hline
State: & $a_{1}$ & $a_{2}$ & $b_{1}$ &$c_{1}$ \\
 \hline
1$S$ &      &         &       & 42.11   \\
2$S$ &      &         &       & 5.26 \\
3$S$ &      &         &       & 1.56 \\
4$S$ &      &         &       & 0.66 \\
5$S$ &      &         &       & 0.34 \\
\hline
1$P$ &26.96 &25.76   &13.51  &4.04  \\
2$P$ &11.98 &11.45   &4.00  &1.20  \\
3$P$ &6.74  &6.44    &1.69  &0.51   \\
4$P$ &4.31  &4.12    &0.86  &0.26   \\
5$P$ &3.00  &2.86    &0.50  &0.15   \\
\hline
1$D$ &11.98 &11.45   &0.80  &0.97   \\
2$D$ &6.74  &6.44    &0.34  &0.41   \\
3$D$ &4.31  &4.12    &0.17  &0.21   \\
4$D$ &3.00  &2.86    &0.10  &0.12   \\
5$D$ &2.20  &2.10    &0.06  &0.08   \\
\hline\hline
\end{tabular}
\end{table*}
\renewcommand\tabcolsep{1.0cm}
\renewcommand{\arraystretch}{0.8}
\begin{table*}[!htbp]
\caption{The spin coupling parameters (MeV) of the $\Omega_{b}$ baryons.   \label{pdm5}}
\begin{tabular}{ccccc}
\hline\hline
State: & $a_{1}$ & $a_{2}$ & $b_{1}$ &$c_{1}$ \\
 \hline
1$S$ &      &         &       & 13.54   \\
2$S$ &      &         &       & 1.69 \\
3$S$ &      &         &       & 0.50\\
4$S$ &      &         &       & 0.21 \\
5$S$ &      &         &       & 0.11 \\
\hline
1$P$ &8.67  &8.28     &4.34   &1.30  \\
2$P$ &3.85  &3.68     &1.29   &0.38  \\
3$P$ &2.17  &2.07     &0.54   &0.16   \\
4$P$ &1.39  &1.32     &0.28   &0.08   \\
5$P$ &0.96  &0.92     &0.16   &0.05   \\
\hline
1$D$ &3.85  &3.68     &0.26   &0.31   \\
2$D$ &2.17  &2.07     &0.11   &0.13   \\
3$D$ &1.39  &1.32     &0.06   &0.07   \\
4$D$ &0.96  &0.92    &0.03   &0.04   \\
5$D$ &0.71  &0.68     &0.02   &0.03   \\
\hline\hline
\end{tabular}
\end{table*}
\begin{table}[htbp]
\caption{The mass spectrum (MeV) of $\Omega_{c}$ baryons are given and compared with different quark models.}\label{ppdm5}
\resizebox{\textwidth}{12cm}{\begin{tabular}{ccccccc}
\hline\hline
{\small State }$J^{P}$ & {\small Baryon} & {\small Mass} &{Ours}&{\small EFG \cite{EFG:C10}}&Ref.{\cite{ShahK:PP888}}&Ref.{\cite{Oudichhyas:PP888}} \\
\hline
$%
\begin{array}{rr}
{\small 1}^{1}{\small S}_{1/2} & {\small 1/2}^{+} \\
{\small 1}^{3}{\small S}_{3/2} & {\small 3/2}^{+}%
\end{array}%
$ & $%
\begin{array}{r}
{\small \Omega_{c}^{0}} \\
{\small \Omega_{c}(2770)^{0}}%
\end{array}%
$ & $%
\begin{array}{r}
{\small 2695.2} \\
{\small 2765.9}%
\end{array}%
$ & $%
\begin{array}{r}
{\small 2699.98} \\
{\small 2763.15}%
\end{array}%
$ & $%
\begin{array}{r}
{\small 2698} \\
{\small 2768}%
\end{array}%
$ & $%
\begin{array}{r}
{\small 2695} \\
{\small 2767}%
\end{array}%
$ & $%
\begin{array}{r}
{\small 2702} \\
{\small 2772}%
\end{array}%
$ \\ $%

\begin{array}{rr}
{\small 2}^{1}{\small S}_{1/2} & {\small 1/2}^{+} \\
{\small 2}^{3}{\small S}_{3/2} & {\small 3/2}^{+}%
\end{array}%
$ & $%
\begin{array}{r}
{\small \Omega_{c}(3185)^{0}} \\
{\small }%
\end{array}%
$ & $%
\begin{array}{r}
{\small 3185.1} \\
{\small }%
\end{array}%
$ & $%
\begin{array}{r}
{\small 3185.20} \\
{\small 3193.09}%
\end{array}%
$ & $%
\begin{array}{r}
{\small 3088} \\
{\small 3123}%
\end{array}%
$ & $%
\begin{array}{r}
{\small 3100} \\
{\small 3126}%
\end{array}%
$ & $%
\begin{array}{r}
{\small 3164} \\
{\small 3197}%
\end{array}%
$ \\ $%

\begin{array}{rr}
{\small 3}^{1}{\small S}_{1/2} & {\small 1/2}^{+} \\
{\small 3}^{3}{\small S}_{3/2} & {\small 3/2}^{+}%
\end{array}%
$ & $%
\begin{array}{r}
{\small} \\
{\small}%
\end{array}%
$ & $%
\begin{array}{r}
{\small } \\
{\small }%
\end{array}%
$ & $%
\begin{array}{r}
{\small 3543.86} \\
{\small 3548.95}%
\end{array}%
$ & $%
\begin{array}{r}
{\small 3489} \\
{\small 3510}%
\end{array}%
$ & $%
\begin{array}{r}
{\small 3436} \\
{\small 3450}%
\end{array}%
$ & $%
\begin{array}{r}
{\small 3566} \\
{\small 3571}%
\end{array}%
$ \\ $%

\begin{array}{rr}
{\small 4}^{1}{\small S}_{1/2} & {\small 1/2}^{+} \\
{\small 4}^{3}{\small S}_{3/2} & {\small 3/2}^{+}%
\end{array}%
$ & $%
\begin{array}{r}
{\small } \\
{\small }%
\end{array}%
$ & $%
\begin{array}{r}
{\small } \\
{\small }%
\end{array}%
$ & $%
\begin{array}{r}
{\small 3847.96} \\
{\small 3848.95}%
\end{array}%
$ & $%
\begin{array}{r}
{\small 3814} \\
{\small 3830}%
\end{array}%
$ & $%
\begin{array}{r}
{\small 3737} \\
{\small 3745}%
\end{array}%
$ & $%
\begin{array}{r}
{\small 3928} \\
{\small 3910}%
\end{array}%
$ \\ $%

\begin{array}{rr}
{\small 5}^{1}{\small S}_{1/2} & {\small 1/2}^{+} \\
{\small 5}^{3}{\small S}_{3/2} & {\small 3/2}^{+}%
\end{array}%
$ & $%
\begin{array}{r}
{\small } \\
{\small }%
\end{array}%
$ & $%
\begin{array}{r}
{\small } \\
{\small }%
\end{array}%
$ & $%
\begin{array}{r}
{\small 4117.37} \\
{\small 4117.88}%
\end{array}%
$ & $%
\begin{array}{r}
{\small 4102} \\
{\small 4114}%
\end{array}%
$ & $%
\begin{array}{r}
{\small 4015} \\
{\small 4021}%
\end{array}%
$ & $%
\begin{array}{r}
{\small 4259} \\
{\small 4222}%
\end{array}%
$ \\
\hline

$
\begin{array}{rr}
{\small 1}^{2}{\small P}_{1/2} & {\small 1/2}^{-} \\
{\small 1}^{4}{\small P}_{1/2} & {\small 1/2}^{-} \\
{\small 1}^{2}{\small P}_{3/2} & {\small 3/2}^{-} \\
{\small 1}^{4}{\small P}_{3/2} & {\small 3/2}^{-} \\
{\small 1}^{4}{\small P}_{5/2} & {\small 5/2}^{-}%
\end{array}%
$ & $%
\begin{array}{r}
{\small \Omega_{c}(3000)^{0}} \\
{\small \Omega_{c}(3050)^{0}} \\
{\small \Omega_{c}(3065)^{0}} \\
{\small \Omega_{c}(3090)^{0}} \\
{\small \Omega_{c}(3120)^{0}}%
\end{array}%
$ & $%
\begin{array}{r}
{\small 3000.41} \\
{\small 3050.19}\\
{\small 3065.54} \\
{\small 3090.10}\\
{\small 3119.10}%
\end{array}%
$ & $%
\begin{array}{r}
{\small 3001.93} \\
{\small 3051.74}\\
{\small 3067.14} \\
{\small 3091.72}\\
{\small 3120.64}%
\end{array}%
$ & $%
\begin{array}{r}
{\small 2966} \\
{\small 3055} \\
{\small 3029} \\
{\small 3054} \\
{\small 3051}%
\end{array}%
$ & $%
\begin{array}{r}
{\small 3011} \\
{\small 2976} \\
{\small 3028} \\
{\small 2993} \\
{\small 2947}%
\end{array}%
$ & $%
\begin{array}{r}
{\small } \\
{\small } \\
{\small 3049} \\
{\small } \\
{\small 3055}%
\end{array}%
$ \\
\hline

$
\begin{array}{rr}
{\small 2}^{2}{\small P}_{1/2} & {\small 1/2}^{-} \\
{\small 2}^{4}{\small P}_{1/2} & {\small 1/2}^{-} \\
{\small 2}^{2}{\small P}_{3/2} & {\small 3/2}^{-} \\
{\small 2}^{4}{\small P}_{3/2} & {\small 3/2}^{-} \\
{\small 2}^{4}{\small P}_{5/2} & {\small 5/2}^{-}%
\end{array}%
$ & $%
\begin{array}{r}
{\small } \\
{\small } \\
{\small } \\
{\small } \\
{\small }%
\end{array}%
$ & $%
\begin{array}{r}
 \\
\\
 \\
\\
\end{array}%
$ & $%
\begin{array}{r}
{\small 3422.48} \\
{\small 3442.70}\\
{\small 3447.59} \\
{\small 3460.73}\\
{\small 3473.23}%
\end{array}%
$ & $%
\begin{array}{r}
{\small 3384} \\
{\small 3435} \\
{\small 3415} \\
{\small 3433} \\
{\small 3427}%
\end{array}%
$ & $%
\begin{array}{r}
{\small 3345} \\
{\small 3315} \\
{\small 3359} \\
{\small 3330} \\
{\small 3290}%
\end{array}%
$ & $%
\begin{array}{r}
{\small } \\
{\small } \\
{\small 3408} \\
{\small } \\
{\small 3393}%
\end{array}%
$ \\
\hline

$
\begin{array}{rr}
{\small 3}^{2}{\small P}_{1/2} & {\small 1/2}^{-} \\
{\small 3}^{4}{\small P}_{1/2} & {\small 1/2}^{-} \\
{\small 3}^{2}{\small P}_{3/2} & {\small 3/2}^{-} \\
{\small 3}^{4}{\small P}_{3/2} & {\small 3/2}^{-} \\
{\small 3}^{4}{\small P}_{5/2} & {\small 5/2}^{-}%
\end{array}%
$ & $%
\begin{array}{r}
{\small } \\
{\small } \\
{\small } \\
{\small } \\
{\small }%
\end{array}%
$ & $%
\begin{array}{r}
\\
\\
\\
\\
\end{array}%
$ & $%
\begin{array}{r}
{\small 3752.49} \\
{\small 3763.38}\\
{\small 3765.54} \\
{\small 3773.58}\\
{\small 3780.50}%
\end{array}%
$ & $%
\begin{array}{r}
{\small 3717} \\
{\small 3754} \\
{\small 3737} \\
{\small 3752} \\
{\small 3744}%
\end{array}%
$ & $%
\begin{array}{r}
{\small 3644} \\
{\small 3620} \\
{\small 3656} \\
{\small 3632} \\
{\small 3601}%
\end{array}%
$ & $%
\begin{array}{r}
{\small } \\
{\small } \\
{\small 3732} \\
{\small } \\
{\small 3700}%
\end{array}%
$ \\
\hline

$
\begin{array}{rr}
{\small 4}^{2}{\small P}_{1/2} & {\small 1/2}^{-} \\
{\small 4}^{4}{\small P}_{1/2} & {\small 1/2}^{-} \\
{\small 4}^{2}{\small P}_{3/2} & {\small 3/2}^{-} \\
{\small 4}^{4}{\small P}_{3/2} & {\small 3/2}^{-} \\
{\small 4}^{4}{\small P}_{5/2} & {\small 5/2}^{-}%
\end{array}%
$ & $%
\begin{array}{r}
{\small } \\
{\small } \\
{\small } \\
{\small } \\
{\small }%
\end{array}%
$ & $%
\begin{array}{r}
 \\
\\
\\
\\
\end{array}%
$ & $%
\begin{array}{r}
{\small 4036.37} \\
{\small 4043.18}\\
{\small 4044.32} \\
{\small 4049.72}\\
{\small 4054.10}%
\end{array}%
$ & $%
\begin{array}{r}
{\small 4009} \\
{\small 4037} \\
{\small 4023} \\
{\small 4036} \\
{\small 4028}%
\end{array}%
$ & $%
\begin{array}{r}
{\small 3926} \\
{\small 3903} \\
{\small 3938} \\
{\small 3915} \\
{\small 3884}%
\end{array}%
$ & $%
\begin{array}{r}
{\small } \\
{\small } \\
{\small 4031} \\
{\small } \\
{\small 3983}%
\end{array}%
$ \\
\hline

$
\begin{array}{rr}
{\small 5}^{2}{\small P}_{1/2} & {\small 1/2}^{-} \\
{\small 5}^{4}{\small P}_{1/2} & {\small 1/2}^{-} \\
{\small 5}^{2}{\small P}_{3/2} & {\small 3/2}^{-} \\
{\small 5}^{4}{\small P}_{3/2} & {\small 3/2}^{-} \\
{\small 5}^{4}{\small P}_{5/2} & {\small 5/2}^{-}%
\end{array}%
$ & $%
\begin{array}{r}
{\small } \\
{\small } \\
{\small } \\
{\small } \\
{\small }%
\end{array}%
$ & $%
\begin{array}{r}
\\
\\
\\
\\
\end{array}%
$ & $%
\begin{array}{r}
{\small 4290.35} \\
{\small 4295.00}\\
{\small 4295.68} \\
{\small 4299.55}\\
{\small 4302.57}%
\end{array}%
$ & $%
\begin{array}{r}
{\small } \\
{\small } \\
{\small } \\
{\small } \\
{\small }%
\end{array}%
$ & $%
\begin{array}{r}
{\small } \\
{\small } \\
{\small } \\
{\small } \\
{\small }%
\end{array}%
$ & $%
\begin{array}{r}
{\small } \\
{\small } \\
{\small 4309} \\
{\small } \\
{\small 4248}%
\end{array}%
$ \\
\hline

$
\begin{array}{rr}
{\small 1}^{4}{\small D}_{1/2} & {\small 1/2}^{+} \\
{\small 1}^{2}{\small D}_{3/2} & {\small 3/2}^{+} \\
{\small 1}^{4}{\small D}_{3/2} & {\small 3/2}^{+} \\
{\small 1}^{2}{\small D}_{5/2} & {\small 5/2}^{+} \\
{\small 1}^{4}{\small D}_{5/2} & {\small 5/2}^{+} \\
{\small 1}^{4}{\small D}_{7/2} & {\small 7/2}^{+}%
\end{array}%
$ & $%
\begin{array}{r}
{\small } \\
{\small \Omega_{c}(3327)^{0}} \\
{\small } \\
{\small } \\
{\small } \\
{\small }%
\end{array}%
$ & $%
\begin{array}{r}
{\small }\\
{\small 3327.1}\\
{\small }\\
{\small }\\
{\small }\\
{\small }%
\end{array}%
$ & $%
\begin{array}{r}
{\small 3308.41} \\
{\small 3326.92}\\
{\small 3342.64}\\
{\small 3356.96}\\
{\small 3373.08}\\
{\small 3397.52}%
\end{array}%
$ & $%
\begin{array}{r}
{\small 3287} \\
{\small 3282} \\
{\small 3298} \\
{\small 3286} \\
{\small 3297} \\
{\small 3283}%
\end{array}%
$ & $%
\begin{array}{r}
{\small 3215} \\
{\small 3231} \\
{\small 3262} \\
{\small 3188} \\
{\small 3173} \\
{\small 3136}%
\end{array}%
$ & $%
\begin{array}{r}
{\small } \\
{\small } \\
{\small } \\
{\small 3360} \\
{\small } \\
{\small 3314}%
\end{array}%
$ \\
\hline

$
\begin{array}{rr}
{\small 2}^{4}{\small D}_{1/2} & {\small 1/2}^{+} \\
{\small 2}^{2}{\small D}_{3/2} & {\small 3/2}^{+} \\
{\small 2}^{4}{\small D}_{3/2} & {\small 3/2}^{+} \\
{\small 2}^{2}{\small D}_{5/2} & {\small 5/2}^{+} \\
{\small 2}^{4}{\small D}_{5/2} & {\small 5/2}^{+} \\
{\small 2}^{4}{\small D}_{7/2} & {\small 7/2}^{+}%
\end{array}%
$ & $%
\begin{array}{r}
{\small } \\
{\small } \\
{\small } \\
{\small } \\
{\small } \\
{\small }%
\end{array}%
$ & $%
\begin{array}{r}
 \\
\\
\\
\\
\\
\end{array}%
$ & $%
\begin{array}{r}
{\small 3659.91} \\
{\small 3670.21}\\
{\small 3679.26}\\
{\small 3687.03} \\
{\small 3696.38}\\
{\small 3709.95}%
\end{array}%
$ & $%
\begin{array}{r}
{\small 3623} \\
{\small 3613} \\
{\small 3627} \\
{\small 3614} \\
{\small 3626} \\
{\small 3611}%
\end{array}%
$ & $%
\begin{array}{r}
{\small 3524} \\
{\small 3538} \\
{\small 3565} \\
{\small 3502} \\
{\small 3488} \\
{\small 3456}%
\end{array}%
$ & $%
\begin{array}{r}
{\small } \\
{\small } \\
{\small } \\
{\small 3680} \\
{\small } \\
{\small 3656}%
\end{array}%
$ \\
\hline

$
\begin{array}{rr}
{\small 3}^{4}{\small D}_{1/2} & {\small 1/2}^{+} \\
{\small 3}^{2}{\small D}_{3/2} & {\small 3/2}^{+} \\
{\small 3}^{4}{\small D}_{3/2} & {\small 3/2}^{+} \\
{\small 3}^{2}{\small D}_{5/2} & {\small 5/2}^{+} \\
{\small 3}^{4}{\small D}_{5/2} & {\small 5/2}^{+} \\
{\small 3}^{4}{\small D}_{7/2} & {\small 7/2}^{+}%
\end{array}%
$ & $%
\begin{array}{r}
{\small } \\
{\small } \\
{\small } \\
{\small } \\
{\small } \\
{\small }%
\end{array}%
$ & $%
\begin{array}{r}
\\
\\
\\
\\
\\
\end{array}%
$ & $%
\begin{array}{r}
{\small 3956.72} \\
{\small 3963.26}\\
{\small 3969.13} \\
{\small 3974.00}\\
{\small 3980.09}\\
{\small 3988.71}%
\end{array}%
$ & $%
\begin{array}{r}
{\small } \\
{\small } \\
{\small } \\
{\small } \\
{\small } \\
{\small }%
\end{array}%
$ & $%
\begin{array}{r}
{\small } \\
{\small } \\
{\small } \\
{\small } \\
{\small } \\
{\small }%
\end{array}%
$ & $%
\begin{array}{r}
{\small } \\
{\small } \\
{\small } \\
{\small 3974} \\
{\small } \\
{\small 3968}%
\end{array}%
$ \\
\hline

$
\begin{array}{rr}
{\small 4}^{4}{\small D}_{1/2} & {\small 1/2}^{+} \\
{\small 4}^{2}{\small D}_{3/2} & {\small 3/2}^{+} \\
{\small 4}^{4}{\small D}_{3/2} & {\small 3/2}^{+} \\
{\small 4}^{2}{\small D}_{5/2} & {\small 5/2}^{+} \\
{\small 4}^{4}{\small D}_{5/2} & {\small 5/2}^{+} \\
{\small 4}^{4}{\small D}_{7/2} & {\small 7/2}^{+}%
\end{array}%
$ & $%
\begin{array}{r}
{\small } \\
{\small } \\
{\small } \\
{\small } \\
{\small } \\
{\small }%
\end{array}%
$ & $%
\begin{array}{r}
\\
\\
\\
\\
\\
\end{array}%
$ & $%
\begin{array}{r}
{\small 4219.44}\\
{\small 4223.96}\\
{\small 4228.08}\\
{\small 4231.41}\\
{\small 4235.69}\\
{\small 4241.64}%
\end{array}%
$ & $%
\begin{array}{r}
{\small } \\
{\small } \\
{\small } \\
{\small } \\
{\small } \\
{\small }%
\end{array}%
$ & $%
\begin{array}{r}
{\small } \\
{\small } \\
{\small } \\
{\small } \\
{\small } \\
{\small }%
\end{array}%
$ & $%
\begin{array}{r}
{\small } \\
{\small } \\
{\small } \\
{\small 4248} \\
{\small } \\
{\small 4258}%
\end{array}%
$ \\
\hline

$
\begin{array}{rr}
{\small 5}^{4}{\small D}_{1/2} & {\small 1/2}^{+} \\
{\small 5}^{2}{\small D}_{3/2} & {\small 3/2}^{+} \\
{\small 5}^{4}{\small D}_{3/2} & {\small 3/2}^{+} \\
{\small 5}^{2}{\small D}_{5/2} & {\small 5/2}^{+} \\
{\small 5}^{4}{\small D}_{5/2} & {\small 5/2}^{+} \\
{\small 5}^{4}{\small D}_{7/2} & {\small 7/2}^{+}%
\end{array}%
$ & $%
\begin{array}{r}
{\small } \\
{\small } \\
{\small } \\
{\small } \\
{\small } \\
{\small }%
\end{array}%
$ & $%
\begin{array}{r}
\\
\\
\\
\\
\\
\end{array}%
$ & $%
\begin{array}{r}
{\small 4458.12} \\
{\small 4461.43}\\
{\small 4464.47}\\
{\small 4466.89} \\
{\small 4470.06}\\
{\small 4474.42}%
\end{array}%
$ & $%
\begin{array}{r}
{\small } \\
{\small } \\
{\small } \\
{\small } \\
{\small } \\
{\small }%
\end{array}%
$ & $%
\begin{array}{r}
{\small } \\
{\small } \\
{\small } \\
{\small } \\
{\small } \\
{\small }%
\end{array}%
$ & $%
\begin{array}{r}
{\small } \\
{\small } \\
{\small } \\
{\small 4505} \\
{\small } \\
{\small 4529}%
\end{array}%
$ \\
\hline\hline
\end{tabular}}
\end{table}
\begin{table*}[!htbp]
\caption{The mass spectrum (MeV) of $\Omega_{b}$ baryons are given and compared with different quark models.}\label{ppdm6}
\resizebox{\textwidth}{12cm}{\begin{tabular}{ccccccc}
\hline\hline
{\small State }$J^{P}$ & {\small Baryon} & {\small Mass} &{Ours}&{\small EFG \cite{EFG:C10}}&Ref.{\cite{Kakayas:PP888}} &Ref.{\cite{OudichhyGan:PP888}}  \\
\hline
$%
\begin{array}{rr}
{\small 1}^{1}{\small S}_{1/2} & {\small 1/2}^{+} \\
{\small 1}^{3}{\small S}_{3/2} & {\small 3/2}^{+}%
\end{array}%
$ & $%
\begin{array}{r}
{\small \Omega_{b}}^{-} \\
{\small }%
\end{array}%
$ & $%
\begin{array}{r}
{\small 6045.2} \\
{\small }%
\end{array}%
$ & $%
\begin{array}{r}
{\small 6040.25} \\
{\small 6060.55}%
\end{array}%
$ & $%
\begin{array}{r}
{\small 6064} \\
{\small 6088}%
\end{array}%
$ & $%
\begin{array}{r}
{\small 6046} \\
{\small 6082}%
\end{array}%
$ & $%
\begin{array}{r}
{\small 6054} \\
{\small 6074}%
\end{array}%
$ \\ $%

\begin{array}{rr}
{\small 2}^{1}{\small S}_{1/2} & {\small 1/2}^{+} \\
{\small 2}^{3}{\small S}_{3/2} & {\small 3/2}^{+}%
\end{array}%
$ & $%
\begin{array}{r}
{\small } \\
{\small }%
\end{array}%
$ & $%
\begin{array}{r}
{\small } \\
{\small }%
\end{array}%
$ & $%
\begin{array}{r}
{\small 6439.49} \\
{\small 6442.03}%
\end{array}%
$ & $%
\begin{array}{r}
{\small 6450} \\
{\small 6461}%
\end{array}%
$ & $%
\begin{array}{r}
{\small 6438} \\
{\small 6462}%
\end{array}%
$ & $%
\begin{array}{r}
{\small 6455} \\
{\small 6481}%
\end{array}%
$ \\ $%

\begin{array}{rr}
{\small 3}^{1}{\small S}_{1/2} & {\small 1/2}^{+} \\
{\small 3}^{3}{\small S}_{3/2} & {\small 3/2}^{+}%
\end{array}%
$ & $%
\begin{array}{r}
{\small} \\
{\small}%
\end{array}%
$ & $%
\begin{array}{r}
{\small } \\
{\small }%
\end{array}%
$ & $%
\begin{array}{r}
{\small 6763.25} \\
{\small 6764.01}%
\end{array}%
$ & $%
\begin{array}{r}
{\small 6804} \\
{\small 6811}%
\end{array}%
$ & $%
\begin{array}{r}
{\small 6740} \\
{\small 6753}%
\end{array}%
$ & $%
\begin{array}{r}
{\small 6832} \\
{\small 6864}%
\end{array}%
$ \\ $%

\begin{array}{rr}
{\small 4}^{1}{\small S}_{1/2} & {\small 1/2}^{+} \\
{\small 4}^{3}{\small S}_{3/2} & {\small 3/2}^{+}%
\end{array}%
$ & $%
\begin{array}{r}
{\small } \\
{\small }%
\end{array}%
$ & $%
\begin{array}{r}
{\small } \\
{\small }%
\end{array}%
$ & $%
\begin{array}{r}
{\small 7045.87} \\
{\small 7046.19}%
\end{array}%
$ & $%
\begin{array}{r}
{\small 7091} \\
{\small 7096}%
\end{array}%
$ & $%
\begin{array}{r}
{\small 7022} \\
{\small 7030}%
\end{array}%
$ & $%
\begin{array}{r}
{\small 7190} \\
{\small 7226}%
\end{array}%
$ \\ $%

\begin{array}{rr}
{\small 5}^{1}{\small S}_{1/2} & {\small 1/2}^{+} \\
{\small 5}^{3}{\small S}_{3/2} & {\small 3/2}^{+}%
\end{array}%
$ & $%
\begin{array}{r}
{\small } \\
{\small }%
\end{array}%
$ & $%
\begin{array}{r}
{\small } \\
{\small }%
\end{array}%
$ & $%
\begin{array}{r}
{\small 7300.17} \\
{\small 7300.33}%
\end{array}%
$ & $%
\begin{array}{r}
{\small 7338} \\
{\small 7343}%
\end{array}%
$ & $%
\begin{array}{r}
{\small 7290} \\
{\small 7296}%
\end{array}%
$ & $%
\begin{array}{r}
{\small 7531} \\
{\small 7572}%
\end{array}%
$ \\
\hline

$
\begin{array}{rr}
{\small 1}^{2}{\small P}_{1/2} & {\small 1/2}^{-} \\
{\small 1}^{4}{\small P}_{1/2} & {\small 1/2}^{-} \\
{\small 1}^{2}{\small P}_{3/2} & {\small 3/2}^{-} \\
{\small 1}^{4}{\small P}_{3/2} & {\small 3/2}^{-} \\
{\small 1}^{4}{\small P}_{5/2} & {\small 5/2}^{-}%
\end{array}%
$ & $%
\begin{array}{r}
{\small \Omega_{b}(6316)^{-}} \\
{\small \Omega_{b}(6330)^{-}} \\
{\small \Omega_{b}(6340)^{-}} \\
{\small \Omega_{b}(6350)^{-}} \\
{\small }%
\end{array}%
$ & $%
\begin{array}{r}
{\small 6315.6} \\
{\small 6333.3}\\
{\small 6339.7} \\
{\small 6349.8}\\
{\small }%
\end{array}%
$ & $%
\begin{array}{r}
{\small 6318.95} \\
{\small 6334.95}\\
{\small 6339.90} \\
{\small 6347.80}\\
{\small 6357.09}%
\end{array}%
$ & $%
\begin{array}{r}
{\small 6330} \\
{\small 6339} \\
{\small 6331} \\
{\small 6340} \\
{\small 6334}%
\end{array}%
$ & $%
\begin{array}{r}
{\small 4344} \\
{\small 4345} \\
{\small 4341} \\
{\small 4343} \\
{\small 4339}%
\end{array}%
$ & $%
\begin{array}{r}
{\small } \\
{\small } \\
{\small 6348} \\
{\small } \\
{\small 6362}%
\end{array}%
$ \\
\hline

$
\begin{array}{rr}
{\small 2}^{2}{\small P}_{1/2} & {\small 1/2}^{-} \\
{\small 2}^{4}{\small P}_{1/2} & {\small 1/2}^{-} \\
{\small 2}^{2}{\small P}_{3/2} & {\small 3/2}^{-} \\
{\small 2}^{4}{\small P}_{3/2} & {\small 3/2}^{-} \\
{\small 2}^{4}{\small P}_{5/2} & {\small 5/2}^{-}%
\end{array}%
$ & $%
\begin{array}{r}
{\small } \\
{\small } \\
{\small } \\
{\small } \\
{\small }%
\end{array}%
$ & $%
\begin{array}{r}
 \\
\\
 \\
\\
\end{array}%
$ & $%
\begin{array}{r}
{\small 6670.62} \\
{\small 6677.12}\\
{\small 6678.69} \\
{\small 6682.91}\\
{\small 6686.93}%
\end{array}%
$ & $%
\begin{array}{r}
{\small 6706} \\
{\small 6710} \\
{\small 6699} \\
{\small 6705} \\
{\small 6700}%
\end{array}%
$ & $%
\begin{array}{r}
{\small 6596} \\
{\small 6597} \\
{\small 6594} \\
{\small 6595} \\
{\small 6592}%
\end{array}%
$ & $%
\begin{array}{r}
{\small } \\
{\small } \\
{\small 6662} \\
{\small } \\
{\small 6653}%
\end{array}%
$ \\
\hline

$
\begin{array}{rr}
{\small 3}^{2}{\small P}_{1/2} & {\small 1/2}^{-} \\
{\small 3}^{4}{\small P}_{1/2} & {\small 1/2}^{-} \\
{\small 3}^{2}{\small P}_{3/2} & {\small 3/2}^{-} \\
{\small 3}^{4}{\small P}_{3/2} & {\small 3/2}^{-} \\
{\small 3}^{4}{\small P}_{5/2} & {\small 5/2}^{-}%
\end{array}%
$ & $%
\begin{array}{r}
{\small } \\
{\small } \\
{\small } \\
{\small } \\
{\small }%
\end{array}%
$ & $%
\begin{array}{r}
\\
\\
\\
\\
\end{array}%
$ & $%
\begin{array}{r}
{\small 6967.16} \\
{\small 6970.66}\\
{\small 6971.35} \\
{\small 6973.94}\\
{\small 6976.16}%
\end{array}%
$ & $%
\begin{array}{r}
{\small 7003} \\
{\small 7009} \\
{\small 6998} \\
{\small 7002} \\
{\small 6996}%
\end{array}%
$ & $%
\begin{array}{r}
{\small 6829} \\
{\small 6830} \\
{\small 6827} \\
{\small 6828} \\
{\small 6826}%
\end{array}%
$ & $%
\begin{array}{r}
{\small } \\
{\small } \\
{\small 6962} \\
{\small } \\
{\small 6689}%
\end{array}%
$ \\
\hline

$
\begin{array}{rr}
{\small 4}^{2}{\small P}_{1/2} & {\small 1/2}^{-} \\
{\small 4}^{4}{\small P}_{1/2} & {\small 1/2}^{-} \\
{\small 4}^{2}{\small P}_{3/2} & {\small 3/2}^{-} \\
{\small 4}^{4}{\small P}_{3/2} & {\small 3/2}^{-} \\
{\small 4}^{4}{\small P}_{5/2} & {\small 5/2}^{-}%
\end{array}%
$ & $%
\begin{array}{r}
{\small } \\
{\small } \\
{\small } \\
{\small } \\
{\small }%
\end{array}%
$ & $%
\begin{array}{r}
 \\
\\
\\
\\
\end{array}%
$ & $%
\begin{array}{r}
{\small 7230.27} \\
{\small 7232.46}\\
{\small 7232.83} \\
{\small 7234.56}\\
{\small 7235.97}%
\end{array}%
$ & $%
\begin{array}{r}
{\small 7257} \\
{\small 7265} \\
{\small 7250} \\
{\small 7258} \\
{\small 7251}%
\end{array}%
$ & $%
\begin{array}{r}
{\small 7044} \\
{\small 7043} \\
{\small 7043} \\
{\small 7043} \\
{\small 7042}%
\end{array}%
$ & $%
\begin{array}{r}
{\small } \\
{\small } \\
{\small 7249} \\
{\small } \\
{\small 7200}%
\end{array}%
$ \\
\hline

$
\begin{array}{rr}
{\small 5}^{2}{\small P}_{1/2} & {\small 1/2}^{-} \\
{\small 5}^{4}{\small P}_{1/2} & {\small 1/2}^{-} \\
{\small 5}^{2}{\small P}_{3/2} & {\small 3/2}^{-} \\
{\small 5}^{4}{\small P}_{3/2} & {\small 3/2}^{-} \\
{\small 5}^{4}{\small P}_{5/2} & {\small 5/2}^{-}%
\end{array}%
$ & $%
\begin{array}{r}
{\small } \\
{\small } \\
{\small } \\
{\small } \\
{\small }%
\end{array}%
$ & $%
\begin{array}{r}
\\
\\
\\
\\
\end{array}%
$ & $%
\begin{array}{r}
{\small 7469.70} \\
{\small 7471.19}\\
{\small 7471.41} \\
{\small 7472.65}\\
{\small 7473.62}%
\end{array}%
$ & $%
\begin{array}{r}
{\small } \\
{\small } \\
{\small } \\
{\small } \\
{\small }%
\end{array}%
$ & $%
\begin{array}{r}
{\small } \\
{\small } \\
{\small } \\
{\small } \\
{\small }%
\end{array}%
$ & $%
\begin{array}{r}
{\small } \\
{\small } \\
{\small 7526} \\
{\small } \\
{\small 7458}%
\end{array}%
$ \\
\hline

$
\begin{array}{rr}
{\small 1}^{4}{\small D}_{1/2} & {\small 1/2}^{+} \\
{\small 1}^{2}{\small D}_{3/2} & {\small 3/2}^{+} \\
{\small 1}^{4}{\small D}_{3/2} & {\small 3/2}^{+} \\
{\small 1}^{2}{\small D}_{5/2} & {\small 5/2}^{+} \\
{\small 1}^{4}{\small D}_{5/2} & {\small 5/2}^{+} \\
{\small 1}^{4}{\small D}_{7/2} & {\small 7/2}^{+}%
\end{array}%
$ & $%
\begin{array}{r}
{\small } \\
{\small } \\
{\small } \\
{\small } \\
{\small } \\
{\small }%
\end{array}%
$ & $%
\begin{array}{r}
 \\
\\
\\
\\
\\
\end{array}%
$ & $%
\begin{array}{r}
{\small 6578.47} \\
{\small 6584.41}\\
{\small 6589.46}\\
{\small 6594.07}\\
{\small 6599.25}\\
{\small 6607.10}%
\end{array}%
$ & $%
\begin{array}{r}
{\small 6540} \\
{\small 6530} \\
{\small 6549} \\
{\small 6520} \\
{\small 6529} \\
{\small 6517}%
\end{array}%
$ & $%
\begin{array}{r}
{\small 6485} \\
{\small 6480} \\
{\small 6482} \\
{\small 6476} \\
{\small 6478} \\
{\small 6472}%
\end{array}%
$ & $%
\begin{array}{r}
{\small } \\
{\small } \\
{\small } \\
{\small 6629} \\
{\small } \\
{\small 6638}%
\end{array}%
$ \\
\hline

$
\begin{array}{rr}
{\small 2}^{4}{\small D}_{1/2} & {\small 1/2}^{+} \\
{\small 2}^{2}{\small D}_{3/2} & {\small 3/2}^{+} \\
{\small 2}^{4}{\small D}_{3/2} & {\small 3/2}^{+} \\
{\small 2}^{2}{\small D}_{5/2} & {\small 5/2}^{+} \\
{\small 2}^{4}{\small D}_{5/2} & {\small 5/2}^{+} \\
{\small 2}^{4}{\small D}_{7/2} & {\small 7/2}^{+}%
\end{array}%
$ & $%
\begin{array}{r}
{\small } \\
{\small } \\
{\small } \\
{\small } \\
{\small } \\
{\small }%
\end{array}%
$ & $%
\begin{array}{r}
 \\
\\
\\
\\
\\
\end{array}%
$ & $%
\begin{array}{r}
{\small 6888.04} \\
{\small 6891.35}\\
{\small 6894.26}\\
{\small 6896.75} \\
{\small 6899.76}\\
{\small 6904.12}%
\end{array}%
$ & $%
\begin{array}{r}
{\small 6857} \\
{\small 6846} \\
{\small 6863} \\
{\small 6837} \\
{\small 6846} \\
{\small 6834}%
\end{array}%
$ & $%
\begin{array}{r}
{\small 6730} \\
{\small 6726} \\
{\small 6727} \\
{\small 6723} \\
{\small 6724} \\
{\small 6720}%
\end{array}%
$ & $%
\begin{array}{r}
{\small } \\
{\small } \\
{\small } \\
{\small 6659} \\
{\small } \\
{\small 6643}%
\end{array}%
$ \\
\hline

$
\begin{array}{rr}
{\small 3}^{4}{\small D}_{1/2} & {\small 1/2}^{+} \\
{\small 3}^{2}{\small D}_{3/2} & {\small 3/2}^{+} \\
{\small 3}^{4}{\small D}_{3/2} & {\small 3/2}^{+} \\
{\small 3}^{2}{\small D}_{5/2} & {\small 5/2}^{+} \\
{\small 3}^{4}{\small D}_{5/2} & {\small 5/2}^{+} \\
{\small 3}^{4}{\small D}_{7/2} & {\small 7/2}^{+}%
\end{array}%
$ & $%
\begin{array}{r}
{\small } \\
{\small } \\
{\small } \\
{\small } \\
{\small } \\
{\small }%
\end{array}%
$ & $%
\begin{array}{r}
\\
\\
\\
\\
\\
\end{array}%
$ & $%
\begin{array}{r}
{\small 7159.80} \\
{\small 7161.90}\\
{\small 7163.79} \\
{\small 7165.35}\\
{\small 7167.31}\\
{\small 7170.08}%
\end{array}%
$ & $%
\begin{array}{r}
{\small } \\
{\small } \\
{\small } \\
{\small } \\
{\small } \\
{\small }%
\end{array}%
$ & $%
\begin{array}{r}
{\small 6956} \\
{\small 6953} \\
{\small 6954} \\
{\small 6951} \\
{\small 6951} \\
{\small 6948}%
\end{array}%
$ & $%
\begin{array}{r}
{\small } \\
{\small } \\
{\small } \\
{\small 6689} \\
{\small } \\
{\small 6648}%
\end{array}%
$ \\
\hline

$
\begin{array}{rr}
{\small 4}^{4}{\small D}_{1/2} & {\small 1/2}^{+} \\
{\small 4}^{2}{\small D}_{3/2} & {\small 3/2}^{+} \\
{\small 4}^{4}{\small D}_{3/2} & {\small 3/2}^{+} \\
{\small 4}^{2}{\small D}_{5/2} & {\small 5/2}^{+} \\
{\small 4}^{4}{\small D}_{5/2} & {\small 5/2}^{+} \\
{\small 4}^{4}{\small D}_{7/2} & {\small 7/2}^{+}%
\end{array}%
$ & $%
\begin{array}{r}
{\small } \\
{\small } \\
{\small } \\
{\small } \\
{\small } \\
{\small }%
\end{array}%
$ & $%
\begin{array}{r}
\\
\\
\\
\\
\\
\end{array}%
$ & $%
\begin{array}{r}
{\small 7405.49}\\
{\small 7406.94}\\
{\small 7408.27}\\
{\small 7409.34}\\
{\small 7410.71}\\
{\small 7412.63}%
\end{array}%
$ & $%
\begin{array}{r}
{\small } \\
{\small } \\
{\small } \\
{\small } \\
{\small } \\
{\small }%
\end{array}%
$ & $%
\begin{array}{r}
{\small 7166} \\
{\small 7164} \\
{\small 7164} \\
{\small 7162} \\
{\small 7162} \\
{\small 7160}%
\end{array}%
$ & $%
\begin{array}{r}
{\small } \\
{\small } \\
{\small } \\
{\small 6719} \\
{\small } \\
{\small 6653}%
\end{array}%
$ \\
\hline

$
\begin{array}{rr}
{\small 5}^{4}{\small D}_{1/2} & {\small 1/2}^{+} \\
{\small 5}^{2}{\small D}_{3/2} & {\small 3/2}^{+} \\
{\small 5}^{4}{\small D}_{3/2} & {\small 3/2}^{+} \\
{\small 5}^{2}{\small D}_{5/2} & {\small 5/2}^{+} \\
{\small 5}^{4}{\small D}_{5/2} & {\small 5/2}^{+} \\
{\small 5}^{4}{\small D}_{7/2} & {\small 7/2}^{+}%
\end{array}%
$ & $%
\begin{array}{r}
{\small } \\
{\small } \\
{\small } \\
{\small } \\
{\small } \\
{\small }%
\end{array}%
$ & $%
\begin{array}{r}
\\
\\
\\
\\
\\
\end{array}%
$ & $%
\begin{array}{r}
{\small 7631.64} \\
{\small 7632.71}\\
{\small 7633.68}\\
{\small 7634.46} \\
{\small 7635.48}\\
{\small 7636.88}%
\end{array}%
$ & $%
\begin{array}{r}
{\small } \\
{\small } \\
{\small } \\
{\small } \\
{\small } \\
{\small }%
\end{array}%
$ & $%
\begin{array}{r}
{\small } \\
{\small } \\
{\small } \\
{\small } \\
{\small } \\
{\small }%
\end{array}%
$ & $%
\begin{array}{r}
{\small } \\
{\small } \\
{\small } \\
{\small 7458} \\
{\small } \\
{\small 6658}%
\end{array}%
$ \\
\hline\hline
\end{tabular}}
\end{table*}

\section{The baryons $\Sigma _{c}$ and $\Sigma _{b}$}

By analyzing the existing experimental data in PDG \cite{Workman:A11}, we explore some patterns of the odd-parity $\Sigma_{Q}$ $(Q=c,b)$ baryons consisting of a light isospin-one nonstrange diquark $(nn=uu, ud, dd)$ in a state of $L$ with respect to the spin-1/2 heavy quark $Q$. So far, the $\Sigma_{Q}$ baryons have been observed in experiments, and the data are available from the Particle Data Group, which provides us with more information to study the mass spectra of the $\Sigma_{Q}$ states.

Ref. \cite{Workman:A11} cites the two masses $M(\Sigma_c,1/2^+)=2452.65$ MeV, $M(\Sigma_c,3/2^+)=2517.4$ MeV for $\Sigma_{c}(2455)^{+}$, $\Sigma_{c}(2520)^{+}$ with $J^{P}=1/2^{+}$ and $3/2^{+}$, respectively, which was discovered and identified as $1S$-wave states by the LHCb experiment. Accordingly, by using Particle Data Group masses, the spin-weighted average mass is obtained by \cite{Vinodkumarai:PP888}
\begin{eqnarray}
\bar M^{\text{spin-weighted}}=\frac{\Sigma(2J+1) M(J)}{\Sigma(2J+1)}=\frac{(2\times2453.75\ \text{MeV}+4\times2517.5\ \text{MeV})}{6}=2496.25\ \text{MeV}. \label{www123}
\end{eqnarray}
As can be seen from Table \ref{tab:Eff-mass11}, the spin-average mass $\bar M(\Sigma_{c}, 1S)$ = 2496.09\ MeV is very close to the experimental value in Eq. (\ref{www123}). As the hyperfine splitting $2517.4\mbox{\,MeV}-2452.65\mbox{\,MeV}=64.75$ $\mbox{\,MeV}$ between $\Sigma_{c}(2455)^{+}$ and $\Sigma_{c}(2520)^{+}$ is regarded as a good reference for comparing the results of our model. For the $\Sigma_{c}$ baryons, comparing the measured masses presented in Table \ref{pp12dm1} with our prediction masses, and the parameters as shown in Table \ref{Table 8}, it is seen that the masses of all these states are compatible with the experimental values (within few MeV). We employ Eq. (\ref{pp4}) to calculate the spin-averaged mass $\bar M$ of $1S$-wave with $L=0$, $n=0$,
\begin{eqnarray}
\bar M(\Sigma_{c}, 1S)=M_{c}+\left( m_{nn}+M_{c}\left( 1-\frac{m_{\text{cur}c}^{2}}{M_{c}^{2}}\right) \right)=2496.09\ \text{MeV} ,  \label{we123}
\end{eqnarray}
as well as the following rough estimate for the parameter $c_{1}$ by Eq. (\ref{scr:pp888}),
\begin{eqnarray}
c_{1}(\Sigma_{c}, 1S)=\frac{M_{c}m_{ss}}{M_{c}m_{nn}}\frac{{N^{\prime}_{c}}}{{N_{c}}}c_{1}(\Omega_{c},1P)=\frac{0.991}{0.745}\frac{(1+3.3)(0+1+1)^{3}}{(0+3.3)(0+0+1)^{3}}4.04\ \text{MeV}=56.02\ \text{MeV},  \label{we123}
\end{eqnarray}
with $c_{1}(\Omega_{c},1P)=4.04$\ MeV given in Eq. (\ref{VVq11}). Note that the heavy quark mass $M_c$ cancels out for charmed baryons.

The $\Sigma_c(2800)$ observed by the Belle Collaboration \cite{Mizuke:PPP888} might be a good candidate for a $1P$-wave state (cf.\ e.g.\ Ref. \cite{KarlinerRP:PP888}). For comparison with the experiment values, we also compute the parameters and the masses of the $\Sigma_{c}$ in $1P$-wave states,
\begin{eqnarray}
&&\bar M=2774.67\ \text{MeV},a_{1}=35.86\ \text{MeV},a_{2}=34.27\ \text{MeV},b_{1}=17.97\ \text{MeV},c_{1}=5.37\ \text{MeV}, \label{Vo1q11}\\
&&M(\Sigma_{c},1P): 2668.86\ \text{MeV}, 2735.11\ \text{MeV}, 2755.59\ \text{MeV}, 2788.31\ \text{MeV}, 2826.76\ \text{MeV}.  \label{Vo11q11}
\end{eqnarray}
Although the Belle Collaboration observed the excited $\Sigma_{c}(2800)$ state in the decay channel $\Lambda^{+}_{c}\pi$ \cite{Measuream:PP888} which mass at $M(\Sigma_{c})$ = 2792\ MeV, the $J^{P}$ has not been determined, making it difficult to determine its properties. The $\Sigma_{c}(2800)$ state is calculated by our model to own the mass 2788.31\ MeV, which is in agreement with the experiment as show in Table \ref{pp12dm1}. Hence, for $\Sigma_c(2800)$ we should advocate the fourth state $|{}^4P_{3/2},3/2^{-}\rangle$ of $1P$-wave. The nature of these states is discussed in Refs. \cite{EFG:C10, ChengChua:PP888}.

In the $\Sigma_{b}$ baryon family, there are four states with masses $M(\Sigma^{+}_{b}, 1/2^{+})=5810.56$\ MeV and $M(\Sigma^{\ast+}_{b}, 3/2^{+})=5830.32$\ MeV in PDG \cite{Workman:A11} for the $\Sigma^{+}_{b}$ and $\Sigma^{\ast+}_{b}$ states, and $M(\Sigma^{-}_{b}, 1/2^{+})=5815.64$\ MeV and $M(\Sigma^{\ast-}_{b}, 3/2^{+})=5834.74$\ MeV for the $\Sigma^{-}_{b}$ and $\Sigma^{\ast-}_{b}$ states, respectively. It should be pointed out that the neutral $1S$-wave $\Sigma^{0}_{b}$, $\Sigma^{\ast0}_{b}$ states are still missing. In addition, $\Sigma_{b}(6097)$ has been measured using fully reconstructed $\Lambda^{0}_{b}\rightarrow\Lambda^{+}_{c}\pi^{-}$ and $\Lambda^{+}_{c}\rightarrow \rho \kappa^{+}_{c}\pi^{+}$ decays in Ref. \cite{collaborationT:PPP888}. In our calculations, $\Sigma_{b}(6097)$ can be a good candidate of $1P$-wave excitations. Therefore, we assign $J^{P}=5/2^{-}$ to the $\Sigma_{b}(6097)$ state. Finally, the spin-averaged mass, the parameters and the mass splitting are given by Eq. (\ref{pp4}) and Eq. (\ref{scr:pp888}) in $1P$-wave ($L=1, n=0$),
\begin{eqnarray}
\bar M(\Sigma_{b}, 1P)&=&M_{b}+\sqrt{\pi\alpha(\Sigma_{b})+\left( m_{nn}+M_{b}\left( 1-\frac{m_{\text{cur}b}^{2}}{M_{b}^{2}}\right) \right)^{2}} \notag \\
&=&6082.84\ \text{MeV} , \\ \label{we123}
a_{1}(\Sigma_{b},1P)&=&\frac{M_{c}m_{ss}}{M_{b}m_{nn}}\frac{(n^{\prime}+L^{\prime}+1)^{2}}{(n+L+1)^{2}}a_{1}(\Omega_{c},1P) \notag \\
&=&\frac{1.44\times0.991}{4.48\times0.745}\frac{(0+1+1)^{2}}{(0+1+1)^{2}}26.96\ \text{MeV} \notag \\
&=&11.53\ \text{MeV},\\
a_{2}(\Sigma_{b},1P)&=&\frac{M_{c}m_{ss}}{M_{b}m_{nn}}\frac{(n^{\prime}+L^{\prime}+1)^{2}}{(n+L+1)^{2}}a_{2}(\Omega_{c},1P) \notag \\
&=&\frac{1.44\times0.991}{4.48\times0.745}\frac{(0+1+1)^{2}}{(0+1+1)^{2}}25.76\ \text{MeV} \notag \\
&=&11.01\ \text{MeV},\\
b_{1}(\Sigma_{b},1P)&=&\frac{M_{c}m_{ss}}{M_{b}m_{nn}}\frac{L^{\prime}(L^{\prime}+\frac{1}{2})(L^{\prime}+1)(n^{\prime}+L^{\prime}+1)^{3}}{L(L+\frac{1}{2})(L+1)(n+L+1)^{3}}b_{1}(\Omega_{c},1P) \notag \\
&=&\frac{1.44\times0.991}{4.48\times0.745}\frac{(1+\frac{1}{2})(1+1)(0+1+1)^{3}}{(1+\frac{1}{2})(1+1)(0+1+1)^{3}}13.51\ \text{MeV} \notag \\
&=&5.78\ \text{MeV},\\
c_{1}(\Sigma_{b},1P)&=&\frac{M_{c}m_{ss}}{M_{b}m_{nn}}\frac{(L^{\prime}+3.3)(n^{\prime}+L^{\prime}+1)^{3}}{(L+3.3)(n+L+1)^{3}}c_{1}(\Omega_{c},1P) \notag \\
&=&\frac{1.44\times0.991}{4.48\times0.745}\frac{(1+3.3)(0+1+1)^{3}}{(1+3.3)(0+1+1)^{3}}4.04\ \text{MeV} \notag \\
&=&1.73\ \text{MeV},
\end{eqnarray}
\begin{eqnarray}
M(\Sigma_{b},1P): 6048.86\ \text{MeV}, 6070.13\ \text{MeV}, 6076.71\ \text{MeV}, 6087.21\ \text{MeV}, 6099.56\ \text{MeV}.  \label{Vo11q11}
\end{eqnarray}
Evidently, $a_{1}$, $a_{2}$, $b_{1}$ reasonably fulfill (i)-(iii), and $c_{1}$ in (iv) becomes a non-vanishing but small value for the highly excited states. We exploit Eq. (\ref{MMM111}) to calculate the mass splitting for the $\Sigma_{b}$ states. The results of the parameters are listed in Table \ref{Table:pp858} and the masses in Table \ref{85dm2}. Under the analysis of the model, these results are consistent with the experimental values.
\renewcommand\tabcolsep{1.0cm}
\renewcommand{\arraystretch}{0.8}
\begin{table*}[!htbp]
\caption{The spin coupling parameters (MeV) of the $\Sigma_{c}$ baryons.   \label{Table 8}}
\begin{tabular}{ccccc}
\hline\hline
State: & $a_{1}$ & $a_{2}$ & $b_{1}$ &$c_{1}$ \\
 \hline
1$S$ &      &         &       & 56.02   \\
2$S$ &      &         &       & 7.00 \\
3$S$ &      &         &       & 2.07 \\
4$S$ &      &         &       & 0.88 \\
5$S$ &      &         &       & 0.45 \\
\hline
1$P$ &35.86  &34.27   &17.97  &5.37  \\
2$P$ &15.94  &15.23   &5.32   &1.59  \\
3$P$ &8.97   &8.57    &2.25   &0.67   \\
4$P$ &5.74   &5.48    &1.15   &0.34   \\
5$P$ &3.98   &3.81    &0.67   &0.20   \\
\hline
1$D$ &15.94  &15.23   &1.06   &1.29   \\
2$D$ &8.97   &8.57    &0.45   &0.55   \\
3$D$ &5.74   &5.48    &0.23   &0.28   \\
4$D$ &3.98   &3.81    &0.13   &0.16   \\
5$D$ &2.93   &2.80    &0.08   &0.10   \\
\hline\hline
\end{tabular}
\end{table*}
\renewcommand\tabcolsep{1.0cm}
\renewcommand{\arraystretch}{0.8}
\begin{table*}[!htbp]
\caption{The spin coupling parameters (MeV) of the $\Sigma_{b}$ baryons.   \label{Table:pp858}}
\begin{tabular}{ccccc}
\hline\hline
State: & $a_{1}$ & $a_{2}$ & $b_{1}$ &$c_{1}$ \\
 \hline
1$S$ &      &         &       & 18.00   \\
2$S$ &      &         &       & 2.25 \\
3$S$ &      &         &       & 0.67 \\
4$S$ &      &         &       & 0.28 \\
5$S$ &      &         &       & 0.14 \\
\hline
1$P$ &11.53  &11.01   &5.78  &1.73  \\
2$P$ &5.12   &4.90    &1.71  &0.51  \\
3$P$ &2.88   &2.75    &0.72  &0.22   \\
4$P$ &1.84   &1.76    &0.37  &0.11   \\
5$P$ &1.28   &1.22    &0.21  &0.06   \\
\hline
1$D$ &5.12   &4.90    &0.34  &0.42   \\
2$D$ &2.88   &2.75    &0.14  &0.18   \\
3$D$ &1.84   &1.76    &0.07  &0.09   \\
4$D$ &1.28   &1.22    &0.04  &0.05   \\
5$D$ &0.94   &0.90    &0.03  &0.03   \\
\hline\hline
\end{tabular}
\end{table*}
\begin{table*}[!htbp]
\caption{The mass spectrum (MeV) of $\Sigma_{c}$ baryons are given and compared with different quark models.}\label{pp12dm1}
\resizebox{\textwidth}{12cm}{\begin{tabular}{ccccccc}
\hline\hline
{\small State }$J^{P}$ & {\small Baryon} & {\small Mass} &{Ours}&{\small EFG \cite{EFG:C10}}&Ref.{\cite{BingKe:PP888}}&Ref.{\cite{ShahK:PP888}}  \\
\hline
$%
\begin{array}{rr}
{\small 1}^{1}{\small S}_{1/2} & {\small 1/2}^{+} \\
{\small 1}^{3}{\small S}_{3/2} & {\small 3/2}^{+}%
\end{array}%
$ & $%
\begin{array}{r}
{\small \Sigma_{c}(2455)^{+}} \\
{\small \Sigma_{c}(2520)^{+}} \\
\end{array}%
$ & $%
\begin{array}{r}
{\small 2452.65} \\
{\small 2517.4}%
\end{array}%
$ & $%
\begin{array}{r}
{\small 2440.07} \\
{\small 2524.10}%
\end{array}%
$ & $%
\begin{array}{r}
{\small 2443} \\
{\small 2519}%
\end{array}%
$ & $%
\begin{array}{r}
{\small 2456} \\
{\small 2515}%
\end{array}%
$ & $%
\begin{array}{r}
{\small 2452} \\
{\small 2518}%
\end{array}%
$ \\ $%

\begin{array}{rr}
{\small 2}^{1}{\small S}_{1/2} & {\small 1/2}^{+} \\
{\small 2}^{3}{\small S}_{3/2} & {\small 3/2}^{+}%
\end{array}%
$ & $%
\begin{array}{r}
{\small } \\
{\small }%
\end{array}%
$ & $%
\begin{array}{r}
{\small } \\
{\small }%
\end{array}%
$ & $%
\begin{array}{r}
{\small 2857.00} \\
{\small 2867.50}%
\end{array}%
$ & $%
\begin{array}{r}
{\small 2901} \\
{\small 2936}%
\end{array}%
$ & $%
\begin{array}{r}
{\small 2850} \\
{\small 2876}%
\end{array}%
$ & $%
\begin{array}{r}
{\small 2891} \\
{\small 2917}%
\end{array}%
$ \\ $%

\begin{array}{rr}
{\small 3}^{1}{\small S}_{1/2} & {\small 1/2}^{+} \\
{\small 3}^{3}{\small S}_{3/2} & {\small 3/2}^{+}%
\end{array}%
$ & $%
\begin{array}{r}
{\small} \\
{\small}%
\end{array}%
$ & $%
\begin{array}{r}
{\small } \\
{\small }%
\end{array}%
$ & $%
\begin{array}{r}
{\small 3152.63} \\
{\small 3155.75}%
\end{array}%
$ & $%
\begin{array}{r}
{\small 3271} \\
{\small 3293}%
\end{array}%
$ & $%
\begin{array}{r}
{\small 3091 } \\
{\small 3109 }%
\end{array}%
$ & $%
\begin{array}{r}
{\small 3261} \\
{\small 3274}%
\end{array}%
$ \\ $%

\begin{array}{rr}
{\small 4}^{1}{\small S}_{1/2} & {\small 1/2}^{+} \\
{\small 4}^{3}{\small S}_{3/2} & {\small 3/2}^{+}%
\end{array}%
$ & $%
\begin{array}{r}
{\small } \\
{\small }%
\end{array}%
$ & $%
\begin{array}{r}
{\small } \\
{\small }%
\end{array}%
$ & $%
\begin{array}{r}
{\small 3401.95} \\
{\small 3403.26}%
\end{array}%
$ & $%
\begin{array}{r}
{\small 3581} \\
{\small 3598}%
\end{array}%
$ & $%
\begin{array}{r}
{\small } \\
{\small }%
\end{array}%
$ & $%
\begin{array}{r}
{\small 3593} \\
{\small 3601}%
\end{array}%
$ \\ $%

\begin{array}{rr}
{\small 5}^{1}{\small S}_{1/2} & {\small 1/2}^{+} \\
{\small 5}^{3}{\small S}_{3/2} & {\small 3/2}^{+}%
\end{array}%
$ & $%
\begin{array}{r}
{\small } \\
{\small }%
\end{array}%
$ & $%
\begin{array}{r}
{\small } \\
{\small }%
\end{array}%
$ & $%
\begin{array}{r}
{\small 3622.47} \\
{\small 3623.14}%
\end{array}%
$ & $%
\begin{array}{r}
{\small 3861} \\
{\small 3873}%
\end{array}%
$ & $%
\begin{array}{r}
{\small } \\
{\small}%
\end{array}%
$ & $%
\begin{array}{r}
{\small 3900} \\
{\small 3906}%
\end{array}%
$ \\
\hline

$
\begin{array}{rr}
{\small 1}^{2}{\small P}_{1/2} & {\small 1/2}^{-} \\
{\small 1}^{4}{\small P}_{1/2} & {\small 1/2}^{-} \\
{\small 1}^{2}{\small P}_{3/2} & {\small 3/2}^{-} \\
{\small 1}^{4}{\small P}_{3/2} & {\small 3/2}^{-} \\
{\small 1}^{4}{\small P}_{5/2} & {\small 5/2}^{-}%
\end{array}%
$ & $%
\begin{array}{r}
{\small } \\
{\small } \\
{\small } \\
{\small \Sigma_{c}(2800)^{+}} \\
{\small } \\
\end{array}%
$ & $%
\begin{array}{r}
{\small } \\
{\small } \\
{\small } \\
{\small 2792} \\
{\small } \\%
\end{array}%
$ & $%
\begin{array}{r}
{\small 2668.86} \\
{\small 2735.11}\\
{\small 2755.59} \\
{\small 2788.31}\\
{\small 2826.76}%
\end{array}%
$ & $%
\begin{array}{r}
{\small 2713} \\
{\small 2799} \\
{\small 2773} \\
{\small 2798} \\
{\small 2789} \\
\end{array}%
$ & $%
\begin{array}{r}
{\small 2702} \\
{\small 2765} \\
{\small 2785} \\
{\small 2798} \\
{\small 2790} \\
\end{array}%
$ & $%
\begin{array}{r}
{\small 2809} \\
{\small 2755} \\
{\small 2835} \\
{\small 2782} \\
{\small 2710} \\
\end{array}%
$ \\
\hline

$
\begin{array}{rr}
{\small 2}^{2}{\small P}_{1/2} & {\small 1/2}^{-} \\
{\small 2}^{4}{\small P}_{1/2} & {\small 1/2}^{-} \\
{\small 2}^{2}{\small P}_{3/2} & {\small 3/2}^{-} \\
{\small 2}^{4}{\small P}_{3/2} & {\small 3/2}^{-} \\
{\small 2}^{4}{\small P}_{5/2} & {\small 5/2}^{-}%
\end{array}%
$ & $%
\begin{array}{r}
{\small } \\
{\small } \\
{\small } \\
{\small } \\
{\small } \\
\end{array}%
$ & $%
\begin{array}{r}
 \\
\\
 \\
\\
\end{array}%
$ & $%
\begin{array}{r}
{\small 3037.05} \\
{\small 3063.95}\\
{\small 3070.46} \\
{\small 3087.94}\\
{\small 3104.56}%
\end{array}%
$ & $%
\begin{array}{r}
{\small 3125} \\
{\small 3172} \\
{\small 3151} \\
{\small 3172} \\
{\small 3161} \\
\end{array}%
$ & $%
\begin{array}{r}
{\small 2971} \\
{\small 3018} \\
{\small 3036} \\
{\small 3044} \\
{\small 3040} \\
\end{array}%
$ & $%
\begin{array}{r}
{\small 3174} \\
{\small 3128} \\
{\small 3196} \\
{\small 3151} \\
{\small 3090} \\
\end{array}%
$ \\
\hline

$
\begin{array}{rr}
{\small 3}^{2}{\small P}_{1/2} & {\small 1/2}^{-} \\
{\small 3}^{4}{\small P}_{1/2} & {\small 1/2}^{-} \\
{\small 3}^{2}{\small P}_{3/2} & {\small 3/2}^{-} \\
{\small 3}^{4}{\small P}_{3/2} & {\small 3/2}^{-} \\
{\small 3}^{4}{\small P}_{5/2} & {\small 5/2}^{-}%
\end{array}%
$ & $%
\begin{array}{r}
{\small } \\
{\small } \\
{\small } \\
{\small } \\
{\small } \\
\end{array}%
$ & $%
\begin{array}{r}
\\
\\
\\
\\
\end{array}%
$ & $%
\begin{array}{r}
{\small 3314.88} \\
{\small 3329.38}\\
{\small 3332.25} \\
{\small 3342.95}\\
{\small 3352.15}%
\end{array}%
$ & $%
\begin{array}{r}
{\small 3455} \\
{\small 3488} \\
{\small 3469} \\
{\small 3486} \\
{\small 3475} \\
\end{array}%
$ & $%
\begin{array}{r}
{\small } \\
{\small } \\
{\small } \\
{\small } \\
{\small } \\
\end{array}%
$ & $%
\begin{array}{r}
{\small 3505} \\
{\small 3465} \\
{\small 3525} \\
{\small 3485} \\
{\small 3433} \\
\end{array}%
$ \\
\hline

$
\begin{array}{rr}
{\small 4}^{2}{\small P}_{1/2} & {\small 1/2}^{-} \\
{\small 4}^{4}{\small P}_{1/2} & {\small 1/2}^{-} \\
{\small 4}^{2}{\small P}_{3/2} & {\small 3/2}^{-} \\
{\small 4}^{4}{\small P}_{3/2} & {\small 3/2}^{-} \\
{\small 4}^{4}{\small P}_{5/2} & {\small 5/2}^{-}%
\end{array}%
$ & $%
\begin{array}{r}
{\small } \\
{\small } \\
{\small } \\
{\small } \\
{\small } \\
\end{array}%
$ & $%
\begin{array}{r}
 \\
\\
\\
\\
\end{array}%
$ & $%
\begin{array}{r}
{\small 3550.56} \\
{\small 3559.61}\\
{\small 3561.13} \\
{\small 3568.32}\\
{\small 3574.14}%
\end{array}%
$ & $%
\begin{array}{r}
{\small 3743} \\
{\small 3770} \\
{\small 3753} \\
{\small 3768} \\
{\small 3757} \\
\end{array}%
$ & $%
\begin{array}{r}
{\small } \\
{\small } \\
{\small } \\
{\small } \\
{\small } \\
\end{array}%
$ & $%
\begin{array}{r}
{\small 3814} \\
{\small 3777} \\
{\small 3832} \\
{\small 2796} \\
{\small 3747} \\
\end{array}%
$ \\
\hline

$
\begin{array}{rr}
{\small 5}^{2}{\small P}_{1/2} & {\small 1/2}^{-} \\
{\small 5}^{4}{\small P}_{1/2} & {\small 1/2}^{-} \\
{\small 5}^{2}{\small P}_{3/2} & {\small 3/2}^{-} \\
{\small 5}^{4}{\small P}_{3/2} & {\small 3/2}^{-} \\
{\small 5}^{4}{\small P}_{5/2} & {\small 5/2}^{-}%
\end{array}%
$ & $%
\begin{array}{r}
{\small } \\
{\small } \\
{\small } \\
{\small } \\
{\small } \\
\end{array}%
$ & $%
\begin{array}{r}
\\
\\
\\
\\
\end{array}%
$ & $%
\begin{array}{r}
{\small 3760.07} \\
{\small 3766.26}\\
{\small 3767.17} \\
{\small 3772.32}\\
{\small 3776.33}%
\end{array}%
$ & $%
\begin{array}{r}
{\small } \\
{\small } \\
{\small } \\
{\small } \\
{\small } \\
\end{array}%
$ \\
\hline

$
\begin{array}{rr}
{\small 1}^{4}{\small D}_{1/2} & {\small 1/2}^{+} \\
{\small 1}^{2}{\small D}_{3/2} & {\small 3/2}^{+} \\
{\small 1}^{4}{\small D}_{3/2} & {\small 3/2}^{+} \\
{\small 1}^{2}{\small D}_{5/2} & {\small 5/2}^{+} \\
{\small 1}^{4}{\small D}_{5/2} & {\small 5/2}^{+} \\
{\small 1}^{4}{\small D}_{7/2} & {\small 7/2}^{+}%
\end{array}%
$ & $%
\begin{array}{r}
{\small } \\
{\small } \\
{\small } \\
{\small } \\
{\small } \\
{\small }%
\end{array}%
$ & $%
\begin{array}{r}
 \\
\\
\\
\\
\\
\end{array}%
$ & $%
\begin{array}{r}
{\small 2933.33} \\
{\small 2957.94}\\
{\small 2978.85}\\
{\small 2997.90}\\
{\small 3019.35}\\
{\small 3051.86}%
\end{array}%
$ & $%
\begin{array}{r}
{\small 3041} \\
{\small 3040} \\
{\small 3043} \\
{\small 3023} \\
{\small 3038} \\
{\small 3013}%
\end{array}%
$ & $%
\begin{array}{r}
{\small 2949} \\
{\small 2952} \\
{\small 2964} \\
{\small 2942} \\
{\small 2963} \\
{\small 2943}%
\end{array}%
$ & $%
\begin{array}{r}
{\small 3036} \\
{\small 3112} \\
{\small 3061} \\
{\small 2993} \\
{\small 2968} \\
{\small 2909}%
\end{array}%
$ \\
\hline

$
\begin{array}{rr}
{\small 2}^{4}{\small D}_{1/2} & {\small 1/2}^{+} \\
{\small 2}^{2}{\small D}_{3/2} & {\small 3/2}^{+} \\
{\small 2}^{4}{\small D}_{3/2} & {\small 3/2}^{+} \\
{\small 2}^{2}{\small D}_{5/2} & {\small 5/2}^{+} \\
{\small 2}^{4}{\small D}_{5/2} & {\small 5/2}^{+} \\
{\small 2}^{4}{\small D}_{7/2} & {\small 7/2}^{+}%
\end{array}%
$ & $%
\begin{array}{r}
{\small } \\
{\small } \\
{\small } \\
{\small } \\
{\small } \\
{\small }%
\end{array}%
$ & $%
\begin{array}{r}
 \\
\\
\\
\\
\\
\end{array}%
$ & $%
\begin{array}{r}
{\small 3233.06} \\
{\small 3246.75}\\
{\small 3258.79}\\
{\small 3269.13} \\
{\small 3281.56}\\
{\small 3299.62}%
\end{array}%
$ & $%
\begin{array}{r}
{\small 3370} \\
{\small 3364} \\
{\small 3366} \\
{\small 3349} \\
{\small 3365} \\
{\small 3342}%
\end{array}%
$ & $%
\begin{array}{r}
{\small } \\
{\small } \\
{\small } \\
{\small } \\
{\small } \\
{\small }%
\end{array}%
$ & $%
\begin{array}{r}
{\small 3376} \\
{\small 3398} \\
{\small 3442} \\
{\small 3316} \\
{\small 3339} \\
{\small 3265}%
\end{array}%
$ \\
\hline

$
\begin{array}{rr}
{\small 3}^{4}{\small D}_{1/2} & {\small 1/2}^{+} \\
{\small 3}^{2}{\small D}_{3/2} & {\small 3/2}^{+} \\
{\small 3}^{4}{\small D}_{3/2} & {\small 3/2}^{+} \\
{\small 3}^{2}{\small D}_{5/2} & {\small 5/2}^{+} \\
{\small 3}^{4}{\small D}_{5/2} & {\small 5/2}^{+} \\
{\small 3}^{4}{\small D}_{7/2} & {\small 7/2}^{+}%
\end{array}%
$ & $%
\begin{array}{r}
{\small } \\
{\small } \\
{\small } \\
{\small } \\
{\small } \\
{\small }%
\end{array}%
$ & $%
\begin{array}{r}
\\
\\
\\
\\
\\
\end{array}%
$ & $%
\begin{array}{r}
{\small 3481.42} \\
{\small 3490.12}\\
{\small 3497.93} \\
{\small 3504.40}\\
{\small 3512.51}\\
{\small 3523.98}%
\end{array}%
$ & $%
\begin{array}{r}
{\small } \\
{\small } \\
{\small } \\
{\small } \\
{\small } \\
{\small }%
\end{array}%
$ \\
\hline

$
\begin{array}{rr}
{\small 4}^{4}{\small D}_{1/2} & {\small 1/2}^{+} \\
{\small 4}^{2}{\small D}_{3/2} & {\small 3/2}^{+} \\
{\small 4}^{4}{\small D}_{3/2} & {\small 3/2}^{+} \\
{\small 4}^{2}{\small D}_{5/2} & {\small 5/2}^{+} \\
{\small 4}^{4}{\small D}_{5/2} & {\small 5/2}^{+} \\
{\small 4}^{4}{\small D}_{7/2} & {\small 7/2}^{+}%
\end{array}%
$ & $%
\begin{array}{r}
{\small } \\
{\small } \\
{\small } \\
{\small } \\
{\small } \\
{\small }%
\end{array}%
$ & $%
\begin{array}{r}
\\
\\
\\
\\
\\
\end{array}%
$ & $%
\begin{array}{r}
{\small 3699.28}\\
{\small 3705.30}\\
{\small 3710.77}\\
{\small 3715.20}\\
{\small 3720.89}\\
{\small 3728.81}%
\end{array}%
$ & $%
\begin{array}{r}
{\small } \\
{\small } \\
{\small } \\
{\small } \\
{\small } \\
{\small }%
\end{array}%
$ \\
\hline

$
\begin{array}{rr}
{\small 5}^{4}{\small D}_{1/2} & {\small 1/2}^{+} \\
{\small 5}^{2}{\small D}_{3/2} & {\small 3/2}^{+} \\
{\small 5}^{4}{\small D}_{3/2} & {\small 3/2}^{+} \\
{\small 5}^{2}{\small D}_{5/2} & {\small 5/2}^{+} \\
{\small 5}^{4}{\small D}_{5/2} & {\small 5/2}^{+} \\
{\small 5}^{4}{\small D}_{7/2} & {\small 7/2}^{+}%
\end{array}%
$ & $%
\begin{array}{r}
{\small } \\
{\small } \\
{\small } \\
{\small } \\
{\small } \\
{\small }%
\end{array}%
$ & $%
\begin{array}{r}
\\
\\
\\
\\
\\
\end{array}%
$ & $%
\begin{array}{r}
{\small 3896.23} \\
{\small 3900.64}\\
{\small 3904.68}\\
{\small 3907.90} \\
{\small 3912.12}\\
{\small 3917.92}%
\end{array}%
$ & $%
\begin{array}{r}
{\small } \\
{\small } \\
{\small } \\
{\small } \\
{\small } \\
{\small }%
\end{array}%
$ \\
\hline\hline
\end{tabular}}
\end{table*}
\begin{table*}[!htbp]
\caption{The mass spectrum (MeV) of $\Sigma_{b}$ baryons are given and compared with different quark models.}\label{85dm2}
\resizebox{\textwidth}{12cm}{\begin{tabular}{ccccccc}
\hline\hline
{\small State }$J^{P}$ & {\small Baryon} & {\small Mass} &{Ours}&{\small EFG \cite{EFG:C10}} &Ref.{\cite{adiyae:PP888}} &Ref.{\cite{OudichhyGan:PP888}}\\
\hline
$%
\begin{array}{rr}
{\small 1}^{1}{\small S}_{1/2} & {\small 1/2}^{+} \\
{\small 1}^{3}{\small S}_{3/2} & {\small 3/2}^{+}%
\end{array}%
$ & $%
\begin{array}{r}
{\small \Sigma_{b}^{+}} \\
{\small \Sigma_{b}}^{\ast+}%
\end{array}%
$ & $%
\begin{array}{r}
{\small 5810.56} \\
{\small 5830.32}%
\end{array}%
$ & $%
\begin{array}{r}
{\small 5801.27} \\
{\small 5828.25}%
\end{array}%
$ & $%
\begin{array}{r}
{\small 5808} \\
{\small 5834}%
\end{array}%
$ & $%
\begin{array}{r}
{\small 5811} \\
{\small 5832}%
\end{array}%
$ & $%
\begin{array}{r}
{\small 5811} \\
{\small 5830}%
\end{array}%
$ \\ $%

\begin{array}{rr}
{\small 2}^{1}{\small S}_{1/2} & {\small 1/2}^{+} \\
{\small 2}^{3}{\small S}_{3/2} & {\small 3/2}^{+}%
\end{array}%
$ & $%
\begin{array}{r}
{\small } \\
{\small }%
\end{array}%
$ & $%
\begin{array}{r}
{\small } \\
{\small }%
\end{array}%
$ & $%
\begin{array}{r}
{\small 6167.69} \\
{\small 6171.06}%
\end{array}%
$ & $%
\begin{array}{r}
{\small 6213} \\
{\small 6226}%
\end{array}%
$ & $%
\begin{array}{r}
{\small 6262} \\
{\small 6278}%
\end{array}%
$ & $%
\begin{array}{r}
{\small 6275} \\
{\small 6291}%
\end{array}%
$ \\ $%

\begin{array}{rr}
{\small 3}^{1}{\small S}_{1/2} & {\small 1/2}^{+} \\
{\small 3}^{3}{\small S}_{3/2} & {\small 3/2}^{+}%
\end{array}%
$ & $%
\begin{array}{r}
{\small} \\
{\small}%
\end{array}%
$ & $%
\begin{array}{r}
{\small } \\
{\small }%
\end{array}%
$ & $%
\begin{array}{r}
{\small 6458.62} \\
{\small 6459.62}%
\end{array}%
$ & $%
\begin{array}{r}
{\small 6575} \\
{\small 6583}%
\end{array}%
$ & $%
\begin{array}{r}
{\small 6605} \\
{\small 6614}%
\end{array}%
$ & $%
\begin{array}{r}
{\small 6707} \\
{\small 6720}%
\end{array}%
$ \\ $%

\begin{array}{rr}
{\small 4}^{1}{\small S}_{1/2} & {\small 1/2}^{+} \\
{\small 4}^{3}{\small S}_{3/2} & {\small 3/2}^{+}%
\end{array}%
$ & $%
\begin{array}{r}
{\small } \\
{\small }%
\end{array}%
$ & $%
\begin{array}{r}
{\small } \\
{\small }%
\end{array}%
$ & $%
\begin{array}{r}
{\small 6711.06} \\
{\small 6711.48}%
\end{array}%
$ & $%
\begin{array}{r}
{\small 6869} \\
{\small 6876}%
\end{array}%
$ & $%
\begin{array}{r}
{\small 6927} \\
{\small 6933}%
\end{array}%
$ & $%
\begin{array}{r}
{\small 7113} \\
{\small 7124}%
\end{array}%
$ \\ $%

\begin{array}{rr}
{\small 5}^{1}{\small S}_{1/2} & {\small 1/2}^{+} \\
{\small 5}^{3}{\small S}_{3/2} & {\small 3/2}^{+}%
\end{array}%
$ & $%
\begin{array}{r}
{\small } \\
{\small }%
\end{array}%
$ & $%
\begin{array}{r}
{\small } \\
{\small }%
\end{array}%
$ & $%
\begin{array}{r}
{\small 6937.48} \\
{\small 6937.70}%
\end{array}%
$ & $%
\begin{array}{r}
{\small 7124} \\
{\small 7129}%
\end{array}%
$ & $%
\begin{array}{r}
{\small 7231} \\
{\small 7235}%
\end{array}%
$ & $%
\begin{array}{r}
{\small 7497} \\
{\small 7506}%
\end{array}%
$ \\
\hline

$
\begin{array}{rr}
{\small 1}^{2}{\small P}_{1/2} & {\small 1/2}^{-} \\
{\small 1}^{4}{\small P}_{1/2} & {\small 1/2}^{-} \\
{\small 1}^{2}{\small P}_{3/2} & {\small 3/2}^{-} \\
{\small 1}^{4}{\small P}_{3/2} & {\small 3/2}^{-} \\
{\small 1}^{4}{\small P}_{5/2} & {\small 5/2}^{-}%
\end{array}%
$ & $%
\begin{array}{r}
{\small } \\
{\small } \\
{\small } \\
{\small } \\
{\small \Sigma_{b}(6097)^{-}}%
\end{array}%
$ & $%
\begin{array}{r}
{\small } \\
{\small } \\
{\small } \\
{\small } \\
{\small 6098.0}%
\end{array}%
$ & $%
\begin{array}{r}
{\small 6048.86} \\
{\small 6070.13}\\
{\small 6076.71} \\
{\small 6087.21}\\
{\small 6099.56}%
\end{array}%
$ & $%
\begin{array}{r}
{\small 6095} \\
{\small 6101} \\
{\small 6087} \\
{\small 6096} \\
{\small 6084}%
\end{array}%
$ & $%
\begin{array}{r}
{\small 6104} \\
{\small 6106} \\
{\small 6100} \\
{\small 6102} \\
{\small 6097}%
\end{array}%
$ & $%
\begin{array}{r}
{\small } \\
{\small } \\
{\small 6105} \\
{\small } \\
{\small 6118}%
\end{array}%
$ \\
\hline

$
\begin{array}{rr}
{\small 2}^{2}{\small P}_{1/2} & {\small 1/2}^{-} \\
{\small 2}^{4}{\small P}_{1/2} & {\small 1/2}^{-} \\
{\small 2}^{2}{\small P}_{3/2} & {\small 3/2}^{-} \\
{\small 2}^{4}{\small P}_{3/2} & {\small 3/2}^{-} \\
{\small 2}^{4}{\small P}_{5/2} & {\small 5/2}^{-}%
\end{array}%
$ & $%
\begin{array}{r}
{\small } \\
{\small } \\
{\small } \\
{\small } \\
{\small }%
\end{array}%
$ & $%
\begin{array}{r}
 \\
\\
 \\
\\
\end{array}%
$ & $%
\begin{array}{r}
{\small 6371.29} \\
{\small 6379.93}\\
{\small 6382.02} \\
{\small 6387.63}\\
{\small 6392.96}%
\end{array}%
$ & $%
\begin{array}{r}
{\small 6430} \\
{\small 6440} \\
{\small 6424} \\
{\small 6430} \\
{\small 6421}%
\end{array}%
$ & $%
\begin{array}{r}
{\small 6355} \\
{\small 6356} \\
{\small 6353} \\
{\small 6354} \\
{\small 6351}%
\end{array}%
$ & $%
\begin{array}{r}
{\small } \\
{\small } \\
{\small 6506} \\
{\small } \\
{\small 6489}%
\end{array}%
$ \\
\hline

$
\begin{array}{rr}
{\small 3}^{2}{\small P}_{1/2} & {\small 1/2}^{-} \\
{\small 3}^{4}{\small P}_{1/2} & {\small 1/2}^{-} \\
{\small 3}^{2}{\small P}_{3/2} & {\small 3/2}^{-} \\
{\small 3}^{4}{\small P}_{3/2} & {\small 3/2}^{-} \\
{\small 3}^{4}{\small P}_{5/2} & {\small 5/2}^{-}%
\end{array}%
$ & $%
\begin{array}{r}
{\small } \\
{\small } \\
{\small } \\
{\small } \\
{\small }%
\end{array}%
$ & $%
\begin{array}{r}
\\
\\
\\
\\
\end{array}%
$ & $%
\begin{array}{r}
{\small 6638.42} \\
{\small 6643.07}\\
{\small 6644.00} \\
{\small 6647.43}\\
{\small 6650.38}%
\end{array}%
$ & $%
\begin{array}{r}
{\small 6742} \\
{\small 6756} \\
{\small 6736} \\
{\small 6742} \\
{\small 6732}%
\end{array}%
$ & $%
\begin{array}{r}
{\small 6578} \\
{\small 6579} \\
{\small 6577} \\
{\small 6577} \\
{\small 6575}%
\end{array}%
$ & $%
\begin{array}{r}
{\small } \\
{\small } \\
{\small 6884} \\
{\small } \\
{\small 6840}%
\end{array}%
$ \\
\hline

$
\begin{array}{rr}
{\small 4}^{2}{\small P}_{1/2} & {\small 1/2}^{-} \\
{\small 4}^{4}{\small P}_{1/2} & {\small 1/2}^{-} \\
{\small 4}^{2}{\small P}_{3/2} & {\small 3/2}^{-} \\
{\small 4}^{4}{\small P}_{3/2} & {\small 3/2}^{-} \\
{\small 4}^{4}{\small P}_{5/2} & {\small 5/2}^{-}%
\end{array}%
$ & $%
\begin{array}{r}
{\small } \\
{\small } \\
{\small } \\
{\small } \\
{\small }%
\end{array}%
$ & $%
\begin{array}{r}
 \\
\\
\\
\\
\end{array}%
$ & $%
\begin{array}{r}
{\small 6873.75} \\
{\small 6876.66}\\
{\small 6877.15} \\
{\small 6879.45}\\
{\small 6881.31}%
\end{array}%
$ & $%
\begin{array}{r}
{\small 7008} \\
{\small 7024} \\
{\small 7003} \\
{\small 7009} \\
{\small 6999}%
\end{array}%
$ & $%
\begin{array}{r}
{\small 6778} \\
{\small 6779} \\
{\small 6777} \\
{\small 6778} \\
{\small 6776}%
\end{array}%
$ & $%
\begin{array}{r}
{\small } \\
{\small } \\
{\small 7242} \\
{\small } \\
{\small 7174}%
\end{array}%
$ \\
\hline

$
\begin{array}{rr}
{\small 5}^{2}{\small P}_{1/2} & {\small 1/2}^{-} \\
{\small 5}^{4}{\small P}_{1/2} & {\small 1/2}^{-} \\
{\small 5}^{2}{\small P}_{3/2} & {\small 3/2}^{-} \\
{\small 5}^{4}{\small P}_{3/2} & {\small 3/2}^{-} \\
{\small 5}^{4}{\small P}_{5/2} & {\small 5/2}^{-}%
\end{array}%
$ & $%
\begin{array}{r}
{\small } \\
{\small } \\
{\small } \\
{\small } \\
{\small }%
\end{array}%
$ & $%
\begin{array}{r}
\\
\\
\\
\\
\end{array}%
$ & $%
\begin{array}{r}
{\small 7087.08} \\
{\small 7089.06}\\
{\small 7089.35} \\
{\small 7091.01}\\
{\small 7092.30}%
\end{array}%
$ & $%
\begin{array}{r}
{\small } \\
{\small } \\
{\small } \\
{\small } \\
{\small }%
\end{array}%
$ & $%
\begin{array}{r}
{\small } \\
{\small } \\
{\small } \\
{\small } \\
{\small }%
\end{array}%
$ & $%
\begin{array}{r}
{\small } \\
{\small } \\
{\small 7583} \\
{\small } \\
{\small 7493}%
\end{array}%
$ \\
\hline

$
\begin{array}{rr}
{\small 1}^{4}{\small D}_{1/2} & {\small 1/2}^{+} \\
{\small 1}^{2}{\small D}_{3/2} & {\small 3/2}^{+} \\
{\small 1}^{4}{\small D}_{3/2} & {\small 3/2}^{+} \\
{\small 1}^{2}{\small D}_{5/2} & {\small 5/2}^{+} \\
{\small 1}^{4}{\small D}_{5/2} & {\small 5/2}^{+} \\
{\small 1}^{4}{\small D}_{7/2} & {\small 7/2}^{+}%
\end{array}%
$ & $%
\begin{array}{r}
{\small } \\
{\small } \\
{\small } \\
{\small } \\
{\small } \\
{\small }%
\end{array}%
$ & $%
\begin{array}{r}
 \\
\\
\\
\\
\\
\end{array}%
$ & $%
\begin{array}{r}
{\small 6285.89} \\
{\small 6293.79}\\
{\small 6300.50}\\
{\small 6306.62}\\
{\small 6313.51}\\
{\small 6323.94}%
\end{array}%
$ & $%
\begin{array}{r}
{\small 6311} \\
{\small 6285} \\
{\small 6326} \\
{\small 6270} \\
{\small 6284} \\
{\small 6260}%
\end{array}%
$ & $%
\begin{array}{r}
{\small 6303} \\
{\small 6298} \\
{\small 6300} \\
{\small 6294} \\
{\small 6295} \\
{\small 6290}%
\end{array}%
$ & $%
\begin{array}{r}
{\small } \\
{\small } \\
{\small } \\
{\small 6386} \\
{\small } \\
{\small 6393}%
\end{array}%
$ \\
\hline

$
\begin{array}{rr}
{\small 2}^{4}{\small D}_{1/2} & {\small 1/2}^{+} \\
{\small 2}^{2}{\small D}_{3/2} & {\small 3/2}^{+} \\
{\small 2}^{4}{\small D}_{3/2} & {\small 3/2}^{+} \\
{\small 2}^{2}{\small D}_{5/2} & {\small 5/2}^{+} \\
{\small 2}^{4}{\small D}_{5/2} & {\small 5/2}^{+} \\
{\small 2}^{4}{\small D}_{7/2} & {\small 7/2}^{+}%
\end{array}%
$ & $%
\begin{array}{r}
{\small } \\
{\small } \\
{\small } \\
{\small } \\
{\small } \\
{\small }%
\end{array}%
$ & $%
\begin{array}{r}
 \\
\\
\\
\\
\\
\end{array}%
$ & $%
\begin{array}{r}
{\small 6566.15} \\
{\small 6570.54}\\
{\small 6574.41}\\
{\small 6577.73} \\
{\small 6581.72}\\
{\small 6587.52}%
\end{array}%
$ & $%
\begin{array}{r}
{\small 6636} \\
{\small 6612} \\
{\small 6647} \\
{\small 6598} \\
{\small 6612} \\
{\small 6590}%
\end{array}%
$ & $%
\begin{array}{r}
{\small 6533} \\
{\small 6529} \\
{\small 6530} \\
{\small 6526} \\
{\small 6527} \\
{\small 6524}%
\end{array}%
$ & $%
\begin{array}{r}
{\small } \\
{\small } \\
{\small } \\
{\small 6778} \\
{\small } \\
{\small 6751}%
\end{array}%
$ \\
\hline

$
\begin{array}{rr}
{\small 3}^{4}{\small D}_{1/2} & {\small 1/2}^{+} \\
{\small 3}^{2}{\small D}_{3/2} & {\small 3/2}^{+} \\
{\small 3}^{4}{\small D}_{3/2} & {\small 3/2}^{+} \\
{\small 3}^{2}{\small D}_{5/2} & {\small 5/2}^{+} \\
{\small 3}^{4}{\small D}_{5/2} & {\small 5/2}^{+} \\
{\small 3}^{4}{\small D}_{7/2} & {\small 7/2}^{+}%
\end{array}%
$ & $%
\begin{array}{r}
{\small } \\
{\small } \\
{\small } \\
{\small } \\
{\small } \\
{\small }%
\end{array}%
$ & $%
\begin{array}{r}
\\
\\
\\
\\
\\
\end{array}%
$ & $%
\begin{array}{r}
{\small 6809.93} \\
{\small 6812.72}\\
{\small 6815.23} \\
{\small 6817.31}\\
{\small 6819.91}\\
{\small 6823.59}%
\end{array}%
$ & $%
\begin{array}{r}
{\small } \\
{\small } \\
{\small } \\
{\small } \\
{\small } \\
{\small }%
\end{array}%
$ & $%
\begin{array}{r}
{\small 6738} \\
{\small 6736} \\
{\small 6736} \\
{\small 6734} \\
{\small 6735} \\
{\small 6733}%
\end{array}%
$ & $%
\begin{array}{r}
{\small } \\
{\small } \\
{\small } \\
{\small 7148} \\
{\small } \\
{\small 7091}%
\end{array}%
$ \\
\hline

$
\begin{array}{rr}
{\small 4}^{4}{\small D}_{1/2} & {\small 1/2}^{+} \\
{\small 4}^{2}{\small D}_{3/2} & {\small 3/2}^{+} \\
{\small 4}^{4}{\small D}_{3/2} & {\small 3/2}^{+} \\
{\small 4}^{2}{\small D}_{5/2} & {\small 5/2}^{+} \\
{\small 4}^{4}{\small D}_{5/2} & {\small 5/2}^{+} \\
{\small 4}^{4}{\small D}_{7/2} & {\small 7/2}^{+}%
\end{array}%
$ & $%
\begin{array}{r}
{\small } \\
{\small } \\
{\small } \\
{\small } \\
{\small } \\
{\small }%
\end{array}%
$ & $%
\begin{array}{r}
\\
\\
\\
\\
\\
\end{array}%
$ & $%
\begin{array}{r}
{\small 7029.26}\\
{\small 7031.19}\\
{\small 7032.95}\\
{\small 7034.37}\\
{\small 7036.20}\\
{\small 7038.75}%
\end{array}%
$ & $%
\begin{array}{r}
{\small } \\
{\small } \\
{\small } \\
{\small } \\
{\small } \\
{\small }%
\end{array}%
$ & $%
\begin{array}{r}
{\small 6923} \\
{\small 6922} \\
{\small 6922} \\
{\small 6921} \\
{\small 6921} \\
{\small 6920}%
\end{array}%
$ & $%
\begin{array}{r}
{\small } \\
{\small } \\
{\small } \\
{\small 7501} \\
{\small } \\
{\small 7415}%
\end{array}%
$ \\
\hline

$
\begin{array}{rr}
{\small 5}^{4}{\small D}_{1/2} & {\small 1/2}^{+} \\
{\small 5}^{2}{\small D}_{3/2} & {\small 3/2}^{+} \\
{\small 5}^{4}{\small D}_{3/2} & {\small 3/2}^{+} \\
{\small 5}^{2}{\small D}_{5/2} & {\small 5/2}^{+} \\
{\small 5}^{4}{\small D}_{5/2} & {\small 5/2}^{+} \\
{\small 5}^{4}{\small D}_{7/2} & {\small 7/2}^{+}%
\end{array}%
$ & $%
\begin{array}{r}
{\small } \\
{\small } \\
{\small } \\
{\small } \\
{\small } \\
{\small }%
\end{array}%
$ & $%
\begin{array}{r}
\\
\\
\\
\\
\\
\end{array}%
$ & $%
\begin{array}{r}
{\small 7230.55} \\
{\small 7231.97}\\
{\small 7233.27}\\
{\small 7234.30} \\
{\small 7235.66}\\
{\small 7237.52}%
\end{array}%
$ & $%
\begin{array}{r}
{\small } \\
{\small } \\
{\small } \\
{\small } \\
{\small } \\
{\small }%
\end{array}%
$ & $%
\begin{array}{r}
{\small } \\
{\small } \\
{\small } \\
{\small } \\
{\small } \\
{\small }%
\end{array}%
$ & $%
\begin{array}{r}
{\small } \\
{\small } \\
{\small } \\
{\small 7837} \\
{\small } \\
{\small 7526}%
\end{array}%
$ \\
\hline\hline
\end{tabular}}
\end{table*}

\section{The baryons $\Xi^{\prime}_{c}$ and $\Xi^{\prime}_{b}$}

In this section, based on our scheme, a similar method can be applied to the excited $\Xi'_{Q}$ ($csn$ or $bsn$) baryon systems in order to analyze their masses and parameters. For the $\Xi^{\prime}_{c}$ baryon system, the ($S$-wave) ground states with the spin-parity $J^{P}=1/2^{+}$ and $J^{P}=3/2^{+}$ correspond to $\Xi^{\prime 0}_{c}$ and $\Xi_{c}(2645)^{0}$, as the masses at $M(\Xi^{\prime 0}_{c}, 1/2^{+})=2578.7$\ MeV and $M(\Xi_{c}^{0}, 3/2^{+})=2646.16$\ MeV listed by the PDG \cite{Workman:A11} have been established. In this work, we use the scaling relations to calculate the spin coupling parameters $a_{1}$, $a_{2}$, $b_{1}$, $c_{1}$ as shown in Table \ref{Table:p123} for the $\Xi^{\prime}_{c}$ baryons. The mass results are listed in Table \ref{854dm323} and compared with other models.

As the classification of the $P$-wave states $(L=1)$ is similar to the other charm baryons, we use Eq. (\ref{MMM111}) to calculate the mass splitting for the $\Xi^{\prime}_{c}$ states. By analyzing the model results in Table \ref{854dm323}, we find that $\Xi_{c}(2923)^{0}$ and $\Xi_{c}(2930)^{0}$ with the spin-parity $J^{P}=3/2^{-}$ might be good candidates for $P$ states of the $\Xi^{\prime}_{c}$ baryons. The masse $M(\Xi_{c}(2923)^{0})=2907.21$\ MeV is only 15.83\ MeV lower than the mass of the state $\Xi_{c}(2923)^{0}$, and lower than $M(\Xi_{c}(2930)^{0})=2935.17$\ MeV, compared with experimental values within a reasonable range. For a more detailed analysis the $\Xi^{\prime}_{c}$ baryons see also Refs. \cite{BingKe:PP888, Shahikk:PP888}.

In addition, the state $\Xi_{c}(3123)$ was also confirmed by the BaBar Collaboration \cite{AubertColl:PP888}, with a mass $M({\Xi_{c}}^{+})=3122.9$\ MeV listed in PDG \cite{Workman:A11}. From the analysis of our data in Table \ref{854dm323} we infer that the mass shifts of about 22 MeV in the $1D$-wave are relatively small. In the past, the quantum number of $\Xi_{c}(3123)$ was not determined. In our frame, it is possible to determine $\Xi_{c}(3123)$ as the second
state with $J^{P}=3/2^{+}$ or mixed with the first state, which can be a good candidate for a $1D$ state of the $\Xi^{\prime}_{c}$ baryons.

For the $\Xi^{\prime}_{b}$ baryon system, in 2015 the LHCb Collaboration observed two new charged states ${\Xi^{\prime}_{b}(5935)}^{-}$ and ${\Xi^{\ast}_{b}(5955)}^{-}$ in the decay channel $\Xi^{\prime 0}_{b}\pi^{-}$ \cite{Aaijolla:PPP888}. The masses $M({\Xi^{\prime-}_{b}}, 1/2^{+})=5935.02$\ MeV and $M({\Xi^{\ast-}_{b}}, 1/2^{+})=5955.33$\ MeV were proposed to be the ground states $\Xi^{\prime -}_{b}$ and $\Xi^{\ast -}_{b}$ with the $J^{P}=1/2^{+}$ and $J^{P}=3/2^{+}$, respectively. In our work, the ground states ${\Xi^{\prime}_{b}(5935)}^{-}$ and ${\Xi^{\ast}_{b}(5955)}^{-}$ in Table \ref{8225dm44} are in good agreement with other theoretical predictions as well as experimental measurements (see Ref. \cite{Aaijolla:PPP888}).

$\Xi_b(6227)$ which was found in both $\Lambda^{0}_{b}K^{-}$ and $\Xi^{0}_{b}\pi^{-}$ channels \cite{Aaijoll:PP888}, is identified in our model with the second excitation of the
$\Xi^{\prime}_b$ baryons corresponding to $L=1,n=0$ and $J^P=1/2^-$,
\begin{eqnarray}
&&\bar M=6244.84\ \text{MeV},\ a_{1}=9.85\ \text{MeV}, a_{2}=9.41\ \text{MeV}, b_{1}=4.94\ \text{MeV}, c_{1}=1.48\ \text{MeV}, \label{V2q11} \\
&&M(\Xi^{\prime}_{b},1P): 6215.82\ \text{MeV}, 6233.90\ \text{MeV}, 6239.61\ \text{MeV}, 6248.59\ \text{MeV}, 6259.13\ \text{MeV}.  \label{V22q11}
\end{eqnarray}
The predicted masses are compatible with the experimental values, closer to the second state or mixed with the first state. The same conclusion holds for the masses of the $\Xi^{\prime}_{b}$ baryons as shown in Table \ref{85bnh22} and Table \ref{8225dm44}. The latter can be inquired also for a discussion of $\Xi_b(6227)$ in different models \cite{Kakayas:PP888, OudichhyGan:PP888} and the well-matching with the experiment.
\renewcommand\tabcolsep{1.0cm}
\renewcommand{\arraystretch}{0.8}
\begin{table*}[!htbp]
\caption{The spin coupling parameters (MeV) of the $\Xi_{c}'$ baryons.   \label{Table:p123}}
\begin{tabular}{ccccc}
\hline\hline
State: & $a_{1}$ & $a_{2}$ & $b_{1}$ &$c_{1}$ \\
 \hline
1$S$ &      &         &       & 47.86   \\
2$S$ &      &         &       & 5.98 \\
3$S$ &      &         &       & 1.77 \\
4$S$ &      &         &       & 0.75 \\
5$S$ &      &         &       & 0.38 \\
\hline
1$P$ &30.64   &29.28   &15.35  &4.59  \\
2$P$ &13.62   &13.01   &4.55   &1.36  \\
3$P$ &7.66    &7.32    &1.92   &0.57   \\
4$P$ &4.90    &4.68    &0.98   &0.29   \\
5$P$ &3.40    &3.25    &0.57   &0.17   \\
\hline
1$D$ &13.62   &13.01   &0.91   &1.10   \\
2$D$ &7.66    &7.32    &0.38   &0.47   \\
3$D$ &4.90    &4.68    &0.20   &0.24   \\
4$D$ &3.40    &3.25    &0.11   &0.14   \\
5$D$ &2.50    &2.39    &0.07   &0.09   \\
\hline\hline
\end{tabular}
\end{table*}
\renewcommand\tabcolsep{1.0cm}
\renewcommand{\arraystretch}{0.8}
\begin{table*}[!htbp]
\caption{The spin coupling parameters (MeV) of the $\Xi_{b}'$ baryons.   \label{85bnh22}}
\begin{tabular}{ccccc}
\hline\hline
State: & $a_{1}$ & $a_{2}$ & $b_{1}$ &$c_{1}$ \\
 \hline
1$S$ &      &         &       & 15.38   \\
2$S$ &      &         &       & 1.92 \\
3$S$ &      &         &       & 0.57 \\
4$S$ &      &         &       & 0.24 \\
5$S$ &      &         &       & 0.12 \\
\hline
1$P$ &9.85 &9.41     &4.94  &1.48  \\
2$P$ &4.38 &4.18     &1.46  &0.44  \\
3$P$ &2.46 &2.35     &0.62  &0.18   \\
4$P$ &1.58 &1.51     &0.32  &0.09   \\
5$P$ &1.09 &1.05     &0.18  &0.05   \\
\hline
1$D$ &4.38 &4.18     &0.29  &0.35   \\
2$D$ &2.46 &2.35     &0.12  &0.15   \\
3$D$ &1.58 &1.51     &0.06  &0.08   \\
4$D$ &1.09 &1.05     &0.04  &0.04   \\
5$D$ &0.80 &0.77     &0.02  &0.03   \\
\hline\hline
\end{tabular}
\end{table*}
\begin{table*}[!htbp]
\caption{The mass spectrum (MeV) of $\Xi_{c}^{\prime}$ baryons are given and compared with different quark models.}\label{854dm323}
\resizebox{\textwidth}{12cm}{\begin{tabular}{ccccccc}
\hline\hline
{\small State }$J^{P}$ & {\small Baryon} & {\small Mass} &{Ours}&{\small EFG \cite{EFG:C10}}&Ref.{\cite{BingKe:PP888}} &Ref.{\cite{ShahK:PP888}} \\
\hline
$%
\begin{array}{rr}
{\small 1}^{1}{\small S}_{1/2} & {\small 1/2}^{+} \\
{\small 1}^{3}{\small S}_{3/2} & {\small 3/2}^{+}%
\end{array}%
$ & $%
\begin{array}{r}
{\small \Xi_{c}^{\prime 0}} \\
{\small \Xi_{c}(2645)^{0}}%
\end{array}%
$ & $%
\begin{array}{r}
{\small 2578.70} \\
{\small 2646.16}%
\end{array}%
$ & $%
\begin{array}{r}
{\small 2575.23} \\
{\small 2647.02}%
\end{array}%
$ & $%
\begin{array}{r}
{\small 2579} \\
{\small 2649}%
\end{array}%
$ & $%
\begin{array}{r}
{\small 2579} \\
{\small 2649}%
\end{array}%
$ & $%
\begin{array}{r}
{\small 2471} \\
{\small 2647}%
\end{array}%
$ \\ $%

\begin{array}{rr}
{\small 2}^{1}{\small S}_{1/2} & {\small 1/2}^{+} \\
{\small 2}^{3}{\small S}_{3/2} & {\small 3/2}^{+}%
\end{array}%
$ & $%
\begin{array}{r}
{\small } \\
{\small }%
\end{array}%
$ & $%
\begin{array}{r}
{\small } \\
{\small }%
\end{array}%
$ & $%
\begin{array}{r}
{\small 3014.28} \\
{\small 3023.25}%
\end{array}%
$ & $%
\begin{array}{r}
{\small 2983} \\
{\small 3026}%
\end{array}%
$ & $%
\begin{array}{r}
{\small 2977} \\
{\small 3007}%
\end{array}%
$ & $%
\begin{array}{r}
{\small 2937} \\
{\small 3004}%
\end{array}%
$ \\ $%

\begin{array}{rr}
{\small 3}^{1}{\small S}_{1/2} & {\small 1/2}^{+} \\
{\small 3}^{3}{\small S}_{3/2} & {\small 3/2}^{+}%
\end{array}%
$ & $%
\begin{array}{r}
{\small} \\
{\small}%
\end{array}%
$ & $%
\begin{array}{r}
{\small } \\
{\small }%
\end{array}%
$ & $%
\begin{array}{r}
{\small 3334.21} \\
{\small 3336.87}%
\end{array}%
$ & $%
\begin{array}{r}
{\small 3377} \\
{\small 3396}%
\end{array}%
$ & $%
\begin{array}{r}
{\small 3215} \\
{\small 3236}%
\end{array}%
$ & $%
\begin{array}{r}
{\small 3303} \\
{\small 3338}%
\end{array}%
$ \\ $%

\begin{array}{rr}
{\small 4}^{1}{\small S}_{1/2} & {\small 1/2}^{+} \\
{\small 4}^{3}{\small S}_{3/2} & {\small 3/2}^{+}%
\end{array}%
$ & $%
\begin{array}{r}
{\small } \\
{\small }%
\end{array}%
$ & $%
\begin{array}{r}
{\small } \\
{\small }%
\end{array}%
$ & $%
\begin{array}{r}
{\small 3605.41} \\
{\small 3606.54}%
\end{array}%
$ & $%
\begin{array}{r}
{\small 3695} \\
{\small 3709}%
\end{array}%
$ & $%
\begin{array}{r}
{\small } \\
{\small }%
\end{array}%
$ & $%
\begin{array}{r}
{\small 3626} \\
{\small 3646}%
\end{array}%
$ \\ $%

\begin{array}{rr}
{\small 5}^{1}{\small S}_{1/2} & {\small 1/2}^{+} \\
{\small 5}^{3}{\small S}_{3/2} & {\small 3/2}^{+}%
\end{array}%
$ & $%
\begin{array}{r}
{\small } \\
{\small }%
\end{array}%
$ & $%
\begin{array}{r}
{\small } \\
{\small }%
\end{array}%
$ & $%
\begin{array}{r}
{\small 3845.81} \\
{\small 3846.39}%
\end{array}%
$ & $%
\begin{array}{r}
{\small 3978} \\
{\small 3989}%
\end{array}%
$ & $%
\begin{array}{r}
{\small } \\
{\small}%
\end{array}%
$ & $%
\begin{array}{r}
{\small 3921} \\
{\small 3934}%
\end{array}%
$ \\
\hline

$
\begin{array}{rr}
{\small 1}^{2}{\small P}_{1/2} & {\small 1/2}^{-} \\
{\small 1}^{4}{\small P}_{1/2} & {\small 1/2}^{-} \\
{\small 1}^{2}{\small P}_{3/2} & {\small 3/2}^{-} \\
{\small 1}^{4}{\small P}_{3/2} & {\small 3/2}^{-} \\
{\small 1}^{4}{\small P}_{5/2} & {\small 5/2}^{-}%
\end{array}%
$ & $%
\begin{array}{r}
{\small } \\
{\small } \\
{\small \Xi_{c}(2923)^{0}} \\
{\small \Xi_{c}(2930)^{0}} \\
{\small }%
\end{array}%
$ & $%
\begin{array}{r}
{\small }\\
{\small }\\
{\small 2923.04}\\
{\small 2938.55}\\
{\small }%
\end{array}%
$ & $%
\begin{array}{r}
{\small 2833.11} \\
{\small 2889.71}\\
{\small 2907.21} \\
{\small 2935.17}\\
{\small 2968.02}%
\end{array}%
$ & $%
\begin{array}{r}
{\small 2854} \\
{\small 2936} \\
{\small 2912} \\
{\small 2935} \\
{\small 2929}%
\end{array}%
$ & $%
\begin{array}{r}
{\small 2839} \\
{\small 2900} \\
{\small 2921} \\
{\small 2932} \\
{\small 2927}%
\end{array}%
$ & $%
\begin{array}{r}
{\small 2877} \\
{\small 2834} \\
{\small 2899} \\
{\small 2856} \\
{\small 2798}%
\end{array}%
$ \\
\hline

$
\begin{array}{rr}
{\small 2}^{2}{\small P}_{1/2} & {\small 1/2}^{-} \\
{\small 2}^{4}{\small P}_{1/2} & {\small 1/2}^{-} \\
{\small 2}^{2}{\small P}_{3/2} & {\small 3/2}^{-} \\
{\small 2}^{4}{\small P}_{3/2} & {\small 3/2}^{-} \\
{\small 2}^{4}{\small P}_{5/2} & {\small 5/2}^{-}%
\end{array}%
$ & $%
\begin{array}{r}
{\small } \\
{\small } \\
{\small } \\
{\small } \\
{\small }%
\end{array}%
$ & $%
\begin{array}{r}
 \\
\\
 \\
\\
\end{array}%
$ & $%
\begin{array}{r}
{\small 3218.35} \\
{\small 3241.33}\\
{\small 3246.89} \\
{\small 3261.82}\\
{\small 3276.02}%
\end{array}%
$ & $%
\begin{array}{r}
{\small 3267} \\
{\small 3313} \\
{\small 3293} \\
{\small 3311} \\
{\small 3303}%
\end{array}%
$ & $%
\begin{array}{r}
{\small 3094} \\
{\small 3144} \\
{\small 3172} \\
{\small 3165} \\
{\small 3170}%
\end{array}%
$ & $%
\begin{array}{r}
{\small 3222} \\
{\small 3189} \\
{\small 3239} \\
{\small 3206} \\
{\small 3162}%
\end{array}%
$ \\
\hline

$
\begin{array}{rr}
{\small 3}^{2}{\small P}_{1/2} & {\small 1/2}^{-} \\
{\small 3}^{4}{\small P}_{1/2} & {\small 1/2}^{-} \\
{\small 3}^{2}{\small P}_{3/2} & {\small 3/2}^{-} \\
{\small 3}^{4}{\small P}_{3/2} & {\small 3/2}^{-} \\
{\small 3}^{4}{\small P}_{5/2} & {\small 5/2}^{-}%
\end{array}%
$ & $%
\begin{array}{r}
{\small } \\
{\small } \\
{\small } \\
{\small } \\
{\small }%
\end{array}%
$ & $%
\begin{array}{r}
\\
\\
\\
\\
\end{array}%
$ & $%
\begin{array}{r}
{\small 3516.01} \\
{\small 3528.40}\\
{\small 3530.85} \\
{\small 3539.99}\\
{\small 3547.85}%
\end{array}%
$ & $%
\begin{array}{r}
{\small 3598} \\
{\small 3630} \\
{\small 3613} \\
{\small 3628} \\
{\small 3619}%
\end{array}%
$ & $%
\begin{array}{r}
{\small } \\
{\small } \\
{\small } \\
{\small } \\
{\small }%
\end{array}%
$ & $%
\begin{array}{r}
{\small 3544} \\
{\small 3512} \\
{\small 3561} \\
{\small 3528} \\
{\small 3484}%
\end{array}%
$ \\
\hline

$
\begin{array}{rr}
{\small 4}^{2}{\small P}_{1/2} & {\small 1/2}^{-} \\
{\small 4}^{4}{\small P}_{1/2} & {\small 1/2}^{-} \\
{\small 4}^{2}{\small P}_{3/2} & {\small 3/2}^{-} \\
{\small 4}^{4}{\small P}_{3/2} & {\small 3/2}^{-} \\
{\small 4}^{4}{\small P}_{5/2} & {\small 5/2}^{-}%
\end{array}%
$ & $%
\begin{array}{r}
{\small } \\
{\small } \\
{\small } \\
{\small } \\
{\small }%
\end{array}%
$ & $%
\begin{array}{r}
 \\
\\
\\
\\
\end{array}%
$ & $%
\begin{array}{r}
{\small 3770.84} \\
{\small 3778.57}\\
{\small 3779.87} \\
{\small 3786.01}\\
{\small 3790.99}%
\end{array}%
$ & $%
\begin{array}{r}
{\small 3887} \\
{\small 3912} \\
{\small 3898} \\
{\small 3911} \\
{\small 3902}%
\end{array}%
$ & $%
\begin{array}{r}
{\small } \\
{\small } \\
{\small } \\
{\small } \\
{\small }%
\end{array}%
$ & $%
\begin{array}{r}
{\small 3837} \\
{\small 3808} \\
{\small 3851} \\
{\small 3823} \\
{\small 3784}%
\end{array}%
$ \\
\hline

$
\begin{array}{rr}
{\small 5}^{2}{\small P}_{1/2} & {\small 1/2}^{-} \\
{\small 5}^{4}{\small P}_{1/2} & {\small 1/2}^{-} \\
{\small 5}^{2}{\small P}_{3/2} & {\small 3/2}^{-} \\
{\small 5}^{4}{\small P}_{3/2} & {\small 3/2}^{-} \\
{\small 5}^{4}{\small P}_{5/2} & {\small 5/2}^{-}%
\end{array}%
$ & $%
\begin{array}{r}
{\small } \\
{\small } \\
{\small } \\
{\small } \\
{\small }%
\end{array}%
$ & $%
\begin{array}{r}
\\
\\
\\
\\
\end{array}%
$ & $%
\begin{array}{r}
{\small 3998.38} \\
{\small 4003.67}\\
{\small 4004.44} \\
{\small 4008.84}\\
{\small 4012.27}%
\end{array}%
$ & $%
\begin{array}{r}
{\small } \\
{\small } \\
{\small } \\
{\small } \\
{\small }%
\end{array}%
$ \\
\hline

$
\begin{array}{rr}
{\small 1}^{4}{\small D}_{1/2} & {\small 1/2}^{+} \\
{\small 1}^{2}{\small D}_{3/2} & {\small 3/2}^{+} \\
{\small 1}^{4}{\small D}_{3/2} & {\small 3/2}^{+} \\
{\small 1}^{2}{\small D}_{5/2} & {\small 5/2}^{+} \\
{\small 1}^{4}{\small D}_{5/2} & {\small 5/2}^{+} \\
{\small 1}^{4}{\small D}_{7/2} & {\small 7/2}^{+}%
\end{array}%
$ & $%
\begin{array}{r}
{\small } \\
{\small \Xi_{c}(3123)^{+}} \\
{\small } \\
{\small } \\
{\small } \\
{\small }%
\end{array}%
$ & $%
\begin{array}{r}
{\small } \\
{\small 3122.9}\\
{\small }\\
{\small }\\
{\small }\\
{\small }%
\end{array}%
$ & $%
\begin{array}{r}
{\small 3111.88} \\
{\small 3132.91}\\
{\small 3150.77}\\
{\small 3167.05}\\
{\small 3185.38}\\
{\small 3213.14}%
\end{array}%
$ & $%
\begin{array}{r}
{\small 3163} \\
{\small 3160} \\
{\small 3167} \\
{\small 3153} \\
{\small 3166} \\
{\small 3147}%
\end{array}%
$ & $%
\begin{array}{r}
{\small 3075} \\
{\small 3089} \\
{\small 3081} \\
{\small 3091} \\
{\small 3077} \\
{\small 3078}%
\end{array}%
$ & $%
\begin{array}{r}
{\small 3147} \\
{\small 3109} \\
{\small 3090} \\
{\small 3058} \\
{\small 3039} \\
{\small 2995}%
\end{array}%
$ \\
\hline

$
\begin{array}{rr}
{\small 2}^{4}{\small D}_{1/2} & {\small 1/2}^{+} \\
{\small 2}^{2}{\small D}_{3/2} & {\small 3/2}^{+} \\
{\small 2}^{4}{\small D}_{3/2} & {\small 3/2}^{+} \\
{\small 2}^{2}{\small D}_{5/2} & {\small 5/2}^{+} \\
{\small 2}^{4}{\small D}_{5/2} & {\small 5/2}^{+} \\
{\small 2}^{4}{\small D}_{7/2} & {\small 7/2}^{+}%
\end{array}%
$ & $%
\begin{array}{r}
{\small } \\
{\small } \\
{\small } \\
{\small } \\
{\small } \\
{\small }%
\end{array}%
$ & $%
\begin{array}{r}
 \\
\\
\\
\\
\\
\end{array}%
$ & $%
\begin{array}{r}
{\small 3430.60} \\
{\small 3442.30}\\
{\small 3452.58}\\
{\small 3461.41} \\
{\small 3472.04}\\
{\small 3487.47}%
\end{array}%
$ & $%
\begin{array}{r}
{\small 3505} \\
{\small 3497} \\
{\small 3506} \\
{\small 3493} \\
{\small 3504} \\
{\small 3486}%
\end{array}%
$ & $%
\begin{array}{r}
{\small } \\
{\small } \\
{\small } \\
{\small } \\
{\small } \\
{\small }%
\end{array}%
$ & $%
\begin{array}{r}
{\small 3470} \\
{\small 3417} \\
{\small 3434} \\
{\small 3701} \\
{\small 3388} \\
{\small 3330}%
\end{array}%
$ \\
\hline

$
\begin{array}{rr}
{\small 3}^{4}{\small D}_{1/2} & {\small 1/2}^{+} \\
{\small 3}^{2}{\small D}_{3/2} & {\small 3/2}^{+} \\
{\small 3}^{4}{\small D}_{3/2} & {\small 3/2}^{+} \\
{\small 3}^{2}{\small D}_{5/2} & {\small 5/2}^{+} \\
{\small 3}^{4}{\small D}_{5/2} & {\small 5/2}^{+} \\
{\small 3}^{4}{\small D}_{7/2} & {\small 7/2}^{+}%
\end{array}%
$ & $%
\begin{array}{r}
{\small } \\
{\small } \\
{\small } \\
{\small } \\
{\small } \\
{\small }%
\end{array}%
$ & $%
\begin{array}{r}
\\
\\
\\
\\
\\
\end{array}%
$ & $%
\begin{array}{r}
{\small 3697.87} \\
{\small 3705.31}\\
{\small 3711.98} \\
{\small 3717.51}\\
{\small 3724.43}\\
{\small 3734.23}%
\end{array}%
$ & $%
\begin{array}{r}
{\small } \\
{\small } \\
{\small } \\
{\small } \\
{\small } \\
{\small }%
\end{array}%
$ \\
\hline

$
\begin{array}{rr}
{\small 4}^{4}{\small D}_{1/2} & {\small 1/2}^{+} \\
{\small 4}^{2}{\small D}_{3/2} & {\small 3/2}^{+} \\
{\small 4}^{4}{\small D}_{3/2} & {\small 3/2}^{+} \\
{\small 4}^{2}{\small D}_{5/2} & {\small 5/2}^{+} \\
{\small 4}^{4}{\small D}_{5/2} & {\small 5/2}^{+} \\
{\small 4}^{4}{\small D}_{7/2} & {\small 7/2}^{+}%
\end{array}%
$ & $%
\begin{array}{r}
{\small } \\
{\small } \\
{\small } \\
{\small } \\
{\small } \\
{\small }%
\end{array}%
$ & $%
\begin{array}{r}
\\
\\
\\
\\
\\
\end{array}%
$ & $%
\begin{array}{r}
{\small 3933.74}\\
{\small 3938.88}\\
{\small 3943.56}\\
{\small 3947.34}\\
{\small 3952.20}\\
{\small 3958.98}%
\end{array}%
$ & $%
\begin{array}{r}
{\small } \\
{\small } \\
{\small } \\
{\small } \\
{\small } \\
{\small }%
\end{array}%
$ \\
\hline

$
\begin{array}{rr}
{\small 5}^{4}{\small D}_{1/2} & {\small 1/2}^{+} \\
{\small 5}^{2}{\small D}_{3/2} & {\small 3/2}^{+} \\
{\small 5}^{4}{\small D}_{3/2} & {\small 3/2}^{+} \\
{\small 5}^{2}{\small D}_{5/2} & {\small 5/2}^{+} \\
{\small 5}^{4}{\small D}_{5/2} & {\small 5/2}^{+} \\
{\small 5}^{4}{\small D}_{7/2} & {\small 7/2}^{+}%
\end{array}%
$ & $%
\begin{array}{r}
{\small } \\
{\small } \\
{\small } \\
{\small } \\
{\small } \\
{\small }%
\end{array}%
$ & $%
\begin{array}{r}
\\
\\
\\
\\
\\
\end{array}%
$ & $%
\begin{array}{r}
{\small 4147.70} \\
{\small 4151.47}\\
{\small 4154.93}\\
{\small 4157.68} \\
{\small 4161.28}\\
{\small 4166.24}%
\end{array}%
$ & $%
\begin{array}{r}
{\small } \\
{\small } \\
{\small } \\
{\small } \\
{\small } \\
{\small }%
\end{array}%
$ \\
\hline\hline
\end{tabular}}
\end{table*}
\begin{table*}[!htbp]
\caption{The mass spectrum (MeV) of $\Xi_{b}^{\prime}$ baryons are given and compared with different quark models.}\label{8225dm44}
\resizebox{\textwidth}{12cm}{\begin{tabular}{ccccccc}
\hline\hline
{\small State }$J^{P}$ & {\small Baryon} & {\small Mass} &{Ours}&{\small EFG \cite{EFG:C10}} &Ref.{\cite{Kakayas:PP888}} &Ref.{\cite{OudichhyGan:PP888}} \\
\hline
$%
\begin{array}{rr}
{\small 1}^{1}{\small S}_{1/2} & {\small 1/2}^{+} \\
{\small 1}^{3}{\small S}_{3/2} & {\small 3/2}^{+}%
\end{array}%
$ & $%
\begin{array}{r}
{\small \Xi_{b}^{\prime}(5935)^{-}} \\
{\small \Xi_{b}^{\ast}(5955)^{-}}%
\end{array}%
$ & $%
\begin{array}{r}
{\small 5935.02} \\
{\small 5955.33}%
\end{array}%
$ & $%
\begin{array}{r}
{\small 5930.03} \\
{\small 5953.08}%
\end{array}%
$ & $%
\begin{array}{r}
{\small 5936} \\
{\small 5963}%
\end{array}%
$ & $%
\begin{array}{r}
{\small 5935} \\
{\small 5958}%
\end{array}%
$ & $%
\begin{array}{r}
{\small 5935} \\
{\small }%
\end{array}%
$ \\ $%

\begin{array}{rr}
{\small 2}^{1}{\small S}_{1/2} & {\small 1/2}^{+} \\
{\small 2}^{3}{\small S}_{3/2} & {\small 3/2}^{+}%
\end{array}%
$ & $%
\begin{array}{r}
{\small } \\
{\small }%
\end{array}%
$ & $%
\begin{array}{r}
{\small } \\
{\small }%
\end{array}%
$ & $%
\begin{array}{r}
{\small 6341.53} \\
{\small 6344.41}%
\end{array}%
$ & $%
\begin{array}{r}
{\small 6329} \\
{\small 6342}%
\end{array}%
$ & $%
\begin{array}{r}
{\small 6328} \\
{\small 6343}%
\end{array}%
$ & $%
\begin{array}{r}
{\small 6329} \\
{\small }%
\end{array}%
$ \\ $%

\begin{array}{rr}
{\small 3}^{1}{\small S}_{1/2} & {\small 1/2}^{+} \\
{\small 3}^{3}{\small S}_{3/2} & {\small 3/2}^{+}%
\end{array}%
$ & $%
\begin{array}{r}
{\small} \\
{\small}%
\end{array}%
$ & $%
\begin{array}{r}
{\small } \\
{\small }%
\end{array}%
$ & $%
\begin{array}{r}
{\small 6669.60} \\
{\small 6670.46}%
\end{array}%
$ & $%
\begin{array}{r}
{\small 6687} \\
{\small 6695}%
\end{array}%
$ & $%
\begin{array}{r}
{\small 6625} \\
{\small 6634}%
\end{array}%
$ & $%
\begin{array}{r}
{\small 6700} \\
{\small }%
\end{array}%
$ \\ $%

\begin{array}{rr}
{\small 4}^{1}{\small S}_{1/2} & {\small 1/2}^{+} \\
{\small 4}^{3}{\small S}_{3/2} & {\small 3/2}^{+}%
\end{array}%
$ & $%
\begin{array}{r}
{\small } \\
{\small }%
\end{array}%
$ & $%
\begin{array}{r}
{\small } \\
{\small }%
\end{array}%
$ & $%
\begin{array}{r}
{\small 6953.79} \\
{\small 6954.15}%
\end{array}%
$ & $%
\begin{array}{r}
{\small 6978} \\
{\small 6984}%
\end{array}%
$ & $%
\begin{array}{r}
{\small 6902} \\
{\small 6907}%
\end{array}%
$ & $%
\begin{array}{r}
{\small 7051} \\
{\small }%
\end{array}%
$ \\ $%

\begin{array}{rr}
{\small 5}^{1}{\small S}_{1/2} & {\small 1/2}^{+} \\
{\small 5}^{3}{\small S}_{3/2} & {\small 3/2}^{+}%
\end{array}%
$ & $%
\begin{array}{r}
{\small } \\
{\small }%
\end{array}%
$ & $%
\begin{array}{r}
{\small } \\
{\small }%
\end{array}%
$ & $%
\begin{array}{r}
{\small 7208.35} \\
{\small 7208.53}%
\end{array}%
$ & $%
\begin{array}{r}
{\small 7229} \\
{\small 7234}%
\end{array}%
$ & $%
\begin{array}{r}
{\small 7161} \\
{\small 7165}%
\end{array}%
$ & $%
\begin{array}{r}
{\small 7386} \\
{\small }%
\end{array}%
$ \\
\hline

$
\begin{array}{rr}
{\small 1}^{2}{\small P}_{1/2} & {\small 1/2}^{-} \\
{\small 1}^{4}{\small P}_{1/2} & {\small 1/2}^{-} \\
{\small 1}^{2}{\small P}_{3/2} & {\small 3/2}^{-} \\
{\small 1}^{4}{\small P}_{3/2} & {\small 3/2}^{-} \\
{\small 1}^{4}{\small P}_{5/2} & {\small 5/2}^{-}%
\end{array}%
$ & $%
\begin{array}{r}
{\small } \\
{\small \Xi_{b}(6227)^{-}} \\
{\small } \\
{\small } \\
{\small }%
\end{array}%
$ & $%
\begin{array}{r}
{\small } \\
{\small 6227.9} \\
{\small } \\
{\small } \\
{\small }%
\end{array}%
$ & $%
\begin{array}{r}
{\small 6215.82} \\
{\small 6233.90}\\
{\small 6239.61} \\
{\small 6248.59}\\
{\small 6259.13}%
\end{array}%
$ & $%
\begin{array}{r}
{\small 6227} \\
{\small 6233} \\
{\small 6224} \\
{\small 6234} \\
{\small 6226}%
\end{array}%
$ & $%
\begin{array}{r}
{\small 6235} \\
{\small 6237} \\
{\small 6232} \\
{\small 6234} \\
{\small 6229}%
\end{array}%
$ & $%
\begin{array}{r}
{\small } \\
{\small } \\
{\small 6229} \\
{\small } \\
{\small }%
\end{array}%
$ \\
\hline

$
\begin{array}{rr}
{\small 2}^{2}{\small P}_{1/2} & {\small 1/2}^{-} \\
{\small 2}^{4}{\small P}_{1/2} & {\small 1/2}^{-} \\
{\small 2}^{2}{\small P}_{3/2} & {\small 3/2}^{-} \\
{\small 2}^{4}{\small P}_{3/2} & {\small 3/2}^{-} \\
{\small 2}^{4}{\small P}_{5/2} & {\small 5/2}^{-}%
\end{array}%
$ & $%
\begin{array}{r}
{\small } \\
{\small } \\
{\small } \\
{\small } \\
{\small }%
\end{array}%
$ & $%
\begin{array}{r}
 \\
\\
 \\
\\
\end{array}%
$ & $%
\begin{array}{r}
{\small 6574.81} \\
{\small 6582.19}\\
{\small 6583.98} \\
{\small 6588.77}\\
{\small 6593.33}%
\end{array}%
$ & $%
\begin{array}{r}
{\small 6604} \\
{\small 6611} \\
{\small 6598} \\
{\small 6605} \\
{\small 6596}%
\end{array}%
$ & $%
\begin{array}{r}
{\small 6494} \\
{\small 6495} \\
{\small 6492} \\
{\small 6493} \\
{\small 6490}%
\end{array}%
$ & $%
\begin{array}{r}
{\small } \\
{\small } \\
{\small 6605} \\
{\small } \\
{\small }%
\end{array}%
$ \\
\hline

$
\begin{array}{rr}
{\small 3}^{2}{\small P}_{1/2} & {\small 1/2}^{-} \\
{\small 3}^{4}{\small P}_{1/2} & {\small 1/2}^{-} \\
{\small 3}^{2}{\small P}_{3/2} & {\small 3/2}^{-} \\
{\small 3}^{4}{\small P}_{3/2} & {\small 3/2}^{-} \\
{\small 3}^{4}{\small P}_{5/2} & {\small 5/2}^{-}%
\end{array}%
$ & $%
\begin{array}{r}
{\small } \\
{\small } \\
{\small } \\
{\small } \\
{\small }%
\end{array}%
$ & $%
\begin{array}{r}
\\
\\
\\
\\
\end{array}%
$ & $%
\begin{array}{r}
{\small 6874.07} \\
{\small 6878.04}\\
{\small 6878.83} \\
{\small 6881.77}\\
{\small 6884.27}%
\end{array}%
$ & $%
\begin{array}{r}
{\small 6905} \\
{\small 6906} \\
{\small 6897} \\
{\small 6900} \\
{\small 6897}%
\end{array}%
$ & $%
\begin{array}{r}
{\small 6731} \\
{\small 6732} \\
{\small 6729} \\
{\small 6730} \\
{\small 6728}%
\end{array}%
$ & $%
\begin{array}{r}
{\small } \\
{\small } \\
{\small 6961} \\
{\small } \\
{\small }%
\end{array}%
$ \\
\hline

$
\begin{array}{rr}
{\small 4}^{2}{\small P}_{1/2} & {\small 1/2}^{-} \\
{\small 4}^{4}{\small P}_{1/2} & {\small 1/2}^{-} \\
{\small 4}^{2}{\small P}_{3/2} & {\small 3/2}^{-} \\
{\small 4}^{4}{\small P}_{3/2} & {\small 3/2}^{-} \\
{\small 4}^{4}{\small P}_{5/2} & {\small 5/2}^{-}%
\end{array}%
$ & $%
\begin{array}{r}
{\small } \\
{\small } \\
{\small } \\
{\small } \\
{\small }%
\end{array}%
$ & $%
\begin{array}{r}
 \\
\\
\\
\\
\end{array}%
$ & $%
\begin{array}{r}
{\small 7138.00} \\
{\small 7140.48}\\
{\small 7140.90} \\
{\small 7142.87}\\
{\small 7144.47}%
\end{array}%
$ & $%
\begin{array}{r}
{\small 7164} \\
{\small 7174} \\
{\small 7159} \\
{\small 7163} \\
{\small 7156}%
\end{array}%
$ & $%
\begin{array}{r}
{\small 6949} \\
{\small 6950} \\
{\small 6948} \\
{\small 6949} \\
{\small 6947}%
\end{array}%
$ & $%
\begin{array}{r}
{\small } \\
{\small } \\
{\small 7299} \\
{\small } \\
{\small }%
\end{array}%
$ \\
\hline

$
\begin{array}{rr}
{\small 5}^{2}{\small P}_{1/2} & {\small 1/2}^{-} \\
{\small 5}^{4}{\small P}_{1/2} & {\small 1/2}^{-} \\
{\small 5}^{2}{\small P}_{3/2} & {\small 3/2}^{-} \\
{\small 5}^{4}{\small P}_{3/2} & {\small 3/2}^{-} \\
{\small 5}^{4}{\small P}_{5/2} & {\small 5/2}^{-}%
\end{array}%
$ & $%
\begin{array}{r}
{\small } \\
{\small } \\
{\small } \\
{\small } \\
{\small }%
\end{array}%
$ & $%
\begin{array}{r}
\\
\\
\\
\\
\end{array}%
$ & $%
\begin{array}{r}
{\small 7377.26} \\
{\small 7378.96}\\
{\small 7379.20} \\
{\small 7380.61}\\
{\small 7381.72}%
\end{array}%
$ & $%
\begin{array}{r}
{\small } \\
{\small } \\
{\small } \\
{\small } \\
{\small }%
\end{array}%
$ & $%
\begin{array}{r}
{\small } \\
{\small } \\
{\small } \\
{\small } \\
{\small }%
\end{array}%
$ & $%
\begin{array}{r}
{\small } \\
{\small } \\
{\small 7622} \\
{\small } \\
{\small }%
\end{array}%
$ \\
\hline

$
\begin{array}{rr}
{\small 1}^{4}{\small D}_{1/2} & {\small 1/2}^{+} \\
{\small 1}^{2}{\small D}_{3/2} & {\small 3/2}^{+} \\
{\small 1}^{4}{\small D}_{3/2} & {\small 3/2}^{+} \\
{\small 1}^{2}{\small D}_{5/2} & {\small 5/2}^{+} \\
{\small 1}^{4}{\small D}_{5/2} & {\small 5/2}^{+} \\
{\small 1}^{4}{\small D}_{7/2} & {\small 7/2}^{+}%
\end{array}%
$ & $%
\begin{array}{r}
{\small } \\
{\small } \\
{\small } \\
{\small } \\
{\small } \\
{\small }%
\end{array}%
$ & $%
\begin{array}{r}
 \\
\\
\\
\\
\\
\end{array}%
$ & $%
\begin{array}{r}
{\small 6480.78} \\
{\small 6487.54}\\
{\small 6493.27}\\
{\small 6498.50}\\
{\small 6504.38}\\
{\small 6513.30}%
\end{array}%
$ & $%
\begin{array}{r}
{\small 6447} \\
{\small 6431} \\
{\small 6459} \\
{\small 6420} \\
{\small 6432} \\
{\small 6414}%
\end{array}%
$ & $%
\begin{array}{r}
{\small 6380} \\
{\small 6375} \\
{\small 6377} \\
{\small 6371} \\
{\small 6373} \\
{\small 6368}%
\end{array}%
$ & $%
\begin{array}{r}
{\small } \\
{\small } \\
{\small } \\
{\small 6510} \\
{\small } \\
{\small }%
\end{array}%
$ \\
\hline

$
\begin{array}{rr}
{\small 2}^{4}{\small D}_{1/2} & {\small 1/2}^{+} \\
{\small 2}^{2}{\small D}_{3/2} & {\small 3/2}^{+} \\
{\small 2}^{4}{\small D}_{3/2} & {\small 3/2}^{+} \\
{\small 2}^{2}{\small D}_{5/2} & {\small 5/2}^{+} \\
{\small 2}^{4}{\small D}_{5/2} & {\small 5/2}^{+} \\
{\small 2}^{4}{\small D}_{7/2} & {\small 7/2}^{+}%
\end{array}%
$ & $%
\begin{array}{r}
{\small } \\
{\small } \\
{\small } \\
{\small } \\
{\small } \\
{\small }%
\end{array}%
$ & $%
\begin{array}{r}
 \\
\\
\\
\\
\\
\end{array}%
$ & $%
\begin{array}{r}
{\small 6794.07} \\
{\small 6797.83}\\
{\small 6801.13}\\
{\small 6803.97} \\
{\small 6807.38}\\
{\small 6812.33}%
\end{array}%
$ & $%
\begin{array}{r}
{\small 6767} \\
{\small 6751} \\
{\small 6775} \\
{\small 6740} \\
{\small 6751} \\
{\small 6736}%
\end{array}%
$ & $%
\begin{array}{r}
{\small 6632} \\
{\small 6628} \\
{\small 6630} \\
{\small 6625} \\
{\small 6626} \\
{\small 6621}%
\end{array}%
$ & $%
\begin{array}{r}
{\small } \\
{\small } \\
{\small } \\
{\small 6751} \\
{\small } \\
{\small }%
\end{array}%
$ \\
\hline

$
\begin{array}{rr}
{\small 3}^{4}{\small D}_{1/2} & {\small 1/2}^{+} \\
{\small 3}^{2}{\small D}_{3/2} & {\small 3/2}^{+} \\
{\small 3}^{4}{\small D}_{3/2} & {\small 3/2}^{+} \\
{\small 3}^{2}{\small D}_{5/2} & {\small 5/2}^{+} \\
{\small 3}^{4}{\small D}_{5/2} & {\small 5/2}^{+} \\
{\small 3}^{4}{\small D}_{7/2} & {\small 7/2}^{+}%
\end{array}%
$ & $%
\begin{array}{r}
{\small } \\
{\small } \\
{\small } \\
{\small } \\
{\small } \\
{\small }%
\end{array}%
$ & $%
\begin{array}{r}
\\
\\
\\
\\
\\
\end{array}%
$ & $%
\begin{array}{r}
{\small 7067.14} \\
{\small 7069.53}\\
{\small 7071.67} \\
{\small 7073.45}\\
{\small 7075.67}\\
{\small 7078.82}%
\end{array}%
$ & $%
\begin{array}{r}
{\small } \\
{\small } \\
{\small } \\
{\small } \\
{\small } \\
{\small }%
\end{array}%
$ & $%
\begin{array}{r}
{\small 6861} \\
{\small 6859} \\
{\small 6860} \\
{\small 6856} \\
{\small 6857} \\
{\small 6854}%
\end{array}%
$ & $%
\begin{array}{r}
{\small } \\
{\small } \\
{\small } \\
{\small 6984} \\
{\small } \\
{\small }%
\end{array}%
$ \\
\hline

$
\begin{array}{rr}
{\small 4}^{4}{\small D}_{1/2} & {\small 1/2}^{+} \\
{\small 4}^{2}{\small D}_{3/2} & {\small 3/2}^{+} \\
{\small 4}^{4}{\small D}_{3/2} & {\small 3/2}^{+} \\
{\small 4}^{2}{\small D}_{5/2} & {\small 5/2}^{+} \\
{\small 4}^{4}{\small D}_{5/2} & {\small 5/2}^{+} \\
{\small 4}^{4}{\small D}_{7/2} & {\small 7/2}^{+}%
\end{array}%
$ & $%
\begin{array}{r}
{\small } \\
{\small } \\
{\small } \\
{\small } \\
{\small } \\
{\small }%
\end{array}%
$ & $%
\begin{array}{r}
\\
\\
\\
\\
\\
\end{array}%
$ & $%
\begin{array}{r}
{\small 7312.96}\\
{\small 7314.61}\\
{\small 7316.11}\\
{\small 7317.32}\\
{\small 7318.88}\\
{\small 7321.06}%
\end{array}%
$ & $%
\begin{array}{r}
{\small } \\
{\small } \\
{\small } \\
{\small } \\
{\small } \\
{\small }%
\end{array}%
$ & $%
\begin{array}{r}
{\small 7072} \\
{\small 7070} \\
{\small 7071} \\
{\small 7069} \\
{\small 7069} \\
{\small 7067}%
\end{array}%
$ & $%
\begin{array}{r}
{\small } \\
{\small } \\
{\small } \\
{\small 7209} \\
{\small } \\
{\small }%
\end{array}%
$ \\
\hline

$
\begin{array}{rr}
{\small 5}^{4}{\small D}_{1/2} & {\small 1/2}^{+} \\
{\small 5}^{2}{\small D}_{3/2} & {\small 3/2}^{+} \\
{\small 5}^{4}{\small D}_{3/2} & {\small 3/2}^{+} \\
{\small 5}^{2}{\small D}_{5/2} & {\small 5/2}^{+} \\
{\small 5}^{4}{\small D}_{5/2} & {\small 5/2}^{+} \\
{\small 5}^{4}{\small D}_{7/2} & {\small 7/2}^{+}%
\end{array}%
$ & $%
\begin{array}{r}
{\small } \\
{\small } \\
{\small } \\
{\small } \\
{\small } \\
{\small }%
\end{array}%
$ & $%
\begin{array}{r}
\\
\\
\\
\\
\\
\end{array}%
$ & $%
\begin{array}{r}
{\small 7538.65} \\
{\small 7539.77}\\
{\small 7540.88}\\
{\small 7541.76} \\
{\small 7542.92}\\
{\small 7544.51}%
\end{array}%
$ & $%
\begin{array}{r}
{\small } \\
{\small } \\
{\small } \\
{\small } \\
{\small } \\
{\small }%
\end{array}%
$ & $%
\begin{array}{r}
{\small } \\
{\small } \\
{\small } \\
{\small } \\
{\small } \\
{\small }%
\end{array}%
$ & $%
\begin{array}{r}
{\small } \\
{\small } \\
{\small } \\
{\small 7427} \\
{\small } \\
{\small }%
\end{array}%
$ \\
\hline\hline
\end{tabular}}
\end{table*}

\section{Summary}

Stimulated by new excited states found by LHCb, in this paper we study the mass spectra of the heavy baryons and the internal structure. Comparing with the experimental data of discovered singly heavy baryons and with predictions of existing theoretical models, the internal interaction of hadrons and the structure of the $\Sigma_{Q}$, $\Xi^{\prime}_{Q}$ and $\Omega_{Q}$ $(Q=c,b)$ baryons are being explored.

In this work, we use the $JLS$ mixing scheme to study the $S$, $P$ and $D$-wave states of the baryons. To calculate the mass splitting of the singly heavy baryons, we discuss the Regge trajectory and the spin-dependent potential in the quark-diquark picture. In our model, we establish new scaling relations to determine the spin coupling parameters $a_{1}$, $a_{2}$, $b_{1}$, $c_{1}$. The parameters for $1P$-wave states of the $\Omega_{c}$ baryons are treated as the object of the scaling relations. By analyzing the mass spectra of the discovered experimental data in PDG, we predict the mass spectra of several unobserved baryons. In addition, our analysis indicates the two new excited $\Omega_{c}$ states as $2 ^1{S}_{1/2}$ and $1^2{D}_{3/2}$ for $\Omega_{c}(3185)^{0}$ and $\Omega_{c}(3327)^{0}$, respectively. These predictions provide important references for future experimental exploration.

\appendix

\section{$S$-wave}

Analyzing $S$-wave mass splitting with the the orbital angular momentum $L=0$, the singly heavy baryon is considered in the approximation of a system of a single heavy quark and a light diquark, with the heavy quark spin ${S}_{Q}=1/2$ and diquark spin ${S}_{d}=1$, respectively. Therefore, there are two possibilities for the total spin $\mathbf{S}$, one is 1/2 and the other is 3/2. In the scheme of $LS$ coupling, the spin of the diquark $\mathbf{S}_{d}$ and the spin of the heavy quark $\mathbf{S}_{Q}$ couple to give $\mathbf{S}$ $(\mathbf{S}=\mathbf{S}_{d}+\mathbf{S}_{Q})$, before $\mathbf{S}$ is combined with $\mathbf{L}$ to generate the total angular momentum $\mathbf{J}$ $(\mathbf{J}=\mathbf{S}+\mathbf{L})$. We consider $S$-wave ($L = 0$) states in baryons $Qqq$, where the coupling of $L = 0$ with the spin $S = 1/2$ gives states with $J = 1/2$, while coupling with $S = 3/2$ leads to $J = 3/2$. In this case, the first three terms in Eq. (\ref{PP5}) are eliminated, only the last term survives,
\begin{equation}
H_{2}^{SD}=c_{1}\mathbf{S}_{d}\cdot \mathbf{S}_{Q}.  \label{PP6}
\end{equation}
It is very convenient to analyze the influence of spin-spin interaction on the non-trivial
terms for the mass splitting. The matrix elements of $\mathbf{S}_{d}\cdot \mathbf{S}_{Q}$ may be evaluated by explicit construction of states with the third component $S_{3}$ of the total spin given as linear combinations of the states $|S_{d3}, S_{Q3}\rangle$ and calculate the expectation value $\langle\mathbf{S}_d\cdot\mathbf{S}_Q\rangle= [S(S+1)-S_Q(S_Q+1)-S_d(S_d+1)]/2$
of $\mathbf{S}_d\cdot\mathbf{S}_Q$ as the square of the total spin $\mathbf{S}=\mathbf{S}_{Q}+\mathbf{S}_{d}$,
\begin{equation}
\mathbf{S}_{d}\cdot \mathbf{S}_{Q}=\left( \mathbf{S}^{2}-\mathbf{S}_{d}^{2}- \mathbf{S}_{Q}^{2}\right)/2.  \label{PP7}
\end{equation}
The two basis states are
\begin{eqnarray}
|^{2} S_{1/2},S_{3}&=&1/2\rangle =\sqrt{\frac{2}{3}}|1,-\frac{1}{2}\rangle-\sqrt{\frac{1}{3}}|0,\frac{1}{2}\rangle
\notag, \\
|^{4} S_{3/2},S_{3}&=&3/2\rangle =|1,\frac{1}{2}\rangle.  \label{VPP1V}
\end{eqnarray}%
The eigenvalues (two diagonal elements) of $\langle\mathbf{S}_{d}\cdot \mathbf{S}_{Q}\rangle$ in the basis $[^{2}S_{1/2}, ^{4}S_{3/2}]$ can be obtained as
\begin{small}
\begin{eqnarray}
\langle\mathbf{S}_{d}\cdot \mathbf{S}_{Q}\rangle &=& \left[
\begin{array}{cc}
\langle^{2} S_{1/2},S_{3}=1/2|\mathbf{S}_{d}\cdot \mathbf{S}_{Q}|^{2} S_{1/2},S_{3}=1/2\rangle & \langle^{2} S_{1/2},S_{3}=1/2|\mathbf{S}_{d}\cdot \mathbf{S}_{Q}|^{4} S_{3/2},S_{3}=3/2\rangle \\
\langle^{4} S_{3/2},S_{3}=3/2|\mathbf{S}_{d}\cdot \mathbf{S}_{Q}|^{2} S_{1/2},S_{3}=1/2\rangle & \langle^{4} S_{3/2},S_{3}=3/2|\mathbf{S}_{d}\cdot \mathbf{S}_{Q}|^{4} S_{3/2},S_{3}=3/2\rangle
\end{array} \notag
\right]\\
&=& \left[
\begin{array}{cc}
-1 & 0 \\
0 & \frac{1}{2} \label{pp8}
\end{array}
\right].
\end{eqnarray}
\end{small}
Combining with Eqs. (\ref{pp4}) and (\ref{pp8}), the $S$-wave masses of the singly heavy baryons are
\begin{equation}
M=\bar M+c_{1}\left[
\begin{array}{cc}
-1 & 0 \\
0 & \frac{1}{2} \label{pp99}
\end{array}
\right] .
\end{equation}%

\section{$P$-wave}

Let us consider the $P$-wave system with the the orbital angular momentum $L=1$. The spin of the diquark $S_{d}=1$ can be coupled with the heavy quark spin $S_{Q}=1/2$ and $L=1$ to the total angular momentum $J=1/2, 3/2$ or $1/2, 3/2, 5/2$ with negative parity $P=-1$. The expectation value of $\mathbf{L}\cdot \mathbf{S}$ in any coupling scheme is
\begin{eqnarray}
\langle\mathbf{L}\cdot \mathbf{S}\rangle = [J(J+1)-L(L+1)-S(S+1)]/2,
\end{eqnarray}
and the calculation of the operator $\mathbf{L}\cdot\mathbf{S}_{i}$ $(i=Q, d)$ results in
\begin{eqnarray}
\mathbf{L}\cdot\mathbf{S}_{i}=L_{3}S_{i3}+\left(L_{+}S_{i-}+L_{-}S_{i+}\right)/2,
\end{eqnarray}
with raising and lowering operator $L_{\pm}$, $S_{i\pm}$. The expectation values of $\mathbf{L}\cdot\mathbf{S}_{d}$, $\mathbf{L}\cdot\mathbf{S}_{Q}$, $S_{12}$ and $\mathbf{S}_{d}\cdot\mathbf{S}_{Q}$ in Eq. (\ref{PP5}) in the $L-S$ basis can be constructed as linear combinations of the states $|S_{d3}, S_{Q3}, L_{3}\rangle$ of the third components of the respective angular momenta,
\begin{eqnarray}
|^{2} P_{1/2},J_{3}&=&1/2\rangle =\frac{\sqrt{2}}{3}|1,-\frac{1}{2},0\rangle-\frac{1}{3}|0,\frac{1}{2},0\rangle-\frac{\sqrt{2}}{3}|0,-\frac{1}{2},1\rangle+\frac{2}{3}|-1,\frac{1}{2},1\rangle
\notag, \\
|^{4} P_{1/2},J_{3}&=&1/2\rangle =\frac{1}{\sqrt{2}}|1,\frac{1}{2},-1\rangle-\frac{1}{3}|1,-\frac{1}{2},0\rangle-\frac{\sqrt{2}}{3}|0,\frac{1}{2},0\rangle+\frac{1}{3}|0,-\frac{1}{2},1\rangle+\frac{1}{3\sqrt{{2}}}|-1,\frac{1}{2},1\rangle
\notag, \\
|^{2} P_{3/2},J_{3}&=&3/2\rangle =\sqrt{\frac{2}{3}}|1,-\frac{1}{2},1\rangle-\sqrt{\frac{1}{3}}|0,\frac{1}{2},1\rangle
\notag, \\
|^{4} P_{3/2},J_{3}&=&3/2\rangle =\sqrt{\frac{3}{5}}|1,\frac{1}{2},0\rangle-\sqrt{\frac{2}{15}}|1,-\frac{1}{2},1\rangle-\frac{2}{\sqrt{15}}|0,\frac{1}{2},1\rangle
\notag, \\
|^{4} P_{5/2},J_{3}&=&5/2\rangle =|1,\frac{1}{2},1\rangle.   \label{VV}
\end{eqnarray}%
The expectation values of $\langle\mathbf{L}\cdot \mathbf{S}_{i}\rangle$, $\langle S_{12}\rangle$ and $\langle\mathbf{S}_{d}\cdot \mathbf{S}_{Q}\rangle$ are given by
\begin{eqnarray}
\langle\mathbf{L\cdot S}_{d}\rangle_{J=\frac{1}{2}} &=&\left[
\begin{array}{cc}
-\frac{4}{3}& -\frac{\sqrt{2}}{3} \\-\frac{\sqrt{2}}{3} & -\frac{5}{3}
\end{array}%
\right],
\langle\mathbf{L\cdot S}_{Q}\rangle_{J=\frac{1}{2}} =\left[
\begin{array}{cc}
\frac{1}{3} & \frac{\sqrt{2}}{3} \\ \frac{\sqrt{2}}{3} & -\frac{5}{6}
\end{array}%
\right],
\langle S_{12}\rangle_{J=\frac{1}{2}} =\left[
\begin{array}{cc}
0 &\frac{1}{\sqrt{2}} \\ \frac{1}{\sqrt{2}} & -1%
\end{array}%
\right],\notag \\
&&\langle\mathbf{S}_{d}\cdot \mathbf{S}_{Q}\rangle_{J=\frac{1}{2}} =\left[
\begin{array}{cc}
-1 &0 \\ 0 & \frac{1}{2}
\end{array}%
\right],\notag \\
\langle\mathbf{L\cdot S}_{d}\rangle_{J=\frac{3}{2}} &=&\left[
\begin{array}{cc}
\frac{2}{3} &\frac{\sqrt{5}}{3} \\ -\frac{\sqrt{5}}{3} & \frac{2}{3}%
\end{array}%
\right],
\langle\mathbf{L\cdot S}_{Q}\rangle_{J=\frac{3}{2}} =\left[
\begin{array}{cc}
-\frac{1}{6} &\frac{\sqrt{5}}{3} \\ \frac{\sqrt{5}}{3} & -\frac{1}{3}%
\end{array}%
\right],
\langle S_{12}\rangle_{J=\frac{3}{2}} =\left[
\begin{array}{cc}
0 &-\frac{\sqrt{5}}{10} \\ -\frac{\sqrt{5}}{10} & \frac{4}{5}%
\end{array}%
\right],\notag\\
&&\langle\mathbf{S}_{d}\cdot \mathbf{S}_{Q}\rangle_{J=\frac{3}{2}} =\left[
\begin{array}{cc}
-1 & 0 \\ 0 & \frac{1}{2}
\end{array}%
\right],\notag \\
\langle\mathbf{L\cdot S}_{d}\rangle_{J=\frac{5}{2}} &=&1,\quad
\langle\mathbf{L\cdot S}_{Q}\rangle_{J=\frac{5}{2}} =\frac{1}{2},\quad
\langle S_{12}\rangle_{J=\frac{5}{2}} =-\frac{1}{5},\quad
\langle\mathbf{S}_{d}\cdot \mathbf{S}_{Q}\rangle_{J=\frac{5}{2}} =\frac{1}{2}.\quad
\label{In0}
\end{eqnarray}
The matrix forms of these mass shifts are
\begin{eqnarray}
\Delta \mathcal{M}_{J=1/2}&=&\left[
\begin{array}{cc}
\frac{1}{3}(a_{2}-4a_{1}) & \frac{\sqrt{2}}{3}(a_{2}-a_{1})+\frac{b_{1}}{\sqrt{2}%
} \\
\frac{\sqrt{2}}{3}(a_{2}-a_{1})+\frac{b_{1}}{\sqrt{2}} & -\frac{5}{3}(a_{1}+%
\frac{1}{2}a_{2})-b_{1}%
\end{array}%
\right]
+\left[
\begin{array}{cc}
-c_{1} & 0 \\
0 & \frac{1}{2}c_{1}%
\end{array}%
\right] , \notag \\
\Delta \mathcal{M}_{J=3/2}&=&\left[
\begin{array}{cc}
\frac{2}{3}a_{1}-\frac{1}{6}a_{2} & \frac{\sqrt{5}}{3}(a_{2}-a_{1})-\frac{b_{1}}{%
2\sqrt{5}} \\
\frac{\sqrt{5}}{3}(a_{2}-a_{1})-\frac{b_{1}}{2\sqrt{5}} & -\frac{1}{3}%
(2a_{1}+a_{2})+\frac{4b_{1}}{5}%
\end{array}%
\right]+\left[
\begin{array}{cc}
-c_{1} & 0 \\
0 & \frac{1}{2}c_{1}%
\end{array}%
\right] , \notag \\
\Delta \mathcal{M}_{J=5/2}&=&a_{1}+\frac{1}{2}a_{2}-\frac{b_{1}}{5}+\frac{c_{1}}{2}.
\label{M5}
\end{eqnarray}
Diagonalizing the matrices Eq. (\ref{M5}), one can compute the mass shifts $\Delta M(J,j)$ with the total angular momentum $\mathbf{J}$ and the total light-quark angular momentum $\mathbf{j}=\mathbf{L}+\mathbf{S}_{d}$, where ${S}_{d}=1$ is the spin of the diquark, so $j=0, 1, 2$,
\begin{eqnarray}
\Delta M(1/2,0)&=&\frac{1}{4}\left(-6a_{1}-a_{2}-2b_{1}-\sqrt{\Delta_{1}(a_{1},a_{2},b_{1})}\right)+c_{1}\Delta _{3}^{+}(a_{1},a_{2},b_{1}),  \notag \\
\Delta M(1/2,1)&=&\frac{1}{4}\left(-6a_{1}-a_{2}-2b_{1}+\sqrt{\Delta_{1}(a_{1},a_{2},b_{1})}\right)+c_{1}\Delta _{3}^{+}(a_{1},a_{2},b_{1}),  \notag \\
\Delta M(3/2,1)&=&\frac{1}{20}\left(-5a_{2}+8b_{1}-\sqrt{\Delta_{2} (a_{1},a_{2},b_{1})}\right)+c_{1}\Delta
_{4}^{+}(a_{1},a_{2},b_{1}),  \notag \\
\Delta M(3/2,2)&=&\frac{1}{20}\left(-5a_{2}+8b_{1}+\sqrt{\Delta_{2} (a_{1},a_{2},b_{1})}\right)+c_{1}\Delta
_{4}^{+}(a_{1},a_{2},b_{1}),  \notag \\
\Delta M(5/2,2)&=&a_{1}+\frac{a_{2}}{2}-\frac{b_{1}}{5}+\frac{c_{1}}{2},  \label{MM111}
\end{eqnarray}%
where six functions $\Delta _{1,2}(a_{1}, a_{2}, b_{1})$, $\Delta _{3}^{\pm }(a_{1}, a_{2}, b_{1})$ and $\Delta _{4}^{\pm }(a_{1}, a_{2}, b_{1})$ are defined by
\begin{eqnarray}
\Delta_{1}(a_{1},a_{2},b_{1})&=&4(a_{1})^{2}-8a_{1}b_{1}+12(b_{1})^{2}-4a_{1}a_{2}+20b_{1}a_{1}+9(a_{2}) ^{2}, \notag \\
\Delta_{2}(a_{1},a_{2},b_{1})&=&400(a_{1})^{2}-80a_{1}b_{1}+84(b_{1})^{2}-400a_{1}a_{2}-160b_{1}a_{1}+225(a_{2}) ^{2}, \notag\\
\Delta_{3}^{+}(a_{1},a_{2},b_{1})&=&\frac{4-\left(-2-\frac{7a_{2}}{a_{1}}-\frac{6b_{1}}{a_{1}}+\frac{3}{a_{1}}\sqrt{\Delta_{1}(a_{1},a_{2},b_{1})}\right)^{2}/(-2+\frac{2a_{2}}{a_{1}}+\frac{3b_{1}}{a_{1}})^{2}}
{8+\left(-2-\frac{7a_{2}}{a_{1}}-\frac{6b_{1}}{a_{1}}+\frac{3}{a_{1}}\sqrt{\Delta_{1}(a_{1},a_{2},b_{1})}\right)^{2}/(-2+\frac{2a_{2}}{a_{1}}+\frac{3b_{1}}{a_{1}})^{2}}, \notag\\
\Delta _{3}^{-}(a_{1},a_{2},b_{1})&=&\Delta _{3}^{+}\left( \sqrt{\Delta _{1}}%
\rightarrow -\sqrt{\Delta _{1}}\right), \notag\\
\Delta_{4}^{+}(a_{1},a_{2},b_{1})&=&\frac{10-\left(40+\frac{5a_{2}}{a_{1}}-\frac{24b_{1}}{a_{1}}-\frac{3}{a_{1}}\sqrt{\Delta_{2}(a_{1},a_{2},b_{1})}\right)^{2}/(10-\frac{10a_{2}}{a_{1}}+\frac{3b_{1}}{a_{1}})^{2}}
{20+\left(40+\frac{5a_{2}}{a_{1}}-\frac{24b_{1}}{a_{1}}-\frac{3}{a_{1}}\sqrt{\Delta_{2}(a_{1},a_{2},b_{1})}\right)^{2}/(10-\frac{10a_{2}}{a_{1}}+\frac{3b_{1}}{a_{1}})^{2}}, \notag\\
\Delta _{4}^{-}(a_{1},a_{2},b_{1})&=&\Delta _{4}^{+}\left( \sqrt{\Delta _{2}}%
\rightarrow -\sqrt{\Delta _{2}}\right) ,%
\end{eqnarray}
with $\Delta _{3,4}^{-}(a_{1}, a_{2}, b_{1})$ obtained from $\Delta_{3,4}^{+}(a_{1}, a_{2}, b_{1})$ by merely replacing $\sqrt{\Delta _{1,2}}\rightarrow -\sqrt{\Delta _{1,2}}$. The mass spectra of the $P$-wave states for the baryons are
\begin{eqnarray}
M(1/2,0)=\bar M+\Delta M(1/2,0), \notag \\
M(1/2,1)=\bar M+\Delta M(1/2,1), \notag \\
M(3/2,1)=\bar M+\Delta M(3/2,1), \notag \\
M(3/2,2)=\bar M+\Delta M(3/2,2), \notag \\
M(5/2,2)=\bar M+\Delta M(5/2,2). \label{MMM111}
\end{eqnarray}

\section{$D$-wave}

For analyzing the $D$-wave system, the diquark spin $S_{d}=1$ can be coupled with the heavy quark spin $S_{Q}=1/2$ to determine the total spin $S=1/2, 3/2$. Coupling of the orbital angular momentum $L=2$ give six states with the total spin $J=1/2, 3/2, 5/2$ or $3/2, 5/2, 7/2$ with positive parity $P=+1$. The relevant linear combinations of six basis states are
\begin{eqnarray}
|^{4} D_{1/2},J_{3}&=&1/2\rangle =\frac{1}{\sqrt{10}}|1,\frac{1}{2},-1\rangle-\frac{1}{\sqrt{15}}|1,-\frac{1}{2},0\rangle-\sqrt{\frac{2}{15}}|0,\frac{1}{2},0\rangle+\frac{1}{\sqrt{5}}|0,-\frac{1}{2},1\rangle+\frac{1}{\sqrt{10}}|-1,\frac{1}{2},1\rangle \notag \\
&-&\sqrt{\frac{2}{5}}|-1,-\frac{1}{2},2\rangle,
\notag \\
|^{2} D_{3/2},J_{3}&=&3/2\rangle =\sqrt{\frac{2}{15}}|1,-\frac{1}{2},1\rangle-\frac{1}{\sqrt{15}}|0,\frac{1}{2},1\rangle-\frac{2}{\sqrt{15}}|0,-\frac{1}{2},2\rangle+\sqrt{\frac{8}{15}}|-1,\frac{1}{2},2\rangle,
\notag \\
|^{4} D_{3/2},J_{3}&=&3/2\rangle =\frac{1}{\sqrt{5}}|1,\frac{1}{2},0\rangle-\sqrt{\frac{2}{15}}|1,\frac{1}{2},1\rangle-\frac{2}{\sqrt{15}}|0,\frac{1}{2},1\rangle+\frac{2}{\sqrt{15}}|0,-\frac{1}{2},2\rangle+\sqrt{\frac{2}{15}}|-1,\frac{1}{2},2\rangle,
\notag \\
|^{2} D_{5/2},J_{3}&=&5/2\rangle =\sqrt{\frac{2}{3}}|1,-\frac{1}{2},2\rangle-\sqrt{\frac{1}{3}}|0,\frac{1}{2},2\rangle,
\notag \\
|^{4} D_{5/2},J_{3}&=&5/2\rangle =\frac{3}{\sqrt{21}}|1,\frac{1}{2},1\rangle-\frac{2}{\sqrt{21}}|1,-\frac{1}{2},2\rangle-\frac{2\sqrt{2}}{\sqrt{21}}|0,\frac{1}{2},2\rangle,
\notag \\
|^{4} D_{7/2},J_{3}&=&7/2\rangle =|1,\frac{1}{2},2\rangle. \label{VV}
\end{eqnarray}%
The expectation values of $\langle\mathbf{L}\cdot \mathbf{S}_{i}\rangle$ $(i=Q, d)$, $\langle S_{12}\rangle$ and $\langle\mathbf{S}_{d}\cdot \mathbf{S}_{Q}\rangle$ are
\begin{eqnarray}
\langle\mathbf{L\cdot S}_{d}\rangle_{J=\frac{1}{2}} &=&-3,\quad
\langle\mathbf{L\cdot S}_{Q}\rangle_{J=\frac{1}{2}} =-\frac{3}{2},\quad
\langle S_{12}\rangle_{J=\frac{1}{2}} =-1,\quad
\langle\mathbf{S}_{d}\cdot \mathbf{S}_{Q}\rangle_{J=\frac{1}{2}} =\frac{1}{2},\quad \notag \\
\langle\mathbf{L\cdot S}_{d}\rangle_{J=\frac{3}{2}} &=&\left[
\begin{array}{cc}
-2 & -1 \\-1 & -2
\end{array}%
\right],
\langle\mathbf{L\cdot S}_{Q}\rangle_{J=\frac{3}{2}} =\left[
\begin{array}{cc}
\frac{1}{2} & 1 \\ 1 &  -1
\end{array}%
\right],
\langle S_{12}\rangle_{J=\frac{3}{2}} =\left[
\begin{array}{cc}
0 &\frac{1}{2} \\ \frac{1}{2} &0%
\end{array}%
\right],\notag \\
&&\langle\mathbf{S}_{d}\cdot \mathbf{S}_{Q}\rangle_{J=\frac{3}{2}} =\left[
\begin{array}{cc}
-1 &0 \\ 0 & \frac{1}{2}
\end{array}%
\right],\notag \\
\langle\mathbf{L\cdot S}_{d}\rangle_{J=\frac{5}{2}} &=&\left[
\begin{array}{cc}
\frac{4}{3} &-\frac{\sqrt{14}}{3} \\ -\frac{\sqrt{14}}{3} & -\frac{1}{3}%
\end{array}%
\right],
\langle\mathbf{L\cdot S}_{Q}\rangle_{J=\frac{5}{2}} =\left[
\begin{array}{cc}
-\frac{1}{3} &\frac{\sqrt{14}}{3} \\ \frac{\sqrt{14}}{3} & -\frac{1}{6}%
\end{array}%
\right],
\langle S_{12}\rangle_{J=\frac{5}{2}} =\left[
\begin{array}{cc}
0 &-\frac{\sqrt{14}}{14} \\ -\frac{\sqrt{14}}{14} & \frac{5}{7}%
\end{array}%
\right],\notag\\
&&\langle\mathbf{S}_{d}\cdot \mathbf{S}_{Q}\rangle_{J=\frac{5}{2}} =\left[
\begin{array}{cc}
-1 & 0 \\ 0 & \frac{1}{2}
\end{array}%
\right],\notag \\
\langle\mathbf{L\cdot S}_{d}\rangle_{J=\frac{7}{2}} &=&2,\quad
\langle\mathbf{L\cdot S}_{Q}\rangle_{J=\frac{7}{2}} =1,\quad
\langle S_{12}\rangle_{J=\frac{7}{2}} =-\frac{2}{7},\quad
\langle\mathbf{S}_{d}\cdot \mathbf{S}_{Q}\rangle_{J=\frac{7}{2}} =\frac{1}{2}.\quad
\label{In0}
\end{eqnarray}%
The matrix forms of these mass shifts are
\begin{eqnarray}
\Delta \mathcal{M}_{J=1/2}&=&-3a_{1}-\frac{3a_{2}}{2} -
b_{1}+\frac{c_{1}}{2},    \notag \\
\Delta \mathcal{M}_{J=3/2} &=&\left[
\begin{array}{cc}
-2a_{1}+\frac{1}{2}a_{2} & -a_{1}+a_{2}+\frac{1}{2}b_{1} \\
-a_{1}+a_{2}+\frac{1}{2}b_{1} & -2a_{1}-a_{2}%
\end{array}%
\right]
+\left[
\begin{array}{cc}
-c_{1} & 0 \\
0 & \frac{1}{2}c_{1}%
\end{array}%
\right] ,\notag \\
\Delta \mathcal{M}_{J=5/2} &=&\left[
\begin{array}{cc}
\frac{4}{3}a_{1}-\frac{1}{3}a_{2} & -\frac{\sqrt{14}}{3}a_{1}+\frac{\sqrt{14}}{3}a_{2}-\frac{\sqrt{14}}{14}b_{1} \\
-\frac{\sqrt{14}}{3}a_{1}+\frac{\sqrt{14}}{3}a_{2}-\frac{\sqrt{14}}{14}b_{1} & -\frac{1}{3}a_{1}-\frac{1}{6}a_{2}+\frac{5}{7}b_{1}
\end{array}%
\right]
+\left[
\begin{array}{cc}
-c_{1} & 0 \\
0 & \frac{1}{2}c_{1}%
\end{array}%
\right], \notag \\ \label{M3}
\Delta \mathcal{M}_{J=7/2}&=&2a_{1}+a_{2}-\frac{2}{7}b_{1}+\frac{1}{2}c_{1}.
\label{Mm5}
\end{eqnarray}%
Diagonalizing the matrices Eq. (\ref{Mm5}), one can compute six mass shifts $\Delta M(J, j)$, where ${S}_{d}=1$ is the spin of the diquark, so $j=1, 2, 3$,
\begin{eqnarray}
\Delta M(1/2,1)&=&-3a_{1}-\frac{3a_{2}}{2}-b_{1}+\frac{c_{1}}{2},    \notag \\
\Delta M(3/2,1)&=&\frac{1}{4}\left(-8a_{1}-a_{2}-\sqrt{\Theta_{1}(a_{1},a_{2},b_{1})}\right)+c_{1}\Theta _{3}^{+}(a_{1},a_{2},b_{1}),  \notag \\
\Delta M(3/2,2)&=&\frac{1}{4}\left(-8a_{1}-a_{2}+\sqrt{\Theta_{1}(a_{1},a_{2},b_{1})}\right)+c_{1}\Theta
_{3}^{-}(a_{1},a_{2},b_{1}),  \notag \\
\Delta M(5/2,2)&=&\frac{1}{28}\left(14a_{1}-7a_{2}+10b_{1}-\sqrt{\Theta_{2} (a_{1},a_{2},b_{1})}\right)+c_{1}\Theta
_{4}^{+}(a_{1},a_{2},b_{1}),  \notag \\
\Delta M(5/2,3)&=&\frac{1}{28}\left(14a_{1}-7a_{2}+10b_{1}+\sqrt{\Theta_{2}(a_{1},a_{2},b_{1})}\right)+c_{1}\Theta
_{4}^{-}(a_{1},a_{2},b_{1}),  \notag \\
\Delta M(7/2,3)&=&2a_{1}+a_{2}-\frac{2}{7}b_{1}+\frac{c_{1}}{2},  \label{MM222}
\end{eqnarray}%
where six functions $\Theta _{1,2}(a_{1}, a_{2}, b_{1})$, $\Theta _{3}^{\pm }(a_{1}, a_{2}, b_{1})$ and $\Theta _{4}^{\pm }(a_{1}, a_{2}, b_{1})$ are defined by
\begin{eqnarray}
\Theta_{1}(a_{1},a_{2},b_{1})&=&16(a_{1})^{2}-32a_{1}a_{2}+25(a_{2})^{2}-16a_{1}b_{1}+16a_{2}b_{1}+4(b_{1})^{2}, \notag \\
\Theta_{2}(a_{1},a_{2},b_{1})&=&1764(a_{1})^{2}-2548a_{1}a_{2}+1225(a_{2})^{2}+56a_{1}b_{1}-476a_{2}b_{1}+156(b_{1})^{2}, \notag \\
\Theta_{3}^{+}(a_{1},a_{2},b_{1})&=&\frac{2-\left(\frac{3a_{2}}{a_{1}}-\frac{1}{a_{1}}\sqrt{\Theta_{1}(a_{1},a_{2},b_{1})}\right)^{2}/(2-\frac{2a_{2}}{a_{1}}-\frac{b_{1}}{a_{1}})^{2}}
{4+\left(\frac{3a_{2}}{a_{1}}-\frac{1}{a_{1}}\sqrt{\Theta_{1}(a_{1},a_{2},b_{1})}\right)^{2}/(2-\frac{2a_{2}}{a_{1}}-\frac{b_{1}}{a_{1}})^{2}}, \notag\\
\Theta _{3}^{-}(a_{1},a_{2},b_{1})&=&\Theta _{3}^{+}\left( \sqrt{\Theta_{1}}%
\rightarrow -\sqrt{\Theta _{1}}\right), \notag\\
\Theta_{4}^{+}(a_{1},a_{2},b_{1})&=&\frac{28-\left(70-\frac{7a_{2}}{a_{1}}-\frac{30b_{1}}{a_{1}}-\frac{3}{a_{1}}\sqrt{\Theta_{2}(a_{1},a_{2},b_{1})}\right)^{2}/(2-\frac{2a_{2}}{a_{1}}-\frac{b_{1}}{a_{1}})^{2}}
{56+\left(70-\frac{7a_{2}}{a_{1}}-\frac{30b_{1}}{a_{1}}-\frac{3}{a_{1}}\sqrt{\Theta_{2}(a_{1},a_{2},b_{1})}\right)^{2}/(2-\frac{2a_{2}}{a_{1}}-\frac{b_{1}}{a_{1}})^{2}}, \notag\\
\Theta _{4}^{-}(a_{1},a_{2},b_{1})&=&\Theta _{4}^{+}\left( \sqrt{\Theta _{2}}%
\rightarrow -\sqrt{\Theta _{2}}\right) ,%
\end{eqnarray}
with $\Theta _{3,4}^{-}(a_{1}, a_{2}, b_{1})$ obtained from $\Theta_{3,4}^{+}(a_{1}, a_{2}, b_{1})$ by merely replacing $\sqrt{\Theta _{1,2}}\rightarrow -\sqrt{\Theta _{1,2}}$. The mass spectra of the $D$-wave states for the baryons are
\begin{eqnarray}
M(1/2,1)=\bar M+\Delta M(1/2,1), \notag \\
M(3/2,1)=\bar M+\Delta M(3/2,1), \notag \\
M(3/2,2)=\bar M+\Delta M(3/2,2), \notag \\
M(5/2,2)=\bar M+\Delta M(5/2,2), \notag \\
M(5/2,3)=\bar M+\Delta M(5/2,3), \notag \\
M(7/2,3)=\bar M+\Delta M(7/2,3).  \label{MMM222}
\end{eqnarray}

\end{document}